\documentclass[preprint]{aastex}

\slugcomment{ To appear in Ap. J. Suppl.}

\shortauthors{Vestergaard \& Wilkes} 
\shorttitle{Empirical UV Iron Emission Template from I\,Zw\,1}

\newcommand{\lya}{\ifmmode {\rm Ly\,}\alpha \, \else Ly\,$\alpha$\,\fi}
\newcommand{\cstar}{C\,{\sc iii}$^{\star}$}
\newcommand{\hi}{H\,{\sc i}}
\newcommand{\hb}{H\,$\beta$}
\newcommand{\ha}{H\,$\alpha$}

\newcommand{\nii}{N\,{\sc ii}]}
\newcommand{\niii}{N\,{\sc iii}]}
\newcommand{\niv}{N\,{\sc iv}}
\newcommand{\nv}{N\,{\sc v}}
\newcommand{\oi}{O\,{\sc i}}
\newcommand{\oii}{[O{\sc ii}]}
\newcommand{\oiii}{[O\,{\sc iii}]}
\newcommand{\oiiisf}{O\,{\sc iii}]}
\newcommand{\oiv}{O{\sc iv}]}
\newcommand{\ov}{O\,{\sc v}}

\newcommand{\siivoiv}{Si\,{\sc iv}+O\,{\sc iv}]}
\newcommand{\siiv}{Si\,{\sc iv}}
\newcommand{\civ}{C\,{\sc iv}}
\newcommand{\heii}{He\,{\sc ii}}
\newcommand{\aliii}{Al\,{\sc iii}}

\newcommand{\alii}{Al\,{\sc ii}}
\newcommand{\Alii}{Al\,{\sc ii}}

\newcommand{\siiii}{Si\,{\sc iii}}
\newcommand{\siii}{Si\,{\sc ii}}
\newcommand{\sii}{Si\,{\sc i}}
\newcommand{\ciii}{C\,{\sc iii}]}
\newcommand{\cii}{C\,{\sc ii}}
\newcommand{\ci}{C\,{\sc i}}

\newcommand{\caii}{Ca\,{\sc ii}}
\newcommand{\znii}{Zn\,{\sc ii}}

\newcommand{\mgii}{Mg\,{\sc ii}}
\newcommand{\mgi}{Mg\,{\sc i}}

\newcommand{\feii}{Fe\,{\sc ii}}
\newcommand{\feiii}{Fe\,{\sc iii}}
\newcommand{\fev}{Fe\,{\sc v}}
\newcommand{\lam}{$\lambda$}
\newcommand{\kms}{\ifmmode {\rm km\,s}^{-1} \else km\,s$^{-1}$\fi}
\newcommand{\cmcub}{\ifmmode {\rm cm}^{-3} \else cm$^{-3}$\fi}
\newcommand{\ergs}{\ifmmode {\rm ergs\,cm}^{-2}\,{\rm s}^{-1} \else ergs\,cm$^{-
2}$\,s$^{-1}$\fi}
\newcommand{\ergsA}{\ifmmode {\rm ergs\,cm}^{-2}\,{\rm s}^{-1}\,{\rm \AA }^{-1} 
\else ergs\,cm$^{-2}$\,s$^{-1}$\,\AA $^{-1}$\fi}
\newcommand{\ergsHz}{\ifmmode {\rm ergs\,cm}^{-2}\,{\rm s}^{-1}\,{\rm Hz }^{-1} 
\else ergs\,cm$^{-2}$\,s$^{-1}$\,Hz $^{-1}$\fi}
\newcommand{\qso}{quasar}
\newcommand{\qsos}{quasars}
\newcommand{\et}{\mbox{et~al.}\ }
\newcommand{\ie}{\mbox{i.~e.,}\ }
\newcommand{\eg}{\mbox{e.~g.,}\ }

\newcommand{\Eg}{\mbox{E.~g.,}\ }
\newcommand{\HST}{{\it HST}}
\newcommand{\izw}{I\,Zw\,1}
\newcommand{\wav}{wavelength}
\newcommand{\wavs}{wavelengths}
\newcommand{\lsim}{\stackrel{\scriptscriptstyle <}{\scriptstyle {}_\sim}}
\newcommand{\gsim}{\stackrel{\scriptscriptstyle >}{\scriptstyle {}_\sim}}

\begin{document}

\title{An Empirical Ultraviolet Template for Iron Emission in Quasars as Derived 
from I\,Zw\,1\footnote{Based on observations made with the NASA/ESA {\it Hubble 
Space Telescope}, obtained from the data archive at the Space Telescope Science 
Institute.  STScI is operated by the Association of Universities for Research in
Astronomy, Inc. under the NASA contract NAS 5-26555. }}

\author{M. Vestergaard\altaffilmark{2,3,4} and B.J. Wilkes\altaffilmark{2}}

\altaffiltext{2}{Harvard-Smithsonian Center for Astrophysics, 60 Garden Street,
 	Cambridge, MA 02138}
\altaffiltext{3}{The Niels Bohr Institute for Astronomy, Physics and Geophysics,
	Copenhagen University Observatory, Juliane Maries Vej 30, DK-2100 
	Copenhagen \O, Denmark }
\altaffiltext{4}{Current address: Department of Astronomy, The Ohio State
	University, 140 West 18th Avenue, Columbus, OH 43210-1173. 
	Email: vester@astronomy.ohio-state.edu}


\begin{abstract}
We present an empirical template spectrum suitable for fitting and 
subtracting/studying the \feii\ and \feiii\ emission lines  in the 
restframe ultraviolet 
spectra of quasars and active galatic nuclei, the first empirical ultraviolet 
iron template to cover the full range of $\lambda$1250$-$3090\AA . 
Iron emission is often a severe contaminant in optical--ultraviolet spectra of
active galactic nuclei and \qsos. Its presence complicates and limits the
accuracy of measurements
of both strong and weak emission lines and the continuum emission, affecting
studies of line and continuum interrelations, the ionization
structure, and elemental abundances in active galaxies and \qsos.
Despite the wealth of work on modeling the \qso\ \feii\ emission and the
need to account for this emission in observed \qso\ spectra,
there is no ultraviolet template electronically available to aid this process.
The iron template we present is based on {\it Hubble Space Telescope} spectra 
of the Narrow Line Seyfert\,1 galaxy, 
I\,Zwicky\,1 (I\,Zw\,1, $z$\,=\,0.061). The intrinsic narrow lines ($\gsim$ 900 
\kms ) of this source and its rich iron spectrum make the template particularly 
suitable for use with most active galactic nuclei and quasar spectra.

The iron emission spectrum, the line identifications, and the measurements of 
absorption and emission lines are presented and compared with the work of
Laor et~al.  Comments on each individual line feature and the line
fitting are available in the Appendix.
The methods used to develop and apply the template are also described.
We illustrate the application of the derived \feii\ and \feiii\ templates by
fitting and subtracting iron emission from the spectra of four high redshift 
quasars and of the nearby quasar, 3C273, confirming their 
general applicability to active galaxies despite the somewhat unusual properties
of \izw. We briefly discuss the small discrepancies between the observed iron 
emission of these quasars and the ultraviolet template, and compare the 
template with previously published ones.
We discuss the advantages and limitations of the UV \feii\ and \feiii\ 
templates and of the template fitting method.
We conclude that the templates work sufficiently well to be a valuable
and important tool for eliminating and studying the iron emission in active
galaxies, at least until accurate theoretical iron emission models are 
developed.

The \siivoiv \,\lam \,1400 feature in \izw{} is clearly strong relative to 
\civ \,\lam \,1549, and \civ\ and \ciii \,\lam \,1909 are both relatively weak. 
This may partially be due to the higher densities and lower ionization parameter 
prevailing in Narrow Line Seyfert\,1 galaxies, and to the big blue bump 
shifting towards lower energies in more luminous Seyferts, such as \izw .
In \izw\ the narrow line width reveals that \ciii\ is heavily blended with 
\siiii ]\,\lam \,1892, \aliii \,\lam \lam \,1854,1863, and \feiii\ transitions. 
This suggests that the \ciii\ line strength and width may be overestimated in 
many quasar line studies where the lines are broader and deblending is not
possible. This affects density 
estimates of the broad line region.  Photoionization modeling, including all
these line features, and subsequent fitting to the spectra are required to 
estimate the true \ciii\ strength.
We also argue, based on earlier work, that (strong) iron emission may be 
connected with high densities and associated with outflows.

\end{abstract}



%

\section{Introduction} 

Quasar\footnote{``Quasars'' here refers to both the radio-loud and radio-quiet 
subgroups} ultraviolet (hereafter UV) spectra characteristically contain broad 
emission lines originating in the central ($\lsim$ light year) region, the    
broad line region (BLR). The strongest lines are (the resonance lines)
\lya \,\lam \,1216, \siivoiv \lam \,1400, \civ \,\lam \lam \,1548,\,1551, \ciii 
\,\lam \,1909, and \mgii \,\lam \lam \,2796, \,2803.  For a number of years after 
the discovery of \qsos\ these lines and a few iron transitions (Greenstein \& 
Schmidt 1964; Wampler \& Oke 1967) were the only lines detected in their 
UV spectra.  With the availability of high signal-to-noise ratio (S/N), high 
resolution spectra, and in particular data taken with {\it Hopkins Ultraviolet 
Telescope} (\eg Kriss \et 1992; Zheng \et 1995; Zheng, Kriss, \& Davidsen 1996), 
{\it Hubble Space Telescope} (\HST) (\eg Laor \et 1994, 1995; Zheng \et 1997; 
Kriss \et 2000; Kraemer \& Crenshaw 2000), and the Keck telescope (\eg Tran, 
Cohen \& Goodrich 1995; Brotherton \et 1997; Barlow \& Sargent 
1997; Larkin \et 2000; Carson \et 2000) 
it has become obvious that \qso\ spectra contain a plethora of weak lines as well.

\subsection{The Need for Fitting and Removal of Iron Emission}
In order to study \eg the relationship between the continuum and the line
emission, the dynamical structure, the ionization balance and structure, or the
chemical abundances in active galactic nuclei (herafter AGNs) and \qsos\ it is 
important to make reliable measurements of the continuum emission, 
the strong broad lines including the wings of their profiles, and the weak 
emission lines.
In addition to hydrogen, helium and the above-mentioned elements, AGN and \qsos\
also contain atomic and ionic iron, the stable end product of
nucleosynthesis. Due to the large number of electron levels in iron atoms, 
thousands of emission line transitions are distributed throughout the UV and optical 
spectral regions. The weak lines, iron as well as those {\it not} being iron
transitions (hereafter `non-iron' lines), blend together, 
partly due to the associated transitions being very close or overlapping in 
energy, and partly due to the high (presumed) dynamic velocities of the 
broad-line emitting clouds broadening the lines.
Line widths [FWHM, the full width at half peak flux] of up to 
10\,000 \kms\ have been measured (Wilkes 2000). This heavy blending of weaker 
lines, dominated by iron transitions, forms a {\em pseudo-continuum} above the 
intrinsically emitted continuum (see \eg Figs.~\ref{FeCleanfig3C273} and
\ref{FeCleanfigs}) even in high S/N \qso\ spectra (Wills \et 1985; Boroson 
\& Green 1992, hereafter BG92; Wills \et 1995). 
This pseudo-continuum severely complicates the study of both weak non-iron 
features, particularly important for abundance and ionization studies, and the 
wings of strong resonance lines.
The uncertainty introduced into broad emission-line measurements by iron 
emission ``contamination'' can be relatively large: excluding very weak lines 
where it can reach 
$\sim$100\,\%, a rough estimate is 5$-$50\% based on Wilkes (1984) and our own 
line measurements before and after iron emission removal using the template 
presented here (\S~\ref{temp_applic}).  The exact level of uncertainty depends
on the line transition, the line parameter measured and the overall strength of 
the emission line spectrum.  Moreover, the relative strengths of the iron features vary 
greatly from object to object.
These uncertainties are dominated by the difficulty in determining an accurate 
continuum level, one of the main problems in \qso\ line studies, although such 
errors are seldom quoted. Combined, these uncertainties provide a strong 
argument that all \qso\ spectral studies should include iron emission fitting 
and removal. 

These limitations imposed by the contaminating iron emission 
have been known for some time (\eg Wills \& Browne 1986).  A possible 
solution to the continuum level uncertainty is to use only wavelength
ranges which contain pure continuum emission (so-called {\em continuum
windows}; Francis \et 1991) when fitting the continuum. However, few 
such regions exist and they are generally small, especially in the UV 
spectral region, $\sim$\,1000 $-$ 2000\AA . Moreover, in objects with very broad
lines even these line-free regions may be contaminated by broad line-wing 
emission (Boroson, Persson \& Oke 1985; Wills \& Browne 1986). 

\subsection{\bf Earlier Work on Iron Emission \label{earlierFewrk}}

Allowance for the optical and UV iron emission has not extensively been made,
inspite of the increasing recognition of its importance for broad line studies. 
It was rendered a very hard task in the 1980's by the lack of sufficient atomic 
data to allow identification of the iron transitions and high spectral 
resolution to resolve them. 
Earlier work includes identifying the UV iron emission lines in AGN 
[\eg Wills, Netzer \& Wills 1980; Penston 1980; Penston \et 1983 (in stars); 
Hartig \& Baldwin 1986, Johansson \& Jordan 1984]
and modeling this emission (\eg Netzer 1980; Kwan \& Krolik 1981; Netzer \& 
Wills 1983; Wills \et 1985; Collin-Souffrin \et 1986; Penston 1987; 
Collin-Souffrin, Hameury \& Joly 1988; Krolik \& Kallman 1988; Ferland \& Persson 1989;
Dumont \& Collin-Souffrin 1990;  Netzer 1990, and references therein).
Despite extensive efforts the current theoretical models of the optical and UV 
iron emission cannot fully reproduce and explain the observations. Fortunately, 
projects are under way to compute detailed radiative transition 
probabilities of iron atoms and ions (the IRON Project: Hummer \et 1993; 
Nahar, Bautista,\& Pradhan 1997; Nahar et~al.\ 2000, and refererences therein) 
and compute photoionization models of 
AGN which also take the iron emission into account with or without exact
radiative transfer (Sigut \& Pradhan 1998; ``The Kentucky group'': 
Verner \et 1999, and references therein) based on improved atomic data (\eg the 
Opacity Project: Seaton \et 1994). 
Other studies presenting iron line lists and radiative transition probabilities,
which are also used here, include Fuhr, Martin \& Wiese (1988), Ekberg (1993), 
Giridhar \& Arellano Ferro (1995), Nahar (1995), Nahar \& Pradhan (1996), 
Quinet (1996), Quinet, Le Dourneuf, \& Zeippen (1996), and Kurucz \& Bell (1995).
These iron line lists are available on the world wide web.

\subsection{\bf Iron Emission Correction \label{FeCorrection}}

Phillips (1977) was one of the first to compare the spectrum of \izw\ to other 
Seyfert galaxy spectra by broadening the former 
with a Gaussian profile, selected to match the line widths in the latter. 
Phillips did not subtract the iron emission
but confirmed his suspicion that the \feii\ and 
H\,{\sc i} lines have essentially the same widths and profiles. 
Wills \et (1985) simulated the \feii\ emission and absorption in
eight low to intermediate redshift \qsos\ using photoionization modeling of
$\sim$3000 iron lines. In spite of the limitations imposed by the atomic data 
and computing speed at the time, they were able to reproduce the UV-optical 
spectra reasonably well. Some deviations from the observed spectra are, however,
seen (see \S~\ref{template_comp}). 
The photoionization models presented by Wills \et are specific to their 
individual object spectra and were not generated with a general subtraction of
iron emission in AGNs in mind, where a typical iron spectrum is more 
appropriate. 
Boroson, Persson \& Oke (1985) made one of the first attempts to take the 
effects of the iron emission on broad-line studies into direct account.
They estimated the uncertainty associated with line (equivalent width, EW) measurements as a 
function of line widths by broadening an AGN spectrum by a Hanning profile 
(triangular) with a range of widths. They found the measured EWs to weaken with 
increasing FWHM because the continuum level becomes systematically 
overestimated, as the iron lines broaden and blend to form a pseudo-continuum. 
As a consequence, narrow lines incorrectly appear stronger.
This may explain why \qsos , with their broad emission lines were not
recognized early on to be strong iron emitters\footnote{The fact that 
radio-loud \qsos\ were often studied more than radio-quiets, due to their 
easily-detectable, strong radio emission, may also have contributed, because 
radio-quiet \qsos\ show relatively stronger optical iron emission (Peterson, 
Foltz \& Byard 1981; Bergeron \& Kunth 1984; Corbin 1997). } 
(Phillips 1977; Davidson \& Netzer 1979).
One way of correcting for the iron emission is to model it using the profile
information of the \hb\ line, as the isolated \feii\ and \hb\ profiles are 
observed to have both similar widths and profile shapes (Phillips 1977; Laor 
\et 1997b; hereafter L97).  
Thus a synthetic spectrum of the iron emission can be constructed 
by shifting and scaling \hb\ profile templates according to a list of iron line 
positions and relative strengths. However, 
this method depends highly on the atomic data lists to be representative, in 
terms of accuracy and completeness, of the iron emission in AGN and \qsos .  
As mentioned, theoretical studies are not yet able to fully reproduce the 
observed iron emission, but work is in progress (D.\ Verner 1998, private 
communication; Verner \et 1999; see also \S~\ref{template_comp}).

A significant improvement in the accuracy of broad emission line measurements can be obtained 
by using an observed AGN\footnote{The formal distinction between the use of the 
term `AGN' and `\qso ' is only that of the object's luminosity, with \qsos\ 
occupying the most powerful end of the luminosity range of active galactic 
nuclei (M$_V < -$23, V\'eron-Cetty \& V\'eron 1993). As the methods and iron 
templates described here are applicable to AGNs as well, this fact is silently 
assumed in the following, when we refer to \qsos\ only, in addressing the use 
of the template. }   
or \qso\ iron emission template to fit and subtract the iron emission in 
\qso\ spectra before performing line and continuum measurements. The 
benefits of using such a template, containing all the iron transitions 
typically present in \qsos , are manifold. 
Combined with the much improved atomic data, becoming increasingly available 
(\S~\ref{earlierFewrk}), empirical templates provide a powerful tool to study
the observed iron spectrum in terms of theoretical models.
This template method was adopted by BG92, who successfully used an 
optical (4250$-$7000\AA ) iron template based on the nearby, narrow-line Seyfert
1 (NLS1; Osterbrock \& Pogge 1985), \izw , on the Bright Quasar Survey sample. 
Their method is described in \S~\ref{application}.  Corbin \& Boroson (1996) 
use the same method with a 2300$-$3000\AA\ UV iron template of \izw.  
See \S~\ref{template_comp} for a comparison with this template.

The advent of {\it HST} UV \qso\ spectral data allows us to extend the template
method into the UV regime.  We here present, to our knowledge, the first high 
S/N, high resolution, \qso\ empirical UV iron template spectrum ranging from 
restframe \lam \lam 1250 to 3090\AA\ which is applicable to \qso{} data.  
The template is based on {\it HST} (archival) data of \izw\ (L97).  
The method with which the template was generated is an extension of that of 
BG92 and Corbin \& Boroson (1996) including a more detailed 
fitting of the lines which are not iron (see \S~\ref{fetemplt} for details and 
\S~\ref{template_comp} for a discussion).
When optical data between 3000\AA\ and 4250\AA\ become available (L97
present 3000$-$3800\AA\ data), suitable for production of an intermediate wavelength
iron template, the available templates will provide a powerful
tool for consistent modeling of the UV and optical iron emission throughout the
1250\AA\ to 7000\AA\ region of \qso\ spectra.

The procedure of fitting the iron emission using an empirical template assumes
that (1) the iron spectrum of \izw\ is representative of \qso\ iron emission 
spectra and that (2) all \qsos\ have similar iron spectra to within a scaling 
factor and/or a line profile broadening. Not enough is currently known about the
iron emission in AGN to firmly assess the validity of these assumptions. Though 
until theoretical models can explain the observed iron emission more confidently
this template fitting procedure is the best available approach. We test the
method by fitting several representative \qso\ spectra 
(\S~\ref{sample_Fecleaning}) showing that the iron emission 
can be well fitted if allowed the freedom to vary some of the multiplet ratios 
in the empirical template, indicating that the basic assumptions are valid.

The iron emission templates have importance not only for our ability to 
fit and subtract the iron emission in \qso\ spectra, but also as tools 
with which we can study the iron emission strengths themselves. Iron 
is a key coolant emitting $\sim25$\% of the total energy output from the 
BLR (Wills \et 1985; Boller,  Boller, Brandt \& Fink 1996), emphasizing the
importance of including the iron emission in studies of the BLR.

\subsection{The Narrow-Line Seyfert 1, \izw\ }

The spectra of NLS1 galaxies are particularly useful for generating 
empirical iron templates because their strong and rich iron emission 
allows detection of weak iron features and identification of as many iron 
transitions (Figs.~\ref{template} and~\ref{feids}, Table~\ref{fefits}, 
\S~\ref{lineids}) as are typically present in AGN spectra.  The relatively 
narrow width (FWHM\,$\lsim$\,2000\,\kms) of the broad emission lines 
permits the individual non-iron lines to be resolved, isolated, and removed 
from the spectrum. It also allows us to match the line width in most 
iron-contaminated AGN spectra so as to fit and subtract the iron emission 
by broadening the iron template. 

\izw\ (PG\,0050$+$124; $z$\,=\,0.061, see \S~\ref{dataproc}) is classified as a 
NLS1 and has the ``narrow'' broad emission-lines, strong \feii\ emission (\eg 
Phillips 1976; Osterbrock \& Pogge 1985; Lipari, Terlevich, \& Macchetto 1993; 
Pogge 2000; Rodr\'{\i}guez-Ardila, Pastoriza \& Donzelli 2000; Rudy \et 2000; 
but see also Gaskell 2000), steep soft X-ray spectra (Boller \et 1996), and strong
far-IR emission (Halpern \& Oke 1987), typical of this class. Their radio 
properties, however, are similar to those of other Seyfert galaxies (Ulvestad,
Antonucci \& Goodrich 1995).  They are also variable at optical and UV wavelengths
(\eg Zwicky 1971; Giannuzzo \& Stirpe 1996; Rodr\'{\i}guez-Pascual, Mas-Hesse, 
\& Santos-Lleo 1997; Leighly 1999; Miller \et 2000) and the fastest X-ray 
variable AGN known (see discussion by Boller \et 1996).
\izw\ is a well studied object thanks in part to its relative brightness
(Schmidt \& Green 1983), its exceptionally narrow lines, its iron-rich spectrum
(Sargent 1968; Phillips 1976, 1977; Oke \& Lauer 1979; Boroson et~al.\
1985; Halpern \& Oke 1987; BG92; L97), and its richness in low ionization lines 
(Persson \& McGregor 1985; van Groningen 1993; L97; Table~\ref{elids}). \izw\ is 
also an infrared (IR) luminous source (Rieke \& Low 1972; Rieke 1978; Halpern \& Oke 
1987) and has been observed in CO and H$_2$ molecular lines (Barvainis, Alloin \&
Antonucci 1989; Eckart \et 1994).  Some modeling of the observed optical iron 
emission was performed by Phillips (1978). 
\izw\ was observed by {\it IUE} in 1978$-$1982 (Courvoisier \& Paltani 1992; 
Lanzetta, Turnshek \& Sandoval 1993; Paltani \& Courvoisier 1994), and the source 
is included in a number of {\it IUE} studies addressing larger samples of AGN 
(\eg Wu, Boggess \& Gull 1983; Pian \& Treves 1993; Wang, Zhou \&
Gao 1996; Rodr\'{\i}guez-Pascual \et 1997).  Condon, Hutchings \& Gower (1985) 
detected 21\,cm radio emission from \izw\ consistent with emission from a 
late-type host galaxy. 

\izw\ is a particularly 
good choice for an empirical \feii\ template as it is so well studied, 
especially in terms of its optical iron emission 
and was used by BG92 and Corbin \& Boroson (1996) for their 
templates.

Additional candidates of narrow-line, iron-rich AGN suitable for use as iron 
templates may be found among other NLS1s. Possibilities include Mrk\,957 
(5C\,3.100), Ark\,564, 1E\,1226.9$+$1336, E1228$+$123, Mrk\,507 (1748+687), 
Mrk\,42 (1151+465), and the less extreme Mrk\,291 (1552+193), Mrk\,493 
(1557+352) and 1244+026.

\bigskip \noindent
The structure of this paper is as follows: 
\S~\ref{dataproc} addresses the data processing, 
\S~\ref{fetemplt} contains a description of the generation of the template and 
identification of the various spectral lines,
in \S~\ref{template_comp} comparisons are made with other available empirical UV
templates and synthetic \feii\ models, 
\S~\ref{temp_applic} describes the application of the template along with 
initial results, 
\S~\ref{izwidiscuss} is 
dedicated to a discussion of \izw\ and some of its spectral features, and 
\S~\ref{summary} summarizes the main conclusions. Comments on individual 
line features are deferred to Appendix~\ref{linecomments}.

\section{Data Processing \label{dataproc}}

The data presented here consist of {\it HST} Faint Object Spectrograph (FOS) 
archival spectra of \izw . The journal of observation and instrumental setup 
is summarized in Table~\ref{hstobs} for convenience (see also L97).
\placetable{hstobs}

The G130H spectrum was calibrated according to the standard CALFOS procedure 
with updated (1996 March) calibration files, as the flux 
calibration status of the archival data is uncertain\footnote{The data were 
skipped by the pipeline calibration, evident from the data intensity level and 
the missing flux calibration flags in the data headers.}. 
The pipeline calibrations were used for the G190H and G270H spectra.
Multiple observations obtained with the same grating were combined by weighting
with the exposure time to form a single spectrum for each wavelength region. 
A color excess, 
E(B$-$V)\,=\,$N_H /48\times10^{20} {\rm cm^{-2}}$\,=\,0.105\,mag, was 
determined based on the Galactic hydrogen column density, 
$N_H\,=\,5.05\,(\pm 0.1)\,\times 10^{20} {\rm cm^{-2}}$, 
observed\footnote{Note: Errors are not quoted by Stark \et (1992), but the 
applied correction technique for stray radiation was developed by Lockman, 
Jahoda \& McCammon (1986), who quote an error of $\sim$1\,$\times 10^{19} 
{\rm cm^{-2}}$. Using the same correction technique Elvis, Lockman \& Wilkes 
(1989) measure an H\,{\sc I} column density towards \izw\ of 
$N_H\,=\,5.07\,(\pm 0.1)\,\times 10^{20} {\rm cm^{-2}}$
consistent with the Stark \et measurement. 
The slight offset between the two $N_H$ measurements will not
significantly affect the reddening correction.} by Stark \et (1992). 
The spectra were dereddened using the average extinction curve presented by 
Cardelli, Clayton \& Mathis (1989), using
A$_V$\,=\,3.1 \,$\ast$\,E(B$-$V) 
and the IRAF\footnote{IRAF is distributed by the National Optical
Astronomy Observatories, which is operated by the Association of Universities
for Research in Astronomy, Inc.\ (AURA) under cooperative agreement with the
National Science Foundation.} (V2.11)
task `deredden'. Due to the average nature of this 
curve, some residual dust extinction features may be present in the spectra,
especially in the 2200\AA\ region (Fig.~\ref{emissmodel}), where characteristic 
dust extinction features are generally expected (\eg Osterbrock 1989).
Iron emission in this region complicates this identification. 
We corrected for an offset in the wavelength solution, due to the non-zero 
uncertainty in
the wavelength calibration (typical uncertainties are $\sim$0.25\AA\ in G130H, 
$\sim$0.37\AA\ in G190H, $\sim$0.52\AA\ in G270H; Leitherer 1995),
by comparing the observed wavelengths of the galactic interstellar medium (ISM)
absorption lines with their laboratory wavelengths. The absorption lines used are:
\siii \,\lam 1190, \siii \,\lam 1193, \siiii \,\lam 1206, \siii \,\lam 1260, 
\cii \,\lam 1335, \siii \,\lam 1527, \Alii \,\lam 1671, \feii \,\lam 2344, 
\feii \,\lam 2374, \feii \,\lam 2382, \feii \,\lam 2586, \feii \,\lam 2600, 
\mgii \,\lam \lam 2796,\,2803 and \mgi \,\lam 2853 (Table~\ref{alids}).
The applied wavelength offsets are listed in Table~\ref{hstobs}. These offsets, 
consistent with those applied by L97, result in the absorption line positions 
matching the laboratory wavelengths (\eg Savage \et 1993; Morton 1991) to within 
$\pm$0.3\AA{} (rms).

\placetable{hstobs}

The spectra from the individual gratings were co-added to produce 
a full 1141\AA\ -- 3278\AA\ (1075\AA\ -- 3090\AA\ restframe) spectrum. 
The spectra from gratings G190H and G270H were obtained the same day and 
show no difference in the continuum level, and so were co-added without 
scaling either spectrum, using the average flux in the overlapping region.  
The continuum-level in the G130H spectrum, taken 6 months earlier, is 
clearly shifted relative to that of G190H. The G130H spectrum normalization 
is described in \S~\ref{scaling}.
For consistency, the combined, final spectrum was rebinned\footnote{The 
rebinning is necessary in order to combine and simultaneously process the
G130H, G190H, and G270H spectra for generating the template. The resultant
resolution is sufficient for the purpose of generating and applying the 
template.} to match the dispersion of the G270H grating data, 
$\Delta \lambda $\,=\,0.511\AA /pix, the lowest dispersion available in the 
three gratings (Table~\ref{hstobs}).  The spectrum ranges from 1075.2 to 
3089.8\AA{}, has a dispersion of 0.482 \AA /pix and a resolution of 1.86\AA\ 
in the restframe (Leitherer 1995).  The \izw\ spectrum 
is shown in L97 and Figure~\ref{emissmodel} (dotted line).
A redshift of $z$\,=\,0.061, defined by the \mgii\ profile peak, is used
throughout this paper, and is consistent with measurements by Phillips (1976),
Schmidt \& Green (1983), Wu \et (1983), Persson \& McGregor (1985), van 
Groningen (1993), Condon \et (1985) and Smith \et (1997). Condon \et determine a
redshift of 0.061136 to an accuracy of 8$\times$10$^{-6}$ using H\,{\sc i} radio
measurements.

The spectral energy distributions of \qsos\ and AGNs (Elvis \et 1994) span 11 
orders of magnitude in frequency from X-rays to millimeter wavelengths (and to radio 
wavelengths for some objects; Weedman 1986; Peterson 1997).  Observations (\eg Oke,
Shields \& Korycansky 1984; Wills \et 1985)
suggest that the continuum at optical and UV \wavs\ can be approximated by a 
single power-law, F$_{\nu}\,\sim\,\rm \nu^{-\alpha_{\nu}}$.
We thus chose to fit a power-law continuum to the \izw\ restframe spectrum 
before any emission line features were fitted. After the initial completion of the iron 
template, which included careful fittings (\S~\ref{fitmethod}) of the emission 
lines which are not iron, it was realized that due to an STSDAS software 
bug\footnote{the relevant task, ``ADDNEWKEYS'', has now been updated by the
STScI {\it HST} helpdesk.} the archival G130H spectrum was unknowingly 
calibrated with the wrong (pre-costar) flux calibration files. These data were 
then recalibrated\footnote{This recalibration results in a flux level 
$\sim$9\,\% higher than that of L97, because the FOS calibration files were 
updated (1996 March) later than their data processing in 1994.} using the most 
recent (1996 March) calibration files. 
We then fitted a power-law continuum from 1075 to $\sim$1720\AA , independent 
of the power-law continuum at longer wavelengths\footnote{The emission and 
absorption models below 1700\AA\ were then regenerated and new iron templates 
(of \feii\ and \feiii\ emission) were constructed.}. 
The resulting continuum is a broken power-law with the break at 1716\AA: a 
blue continuum slope, $\alpha_{\nu}$\,=\,1.9 and normalization, 
F$_{\lambda}$(1500 \AA) = 3.45$\times$10$^{-14}$ \ergsA, and a red 
continuum slope, $\alpha_{\nu}$\,=\,1.0 and F$_{\lambda}$(1500\AA) = 
3.89$\times$10$^{-14}$ \ergsA. The choice of this continuum 
(Fig.~\ref{emissmodel}) does not affect the application of the template, as 
it will be scaled and broadened to match the target spectrum in the application 
process (\S~\ref{temp_applic}).  The continuum windows used in the fitting 
(\lam \lam 1312$-$1327, 1347$-$1353, 1641$-$1647, 1675$-$1690, 3007$-$3027) 
are specific to the \izw\ spectrum and are different from those suggested by 
Francis \et (1991), which were based on an average AGN spectrum.  
The spectrum of \izw\ is rich in low-ionization lines and weak emission features 
(Fig.~\ref{emissmodel}; L97) which contaminate the ``average'' continuum windows.

\subsection{The G130H Spectrum Scaling and \izw\ Variability \label{scaling}}

The variation in continuum level between the G130H and the G190H+G270H spectra 
taken six months apart is consistent with the known variability of \izw\ (\eg 
Zwicky 1971; Giannuzzo \& Stirpe 1996; Rodr\'{\i}guez-Pascual \et 1997; Leighly
1999).  The 
normalization of the recalibrated G130H spectrum to the G190H+G270H spectrum 
was determined as the ratio of the median flux in the overlapping region
of the spectra  (1.3$\pm$0.56). As the bluest $\sim$40\,\AA\ of the G190H 
spectrum shows
a relative error of 42\% as opposed to 4\% in G130H, the normalized G130H 
spectrum substituted that of G190H in this overlapping region.
However, the G190H flux uncertainty entirely dominates the formal error on
the normalization, and so useful line flux estimates cannot be 
deduced below 1500\,\AA.
In spite of the large formal error, a smooth spectrum and power-law continuum 
resulted in the \lam 1075\,--\,1716\,\AA\ region, implying that the actual
uncertainty is lower. 
As the template is best applied by subdividing the \lam 1075\,--\,3090\AA{}
range before scaling to match the iron emission strength in individual \qso\ 
spectra (\S\S~\ref{application} --~\ref{sample_Fecleaning}), its use is not 
adversely affected by the G130H normalization.
In fact, the poor match of the \feii\ emission in the G130H grating spectrum for 
some objects (see \S~\ref{sample_Fecleaning}) directly shows the need to 
subdivide the template in order to obtain optimum fits to AGN spectra. This is 
because all the individual iron multiplets did not brighten with the same factor 
in \izw\ as did the continuum in the time between the G130H and the G190H 
spectra were observed. If iron multiplet strengths vary in the same object, 
they are also very likely to vary among objects. This is also evident from the 
fact that not all AGNs have \feii\ UV 191, \feiii\ UV34, and/or \feiii\ UV47 
multiplets as strong as does \izw\ (\S~\ref{sample_Fecleaning} and Appendix A).
 Lanzetta \et (1993) present a spectrum of \izw\ representing the average of the
UV spectra observed with the {\em International Ultraviolet Explorer} 
({\it IUE}) from 1978 through 1982. Comparison with the {\it HST} data to 
confirm the scaling of the G130H spectrum is, however, not possible due to the 
widely differing host galaxy contribution\footnote{The {\it IUE} spectrograph
has a large aperture (10\arcsec $\times $ 20\arcsec, Boggess \et 1978) thereby 
sampling the AGN host galaxy.} and to line and continuum variations (cf.\ Wu 
\et 1983; Pian \& Treves 1993; Wang \et 1996; Rodr\'{\i}guez-Pascual \et 1997; 
L97).

The only other UV data available are, to our knowledge, spectropolarimetric data
in the {\it HST} data archives. A comparison with such data requires knowledge 
of the scattering medium and an understanding of the scattered spectrum, which 
are beyond the scope of this work. Hence, it is not possible at present to 
constrain the absolute scaling of G130H spectrum further, though as noted this 
uncertainty does not adversely affect the applicability of the iron template.

\section{The Iron Emission Template \label{fetemplt}}
\subsection{Development of the Template \label{methodCreate}}

Once the data were calibrated, rebinned to a common dispersion, and co-added 
the template was generated using the following procedure:
\begin{itemize}
\item
A power-law continuum was fitted to pure continuum \wav\ regions 
(\S~\ref{dataproc}) in
the spectrum and this continuum fit was subtracted (result is overplotted in 
Fig.~\ref{emissmodel}; dotted line)
\item
All absorption and (non-iron and iron) emission features were identified, 
including the
strong, weak, blended and unblended ones (Fig.~\ref{emissmodel}, 
Tables~\ref{alids} and~\ref{elids}; \S~\ref{abslines}, \S~\ref{lineids})
\item
The non-iron emission features (Fig.~\ref{fitex}) and Galactic/ISM 
absorption features (\S~\ref{abslines}, \S~\ref{fitmethod})were fitted. 
This required a simultaneous fitting of some \feiii\ features 
(Table~\ref{fefits}). Separate absorption and emission models were 
created (Fig.~\ref{emissmodel})
\item
The remaining \feiii\ features were fitted and deblended, as needed 
(Table~\ref{fefits}).  The \feiii\ line emission was isolated and an 
\feiii\ model was created (\S~\ref{fitmethod})
\item 
The absorption and non-iron emission models were subtracted from the 
original spectrum to create an iron template (containing both \feii\ 
and \feiii\ emission; Fig.~\ref{template})
\item
The \feiii\ emission model was subtracted to create a pure \feii\ 
template (Fig.~\ref{feiiimodel})
\item
The pixel values were set to zero in the two iron templates in regions 
containing pure noise residuals left over from subtracting emission and 
absorption feature fits  (Fig.~\ref{template} and~\ref{feiiimodel}).
This prevents introduction of artifacts and noise to 
the target spectra when the template is applied.
\end{itemize}

In the following sections we discuss the fitting and identification steps of
this procedure in more detail. We also compare the template, in 
\S~\ref{template_comp}, with other UV iron templates and theoretical models, 
currently available.

\placefigure{emissmodel}
\placefigure{feiiimodel}
\placefigure{template}

\subsection{Absorption Lines \label{abslines}}

Features were identified as absorption lines (Table~\ref{alids}) when the 
minimum flux in the feature deviated by more than 3\,$\sigma$ ($\sigma$\,=\,rms 
of the fluxes around the average local spectrum level) from the continuum or if 
the position of the feature coincided with that expected as part of a doublet 
line where the strongest line component was already identified.  The 
identification of the specific ion and transition responsible for each 
absorption line was made using the list of typical ISM absorption lines by 
Savage \et (1993) and the line lists by Morton (1991).
All the absorption lines detected and identified in the spectrum are consistent
with Galactic absorption with the exception of the features at 1306\AA\ and 
1310\AA\ which are due to \nv \,\lam \lam\,1238,\,1243 (blueshifted) absorption 
associated with \izw\ (L97).

Absorption features were fitted by multiple Gaussian components assuming a 
constant width for all components of a multiplet. The fit was then subtracted 
from the spectrum.

\placetable{alids}

\subsection{Emission Lines}

\subsubsection{Line Identifications \label{lineids}}

Our identifications of each emission feature are based on 
reference line lists for non-iron (Wilkes 2000, Morton 1991, Verner, Barthel, 
\& Tytler 1994),
and iron transitions [Moore 1950 (\feii\ and \feiii\ multiplets); Penston \et 
1983 (\feii ); Fuhr \et 1988 (\feii\ and \feiii); 
Giridhar \& Arellano Ferro 1995 (\feii ); Nahar 1995 (\feii ); 
Nahar \& Pradhan 1996 (\feiii ); Kurucz \& Bell 1995 (\feii\ and \feiii ); 
D.\ Verner (1996, private communication, \feii )]. 
L97 suggested identification of a large number of features in
the \izw\ spectrum, but that paper was not our main reference as it appeared 
after our work was commenced and after most of the line identifications were
completed. 
The data from L97 are essentially the same data presented here.  Slight 
differences exist as this G130H spectrum is recalibrated with more recent files 
and is renormalized to the level of the G190H spectrum (see \S~\ref{dataproc}).
Certain differences are present between our work and that by L97:
(1) we identify a few additional non-iron features, (2) we suggest 
identifications of individual iron transitions in addition to the multiplet 
identifications in L97, (3) we find slightly different velocity shifts of the 
various line groups, and (4) many of the L97 line measurements deviate, though
by no more than 30\,\% for the stronger and/or isolated lines. 
We briefly discuss (3) and (4) in 
\S~\ref{linevelshifts} and \S~\ref{linestrengths}.

In Appendix~\ref{linecomments} we comment on the individual line features and 
their fits. A thorough discussion of line intensities is given by L97 and not 
repeated here apart from brief discussions of the weak \ciii\ 
(\S~\ref{C3_complex}) and the relatively strong \lam 1400 emission 
(\S~\ref{L1400strength}).

\paragraph{Non-Iron Emission Lines}

The UV {\it HST} spectrum of \izw\  has a sufficiently high S/N and spectral
resolution to permit detection and identification of many weak features in 
addition to the strong, broad emission-lines commonly observed in \qsos .
Table~\ref{elids} lists the detected (non-iron) line features and their 
identification along with basic line parameters; 
see \S~\ref{fitmethod} on line fitting for further details.
L97 also identify most of these lines.  We contribute with a few extra 
identifications and measurements: \siiii $^{\star}$ \,\lam1297, and \siii \, 
\lam \lam \,1527,1533, based on Verner \et (1994), Morton (1991), and 
Wilkes (2000). We are not able to confirm the [\ciii \,\lam 1907 feature 
identified by L97 due to our slightly degraded resolution (\S~\ref{dataproc}).

\placetable{elids}

\placefigure{feids}

\paragraph{Iron Emission Features} 

Figure~\ref{feids} shows suggested identifications of individual, mostly
unblended \feii\ and \feiii\ emission UV multiplets\footnote{A full scale 
version of Fig.~\ref{feids} can be found at 
http://www.astronomy.ohio-state.edu/${\sim}$vester/IronEmission.}.
The identifications
are based on a visual inspection of clearly visible peaks in the spectrum
whose positions and relative strengths were cross-correlated with the 
multiplet table of Moore (1950). This goes a step further than the work
by L97, who mark in their figure~2 only the expected wavelengths of some
of the strongest iron multiplets. The length of the marker of a given 
transition in a given multiplet is proportional to its oscillator strength 
(the `intensity' listed by Moore is a rough measure of the relative oscillator 
strengths in the multiplet; note, that the physical conditions folded with the 
oscillator strengths will give the observed line strengths and ratios). The 
scaling factor is the same for all multiplets (and is arbitrarily chosen).
Note that the transitions in the spectrum sometimes appear blueshifted by 
1-2\,\AA{} relative to the laboratory wavelengths. See \eg \feiii \,UV34 at 
$\sim$1914\,\AA.  Each label contains first the ionization
level, then the UV multiplet number, separated by a hyphen. That is,
`2-104' denotes the \feii\ UV104 multiplet, while `3-158' denotes the
\feiii\ UV158 multiplet, etc. \feii\ and \feiii\ multiplets in the spectrum 
for which the relative transition strengths in the multiplet do not appear
to follow the $\sim$optically thin multiplet strengths listed in Moore (1950)
have an `m' attached to the labeled multiplet number. A `?' indicate that this 
multiplets presence is uncertain. Labels of multiplets which appear 
slightly blended are shown in parentheses.  Square brackets denote multiplets 
whose presence is suggested by the fitting process (see Appendix~A); \feii\ 
\,UV10, which coincide with the geocoronal \lya\ emission, is expected only. 
Heavily blended multiplets are not labeled, especially in the small blue bump 
region from $\sim$2650\,\AA{} to 3090\,\AA{}. The lack of labeled multiplets 
between \lya{} and \civ{} does not denote a lack of \feii\ or \feiii\ emission 
transitions. Most transitions in this range are too weak (relatively) to have 
multiplet numbers assigned. Blueward of \lya{} an accurate identification of 
iron emission multiplets requires advanced modeling of the spectrum due to the 
heavy blending with \lya{} forest and other absorption lines. This is beyond 
the scope of this paper.  L97 mark possible iron multiplets in this region.
No obvious \feii\ or \feiii\ transitions are identified immediately redward
of \civ\ ($\sim$1550 -- 1700\,\AA ). Marziani \et (1996) discuss some \feii\
multiplets in this \wav\ region in earlier \HST\ data of \izw\ (their figure~2).
Some singlet \feii\ and \feiii\ emission features are identified and 
commented on in Appendix A.

\placetable{fefits}

\placefigure{fitex}

\subsubsection{Fitting Procedures\label{fitmethod}}   

The emission lines, listed in Tables~\ref{elids} and \ref{fefits}, were fitted 
using single or multiple Gaussian components, as needed, and then subtracted 
from the spectrum.  The \feiii\  and \feii\ features were separated by use of
the fitted \feiii\ lines in Table~\ref{fefits} to generate two templates, one 
for each ion (\S~\ref{methodCreate}).
The IRAF task `splot' was used for the Gaussian fitting due to its interactive 
nature and the ease with which the fitted components can immediately be 
extracted and compared to the data. `splot' is often thought of as a tool for 
first order estimates but its characteristics allow us to 
constrain our fits faster and more conveniently than other immediately 
available fitting programs, due to their largely non-interactive nature.

Fitting the multiplets and line complexes was an {\it iterative} process. First 
the narrow components were fitted to the visible part of the lines and 
subtracted, then the broad components were fitted to the residuals. The 
broad component fits were then subtracted from the original data permitting an 
improved fit to be obtained for the narrow lines, which in turn were subtracted 
to improve the broad component fit, thereby iterating to obtain an optimal 
solution.  Similar iterations were performed for individual (narrow) components 
in doublets and in regions where the narrow components are somewhat blended 
(\eg \civ , \mgii , and the \ciii\ complex).  When a line complex contains two 
broad components (the \aliii\ and \mgii -doublet complexes, Figs.~\ref{fitex}d 
and~\ref{fitex}g), 
they are often too heavily blended to be well separated and so were fitted by a 
single broad component.
Two broad components were fitted simultaneously to \siiii ] and \ciii , but
as they are heavily blended the fit is not unique (see below; Fig.~\ref{fitex}d;
Appendix~\ref{linecomments}). 

The average line width of the singlet lines is 900 $\pm$150 \kms 
(Tables~\ref{elids} and~\ref{fefits}) consistent with previous studies of the 
line emission in \izw\ (Phillips 1976; BG92; van Groningen 
1993; L97).  However, lines with widths as small as $\sim$ 300 -- 400 \kms\ 
(the spectral resolution limit) are detected, as discussed in 
\S~\ref{linewidths}.
These are often \feii\ and \feiii\ lines (see Table~\ref{fefits}). 

Each Gaussian component is defined by three parameters: position, width, and 
strength. In the heavily blended line complexes the absolute strengths of the 
individual Gaussian components are not well constrained.  The uncertainty in the
intensity of individual lines is estimated at 10$-$50\% depending 
on the width and strength of the line and on how well the data constrain the 
fit. The more blended the component, the more uncertain is the fit. 
Due to the large number of parameters involved in fitting the large line 
complexes, individual component solutions are not unique. In such cases, those 
solutions yielding parameters consistent with other lines were preferred.
We note that, {\bf when the strength of any line fit was not well 
constrained by the observed spectrum, the affected iron residual emission 
was purposefully underestimated in order to avoid overcorrection of the 
\qso\ spectrum to which the template is applied.}
The representation of the entire line complex (\ie the sum of all the 
individual components) was, however, always well constrained by the data and 
the estimated errors are of order a few percent.

Examples of the emission-line fits can be found in Figure~\ref{fitex}.  The 
individual measurements of fitted non-iron and iron lines are listed in 
Tables~\ref{elids} and \ref{fefits}, respectively. 
Measurements are based on the individual Gaussian components making up the line 
profile; sums for the line complexes are also listed. The equivalent widths are
always measured relative to the adopted (global) continuum level 
(\S~\ref{dataproc}).  Very weak iron features were not fitted. 

We emphasize that the multiple Gaussian component fitting is simply a tool,
and we make no assumptions as to the mechanism responsible for the overall 
profile shapes.  Thus, as no physical meaning is associated with the individual 
Gaussian component fits, the absolute and relative fluxes of the fitted 
components cannot be interpreted in terms of physically distinct emitting 
regions. Extreme care should always be exercised when interpreting the 
measurements, particularly of weak features.  Features for which certain 
fitting parameters and/or the line identifications are particularly uncertain 
are mentioned in Appendix~\ref{linecomments}.

\placetable{feids}
\placetable{fefits}

\subsection{Comparison with Earlier UV Iron Templates and Models 
\label{template_comp}}

Despite the wealth of work on modeling the iron emission in \qsos\, there 
is no UV template electronically available which can be applied to observed 
\qso\ spectra, including our own (M. Vestergaard \et 2000, in preparation), 
for removal of the UV \feii\ emission. 
One other empirical UV template is available in the literature: the
Corbin \& Boroson (1996) template covering the iron bump around \mgii . 
Below we briefly comment on how this compares to our template in the 
overlapping region.
A couple of synthetic \feii\ spectra have been generated based on the available 
knowledge of the iron emission mechanism, the quality of the atomic data, and 
the computing facilities at the time (Wills \et 1985; Verner \et 1999). We 
briefly discuss the apparent differences with the current iron template.

\subsubsection{Empirical UV templates}

Corbin \& Boroson (1996) present a \lam \lam 2300\,--\,3000\,\AA\ iron template 
also derived from the {\it HST} spectrum of \izw\ (L97). 
Although a detailed comparison is not possible (the digital spectrum is not
available and Figure~2 in Corbin \& Boroson (1996) shows an already broadened
version of the template), it is clear that differences are present due to their
use of interpolation rather than deblending to remove non-iron features. This
is particularly clear around \mgii\ where interpolation across the \mgii\ line 
likely overestimates the iron emission strength by $\sim$300\,\%; the \feii\
emission level is at most 15\,--\,20\,\% of the peak height of the \feii\ on 
either side of the \mgii\ gap (cf. figure~12 in Verner \et 1999).
In contrast, our approach (\S~\ref{fitmethod}) is to purposefully underestimate 
the iron emission when the strength of the iron and/or the non-iron emission is 
in doubt. We have therefore assumed that the broad component of \mgii, the 
existence of which is evident in Figures.~\ref{emissmodel} and~\ref{fitex}g, 
dominates this \wav\ region, consistent with the findings of L97, and so our 
template shows a gap in the iron emission around \mgii. 

Our sample fitting and subtraction (\S~\ref{sample_Fecleaning}) 
of the iron emission in spectra of high-redshift quasars
using the template presented here yields residual profiles of the \mgii\
line resembling those of the other prominent UV lines, supporting our method.
Our Gaussian fitting of \mgii\ indicates that one will underestimate 
its strength by a factor of $\sim$2 if the broad \mgii\ component
is ignored, as in the interpolation process.

\subsubsection{Synthetic \feii\ Spectra}

Wills \et (1985) present theoretical models of the
1800\,--\,5000\,\AA\ \feii\ emission, the Balmer continuum and lines, and
compare with observed spectra of a small sample of low-redshift quasars
($z \sim$0.12\,--\,0.6).  Their success in simulating the observed spectra 
is generally good, though a few discrepancies in line strengths among 
multiplets remain. Over the past few years the Opacity and IRON projects have
applied the power of modern computers to determining the atomic parameters of
the thousands of possible iron transitions (\eg Hummer \et 1993, Seaton \et 
1994, Nahar \et 1997, 2000, and references therein), so it is worth revisiting
the modeling to see if the deviations of the average Wills \et \feii\ 
models from the \izw\ iron spectrum are now removed. Specifically, Wills \et
predict stronger \feii\ emission between $\sim$1800\,\AA\ and  $\sim$2300\,\AA, 
and weaker \feii\ right around \mgii\ for 3C273. In addition, the relative 
strength of various multiplets appear to differ between their models and the 
\izw\ spectrum.  Variations in physical conditions between the NLS1s, of which 
\izw\ is considered the prototype, and the Wills \et quasars may also explain
some of the differences.

Verner \et (1999) present and discuss their numerical simulations of \feii\ 
emission spectra based on updated iron atomic data, iron line lists (many of 
which are also used here for line identification), and photoionization modeling.
They display the \feii\ emission spectrum for different densities, photon flux,
micro-turbulent velocities below 100\,\kms , and iron abundances. 
Although these line widths are much narrower than those of AGNs and quasars,
complicating a comparison, a few differences between the synthetic spectra of 
Verner \et and Wills \et and our empirical \feii\ and \feiii\ templates can 
immediately be pointed out.
(1) The synthetic iron models contain \feii\ transitions only, whereas the 
\izw\ spectrum also contain \feiii\ transitions. 
(2) The synthetic spectra (Wills \et 1985; Verner \et 1999) show a decrease 
in \feii\ emission above $\sim$2700\,--\,2800\,\AA.  We are not able to see 
this in our template as contamination by the Balmer continuum ($\gsim$ \lam 
\,2800 \,\AA\ especially) is present. 
The {\it HST} spectrum of \izw\  alone does not allow the Balmer continuum 
contribution to be sufficiently constrained. Complete photoionization 
modeling is required to properly account for it.  However, the Balmer 
continuum contribution decreases continuously from $\sim$3600\AA\ down to 
$\sim$2500\AA. Below $\sim$2800\,\AA\ it is relatively weak, and it so should 
not significantly contaminate our 3100\AA -limited iron templates.

Detailed comparison between the \feii\ template and the most recent \feii\ 
models by Verner \et (1999) holds great potential for 
significantly improving our understanding of the iron emitting mechanism and
mapping the physical conditions under which the iron emission is radiated.
Hence, empirical iron templates remain useful even with the emergence
of synthetic iron emission spectra.

\section{\bf Application of the Iron Template \label{temp_applic}}

\subsection{Broadening of the Iron Template \label{tempbroad}}

In order to fit and subtract iron emission from any \qso\ spectrum we need to 
broaden the 
iron template to match the line width of that spectrum. 
This is done by convolving the original template with Gaussian functions of 
different widths thereby creating a grid of spectra with a range of (quantized)
widths. 

Three steps are necessary in the iron template broadening-process:
(1) 
The standard deviation, $\sigma_{conv}$, 
of the convolving Gaussian profile was first estimated using 
\begin{equation}
\sigma_{conv} = {\rm FWHM}_{conv}/2\sqrt{2\,ln\,2} \,=\, 
\sqrt{{\rm FWHM_{\rm QSO}}^2\,-\,{\rm FWHM_{\rm IZw1}}^2}/2\sqrt{2\,ln\,2},
\end{equation}
where FWHM$_{\rm IZw1}$ = 900\,\kms\ is the line width of the \izw\ spectrum. 
(2)
To apply a constant velocity broadening the computations were carried out in 
logarithmic \wav\ space since 
$d\,(log\,\lambda )\,=\,d\,\lambda /\lambda \,=\, d\,v/c$.
The process of rebinning the broadened template back to linear units results in 
a small additional broadening. To ensure that the final broadened template has 
the desired width, we performed broadening simulations on 
artificial data consisting of a single Gaussian line feature of width 
900\,\kms , where the width of the convolving Gaussian profile, $\sigma$, was 
adjusted until the FWHM of the broadened feature, measured in linear \wav\ 
space, matched the desired width to within our measurement errors 
($\sim$1\,\kms ).
(3) The iron template was then convolved (in logarithmic \wav\ space) with a 
Gaussian profile with the adjusted value of $\sigma$ and rebinned back to 
linear units.

We emphasize that because quasar spectral lines are often affected by blending,
it is important to measure the resulting convolved width using a single
artificial spectral line as outlined above, as opposed to estimating the width
on the broadened (blended) template itself. We find blending
effects to overestimate the line widths by $\sim$15\,--\,25\%, based on
simulated line blends. Isolated narrow line cores, like \civ\ and \mgii\ are,
however, expected to be well-determined.

\subsection{Cleaning Quasar Spectra for Iron Emission -- Application of the 
Method \label{application}}

In order to apply this UV iron template to \qso\ spectra, both the line
widths and the iron emission strength of the template must be matched to 
that of the \qso\ spectrum.  We follow the basic method of estimating the 
iron emission strength and subtracting the iron emission used by BG92 on 
the optical (\lam \lam \,4250$-$ 7000\AA ) iron emission of their sample 
\qsos .  This method consists of artificially broadening the iron template 
to a number of widths by convolution with a Gaussian profile 
(\S~\ref{tempbroad}) thereby generating a two-dimensional grid of templates 
with different line widths.  BG92 scaled this ``two-dimensional template'' 
by a number of arbitrary, but fixed, strengths, hence creating a 
three-dimensional grid of templates (with dimensions: wavelength, line 
width, and line strength).  The three-dimensional iron template was 
subtracted from the target spectrum (expanded to a data cube) and the best 
residual (one-dimensional) spectrum was chosen by manual inspection.

Similar to BG92, we created a three-dimensional iron template consisting of five
different scalings of a two-dimensional template, containing a number of line 
widths (in the range 1000\,\kms\ $-$ 15\,000\,\kms\ in steps of 250\,\kms\ is
usually sufficient) which are broadened versions of the original iron template 
(intrinsic width 900\,\kms).  
We also performed a manual inspection of the residual iron-subtracted spectra. 
Our method, however, differs from that of BG92 in the following ways. 
After fitting and subtracting a power-law continuum fit\footnote{Based on our 
experience, optimal fitting of the iron emission in any \qso\ spectrum requires 
a power-law continuum fit, 
since a power-law continuum was initially subtracted from the \izw\ spectrum 
before generating the iron template.} to the iron-contaminated \qso\ spectrum, 
we iterate to determine the optimum parameters as follows:
\begin{itemize}
\item
Determine the iron template (primary) normalization for a given \qso\ spectrum 
interactively from its iron emission strength in pure iron emission windows, 
specified by the user (examples are listed in Table~\ref{fewindows}). There is a
separate normalization for each (1-D) template spectrum with a given broadening. 
\item
Subtract a (three-dimensional) data cube consisting of five scaled, 
two-dimensional templates from the target spectrum. The scalings are (five) 
fractions of the computed
normalization, and can be chosen arbitrarily. A useful range is 50\% $-$ 150\% 
of the normalization factor determined above, though the scalings $\neq$100\% 
are rarely needed except for cross-checks.
\item
Compute the residual flux and $\chi^2$ of all the fits in the iron emission 
windows.
\item
Manually inspect the residual spectra and pick the best (one-dimensional) 
iron-subtracted spectrum.
\item
Add the previously determined continuum fit for the \qso\ back into the 
iron-subtracted spectrum and refit the power-law continuum. 
(In iron-contaminated spectra only a few narrow, 
pure continuum windows may exist.  After a preliminary subtraction of the 
iron emission, larger continuum windows can generally be used).
\item
Subtract this (new) continuum fit from the {\em original} target spectrum.
\item
Determine the new normalizations as in the first item above and repeat all
steps iterating over both continuum setting and iron emission strength until
both fits (continuum and iron emission) converge 
and the final iron-subtracted spectrum is satisfactory.
\end{itemize}
By estimating the template scaling factor interactively, we are not limited to 
an a priori set of quantized scalings (the \qso\ spectra can have any 
normalization and iron emission strength), and we ensure a good starting point 
for the iron spectrum fitting. 
Computations of both the $\chi^2$ and the residual flux in the user-defined iron
emission regions 
help us to determine the  best-fitting iron spectrum objectively.  
Iterating over both the iron emission strength and the continuum setting 
improves the fit to them both, especially in
heavily contaminated \qso\ spectra where it may be hard to define pure continuum
regions over which the underlying global continuum can be fitted. 

\placetable{fewindows}

\placefigure{FeCleanfigs}

\subsection{Sample Iron Emission Subtraction from Quasar Spectra 
\label{sample_Fecleaning}}

We fitted the iron emission in four high-redshift and one low-redshift quasars, 
demonstrating the successful application of the UV iron template. We briefly 
summarize our main conclusions and then comment on these results for each 
individual quasar.
The objects and their spectra were chosen to be representative of typically
available data for the high-redshift quasars and to contain a range in \feii\
emission strength. The 3C273 \HST{} spectrum has the advantage of covering the 
entire UV range.  The iron template is also successfully applied to the
Large Bright Quasar Survey (Forster \et 2001). We note that a perfect fit of
the \izw\ iron emission templates to the iron emission in other AGNs is never
expected. This is owing to the strong dependence of the iron emission spectrum on
the physical conditions in the emitting gas and on the BLR geometry (Netzer
1980). This is manifested in AGN spectra by variations in the strengths of
the different multiplets among individual AGN (see discussion of 3C273 below). 
However, we find the template fitting to work well enough to be a valuable tool 
for eliminating and studying the iron emission in AGNs at least until accurate 
theoretical iron emission models are developed. 

A reasonable match to the iron emission in the quasar spectra presented
here could be obtained by scaling and broadening the combined \feii\ and 
\feiii\ UV templates, as described in \S~\ref{application}. 
Similar to the template, \feii\ emission is rather common around \mgii\ 
and between \civ\ and \ciii.  The need to include one or more of the \feii\ 
UV191, \feiii\ UV34, and \feiii\ UV47 multiplets, which are strong in \izw, 
in the applied template depends on the individual object. 
Improved iron template fits were generally obtained by sub-dividing the 
template and scaling each sub-spectrum (and hence selected groups of 
multiplets) separately.  Good fits were often separately obtained for
the \lya $-$ \civ, \civ $-$ \ciii, \ciii $-$ $\sim$2300\AA, and $\sim$2300 $-$
$\sim$3100\AA\ regions.

The most frequent and significant differences in strengths and multiplet 
ratios of the \feii\ and \feiii\ transitions among the objects considered
here are in the regions shortward of \civ, between $\sim$1900\AA{} and 
$\sim$2300\,\AA, and in the \feii -bump around 2500\AA. 
The (undivided) template generally, but not always, overpredicts the 
\lam 1400 -- 1530\,\AA\ iron emission in these \qsos \footnote{This may 
partly be due to the fact that \izw\ brightened (\S~\ref{scaling}) between 
the G130H and the G190H and G270H spectra were observed, and that the iron 
spectrum does not simply scale with the continuum level.}. A separate 
(re-)scaling of this iron emission is acceptable because we have no 
empirical or simple theoretical constraints on the strength of this 
emission based on the iron emission, say, around \mgii.
The variation in the \lam 1900 -- 2300\,\AA{} emission strength is possibly 
due to differing amounts of dust extinction along the line of sight towards 
individual objects. 
The general weakness of the \lam 2000\,--\,2300\AA\ emission in the template 
suggests that \izw\ may be subject to high levels of dust extinction perhaps 
from its host galaxy and consistent with the observed strong infrared emission 
(Rieke \& Low 1972; Rieke 1978).  The iron multiplet ratios may also (or 
instead) vary significantly in this region among quasars. This discrepancy is 
not unique to the \qsos\ presented here 
(Figs.~\ref{FeCleanfig3C273} and~\ref{FeCleanfigs}), but is also seen in other 
broad-lined quasars (\eg Wills \et 1985; Steidel \& Sargent 1991; Corbin \& 
Boroson 1996; M. Vestergaard \et 2001, in preparation). Residual emission is 
similarly present in the sample iron-subtracted spectrum of Corbin \& Boroson 
(1996; their Figure~2). 

The overall and important conclusion is that a subdivision of the 
$\sim$2000\,\AA\ wide \feii\
and \feiii\ UV templates may often be necessary because the individual multiplet
strengths across the spectrum vary between objects. Similarly, the \feii\ and 
\feiii\ templates may require different scalings as evidenced by the absence of
\feiii\ emission in some sources (\eg Q1451$+$1017; Baldwin \et 1996). 
Also, there is no strong direct coupling expected between the UV \feiii\ and
\feii\ emission. The \feii\ is emitted from the partially ionized hydrogen
zone while the \feiii\ originates in the fully ionized zone. Some correlation 
is expected to exist as both ions are (at least partly) excited by the UV 
continuum, but the
current theoretical models are not accurate enough to firmly establish a 
simple scaling relation between the strong multiplets of UV \feii\ and 
\feiii, if one exists (A. Pradhan 2000, private communication). Therefore, a 
separate scaling of the UV \feii\ and \feiii\ emission templates is allowed 
in order to optimize the fit to the AGN iron emission.
However, this was not necessary for the AGNs presented here.

We first present the fitting results on 3C273 as its \lam \lam 940 $-$ 
3200\,\AA\ {\it HST} spectrum allows the full range of the template to 
be tested on a single AGN.

\paragraph{Q1226$+$023; 3C273 ($z$ = 0.157):}
The 3C273 spectrum (Fig.~\ref{FeCleanfig3C273}) does not contain any obvious 
features of the \feii\ UV191, \feiii\ UV34, and \feiii\ UV47 multiplets so the 
spectrum was fit with templates excluding these features. 
The combined \feii\ and \feiii\ template did a good job, once subdivided into 
five independently scaled segments (at \lam \lam\,1540, 1912, 2423, and 2656\AA). 
A scaling of the full range template to the iron emission redward of \ciii\ and 
in the small blue bump yielded consistent flux levels blueward of \civ\ if no 
\niv] \lam 1486 line is present. We could not distinguish between this fit and
one with a separate scaling and a \niv] line (the latter fit is shown in 
Fig.~\ref{FeCleanfig3C273}).
The fitted iron model in the range between \civ\ and \ciii\ yields residuals 
coinciding with the expected \wavs\ of known lines, such as \heii\ \lam 1640, 
\oiiisf\ \lam 1664, \alii\ \lam 1670, \niii\ \lam 1750, \siii\ \lam 1814, and 
\aliii\ \lam 1857 yielding confidence in the template and the model fit.
We found the \mgii\ profile in the iron subtracted 3C273 spectrum to show 
a slightly blue asymmetric profile.  This profile could be fitted very well 
with two Gaussian functions with the same FWHM of 3800\,\kms\ (one shifted 
$-$3400\,\kms\ from the narrow peak of \mgii; FWHM is consistent with the 
widths of the UV lines and the narrow Balmer lines, $\sim$4000\,\kms). 
This gives us confidence that the iron template in the \mgii\ region is 
representative of the average \feii\ bump in AGNs.

The 3C273 spectrum shows stronger iron emission than \izw\ between 
the \civ\ and \ciii\ lines and in the \lam \lam 2000 $-$ 2300 range 
(Fig.~\ref{FeCleanfig3C273}) relative to the remaining UV iron emission. 
As the \izw\ template does not contain much iron 
emission in the \lam \lam 2100 $-$ 2250\AA{} range, no further attempt to model 
this additional emission is possible with the current template. 
The \izw\ template clearly overestimates the 3C273 iron strength around 2500\AA. 
The emission is rather flat across this region contrary to that in the template. 
Two different attempts to model this region were made 
(Figs.~\ref{FeCleanfig3C273}b and~\ref{FeCleanfig3C273}c). One `model' fits the 
blue part while the other
fits the red part, but neither does a good overall job. The original template
has much stronger emission at 2500\,\AA\ than that shown in 
Fig.~\ref{FeCleanfig3C273}.  In the case of the $\sim$2100\AA\ and  $\sim$2500\AA\
`bumps' a different template is clearly needed to account for this emission. 
Redward of \mgii\ the template shows a similar shape as the observed \feii\ but 
slightly underestimates the emission.  It is likely due to a slight overestimation 
of the underlying continuum level in \izw\ owing to the truncation of its spectrum
at 3089\AA\ rest frame by the FOS G270H grating. 

The power-law\footnote{F$_{\nu} \sim \nu^{-\alpha_{\nu}}$ and F$_{\lambda}
\sim \lambda^{-\alpha_{\lambda}}; \alpha_{\lambda} \,=\,2\,-\,\alpha_{\nu}$
(\eg Weedman 1986).} 
continuum has F$_{\lambda}$(1500 \AA) = 1.608$\times$10$^{-13}$
\ergsA\ and slope, $\alpha_{\lambda}$\,=\,1.67 ($\alpha_{\nu}$\,=\,0.33).
The original spectrum is shifted by $+$1.5$\times$10$^{-13}$ \ergsA\ in 
Fig.~\ref{FeCleanfig3C273}a and by $+$0.25$\times$10$^{-13}$ \ergsA\ in 
Fig.~\ref{FeCleanfig3C273}b. In this figure the residuals from the `blue 
2500\AA' iron model, shown in the middle, represents the original flux level 
of the spectrum. The residuals from the `red 2500\AA' iron model, shown as the 
lowest of the three spectra in panel (b), is shifted by 
$-$0.25$\times$10$^{-13}$ \ergsA.  Figure~\ref{FeCleanfig3C273}c shows the two
2500\AA\ iron models more clearly; the spectrum and the models are continuum
subtracted here.
The iron model has a FWHM of 4000 \kms.
The processing of the \HST\ data is outlined in Appendix~B.

\paragraph{Q0020$+$022 ($z$ = 1.798):}

There is no clear indication of strong iron emission in the 
\lam 1250\,--\,1500\AA\ range and the continuum level is relatively lower than 
the flux level in the \lam 1600\,--\,1900\AA\ range. The flatness of the 
\lam 1930\,--\,2050\AA\ region also indicates very weak iron emission, so
this region was adopted as a continuum region.
The power-law
continuum has normalization, F$_{\lambda}$(1500 \AA) = 1.753$\times$10$^{-16}$
\ergsA\ and slope, $\alpha_{\lambda}$\,=\,1.94 ($\alpha_{\nu}$\,=\,0.06).
The combined \feii\ and \feiii\ template was fitted to the data between
\lam 1530 and 1900\AA\ only (Fig.~\ref{FeCleanfigs}a) yielding residuals
which coincide very well with the expected positions of \heii{} \lam 1640,
O\,{\sc iii}] \lam 1664, \alii{} \lam 1670, \niii{} \lam 1750, and \siii{} \lam 
1814\,\AA. The residual spectrum is shifted by
$-$0.5$\times$10$^{-16}$ \ergsA . The iron model has FWHM of 6500\,\kms.

\paragraph{Q0252$+$016 ($z$ = 2.457):}

An improved fit to the iron emission (Fig.~\ref{FeCleanfigs}b)
could be obtained by including the \feii \,UV\,191 \lam 1786 and 
\feiii \,UV\,34 \lam \lam 1895, 1914, 1926 features in the templates 
and by separately fitting the \lam 1400\,--\,1550\AA\ emission 
which is weaker than in the UV template, but still significant.
It is unclear whether weak \feii\ emission still remains in the blue wing of
\civ{} \lam 1549, though both wings of this line indicate the presence of a
somewhat broad underlying component. 
The power-law continuum has F$_{\lambda}$(1500 \AA) = 2.807$\times$10$^{-16}$ 
\ergsA\ and slope, $\alpha_{\lambda}$\,=\,1.28 ($\alpha_{\nu}$\,=\,0.72).
The residual spectrum is shifted by
$-$1.0$\times$10$^{-16}$ \ergsA . The iron model has FWHM of 5000\,\kms.

\paragraph{Q1629$+$680 ($z$ = 2.478):}

The iron emission is relatively weak (Fig.~\ref{FeCleanfigs}c) compared to the 
other quasars presented here, as indicated by the lower emission level between 
\civ\ and \ciii. Based on the \civ\ line width of FWHM$\sim$4000\, \kms, a 
reasonable
but weak model fit could be made using the combined \feii\ and \feiii\ templates
in which \feii\ UV 191 \lam\,1786 and \feiii\ UV 34 \lam \lam 1895, 1914 were
excluded. One exception is the region between \lam 1400 and \civ, where the UV
template strongly overpredicts the \feii\ emission. In fact, the data does not
indicate noticable \feii\ emission there. Thus, the iron emission was fitted 
using a version of the template in which the \lam 1400\,--\,1550\,\AA\ iron 
emission is excluded.
The power-law continuum has F$_{\lambda}$(1500 \AA) = 8.903$\times$10$^{-17}$ 
\ergsA\ and slope, $\alpha_{\lambda}$\,=\,2.3 ($\alpha_{\nu}$\,=\,$-$0.3).
The original spectrum is shifted by
$+$1.0$\times$10$^{-16}$ \ergsA. The iron model has FWHM of 4000\,\kms.

\paragraph{Q2345$+$061 ($z$ = 1.540):}

A scaling and broadening of the combined \feii\ and \feiii\ templates (with
\feii\ UV 191 \lam \,1786 excluded) to $\sim$2500\,\kms\ offer a reasonable
overall fit to the iron line emission (Fig.~\ref{FeCleanfigs}d). 
Hence, no sub-division of the UV template was performed.  The general shape 
of the iron emission longwards of \mgii\ is reproduced by the template but 
a slight drop in the emission level is seen in the \qso\ data owing to the 
enhanced calibration uncertainty in this region where second order light 
merges with the primary order spectrum. 
The main bumps and wiggles in the \feii\ bump around \mgii\ can be reproduced
with the exception of the \lam $\sim$2300\,--\,2500\,\AA\ region where excess 
emission is clearly present. 
Including the \feiii\ UV\,47 \lam \lam \,2418, 2438 feature in the template 
(shown in Fig.~\ref{FeCleanfigs}d) does not fully account for the iron 
emission in this region. Excess emission is also clearly present at \lam 
$\sim$2050\,--\,2150\,\AA. A good coincidence of the iron 
strength is found for $\sim$1800\,--\,2000\,\AA, based on a scaling of the 
template \feii\ bump around \mgii\ to the data, inspite of the larger noise
level below 2100 \,\AA. The large noise levels at \lam $<$ 1800\,\AA\ renders
the discrepancies of the data with the template fit irrelevant.
The power-law continuum has F$_{\lambda}$(1600 \AA) = 3.62$\times$10$^{-16}$ 
\ergsA\ and slope, $\alpha_{\lambda}$\,=\,1.72 ($\alpha_{\nu}$\,=\,0.28).
The original spectrum is shifted by
$+$1.0$\times$10$^{-16}$ \ergsA. The iron model has FWHM of 2500\,\kms.

\section{Discussion of the \izw\ Spectrum \label{izwidiscuss}}

The spectrum of \izw\ is rich in emission lines, and in particular this high 
S/N {\it HST} spectrum reveals a plethora of weaker lines 
(L97; Fig.~\ref{emissmodel}) not usually seen in \qso\ spectra.
The spectrum has strong low-ionization lines while some of the
higher-ionization lines are weaker than expected for an average \qso\ 
(Francis \et 1991; Zheng \et 1997; Wilkes 2000; L97 and references therein; 
see discussion on 
\siivoiv\ and \civ\ below). Both \oiii \,\lam \,5007 (BG92) and \ciii\ \,\lam 
1909 (this spectrum) are weak. 

While a detailed emission line study is not the subject of this paper, we
briefly discuss our results in relation to earlier work on this object.

\subsection{Line Parameters} 

\subsubsection{Line Widths \label{linewidths}}

Our measurements and line fitting show that the widths of the fitted iron 
emission components are rather narrow; ranging from $\sim$300 to 900 \kms . \Eg 
the 
narrow line core of \feii \,UV\,191 \,\lam \,1786 has FWHM of 550\,\kms\ and 
a similar fit to \feiii \,UV47\,\lam \,2418 yields a FWHM of 715\,\kms\ 
(Table~\ref{fefits}).  This narrow width  may be an intrinsic property of the 
iron emission mechanism or an effect of the continuum level uncertainty (the 
line width will appear narrower if the continuum is set too high). 
Our narrowest line measurements of $\sim$300\,--\,400 \kms\ of the weaker 
non-iron lines are consistent with the findings of van Groningen (1993), who 
concluded, based on work by himself and Phillips (1976), that \izw\ has three 
emitting regions, one of which has a velocity dispersion of $\sim$400\,\kms.

\subsubsection{Line Velocity Shifts \label{linevelshifts}}

The peaks of the broad emission lines are blueshifted (Table~\ref{elids}) relative to the 
systemic redshift ($z_{em}$\,=\,0.061), defined by the \mgii\ doublet 
(Appendix~\ref{linecomments}) and \hi\ radio measurements (Condon \et 1985). 
The high ionization lines have a higher blueshift (average shift 
$\sim$1540$\pm$500\,\kms ) than the low ionization lines 
($\sim$500$\pm$270\,\kms ; see also Table~\ref{elids}, L97, and below 
for exceptions).

A significant amount of blueshifted emission is evident in lines such as \nv , 
the \civ +\siii\ blend, and \heii \,\lam \,1640 (Figs.~\ref{fitex}a, 
\ref{fitex}c, and \ref{emissmodel}, respectively). 
A special case is that of \heii\ where practically all the emission is 
blueshifted, reaching $\sim$1300$-$2200\,\kms\ relative to the \qso\ restframe
(see also L97). The \nv\ line profile can be fitted as two emission components 
originating at different velocity shifts, $\sim \,-$900\,\kms{} and 
$\sim \,-$2400\,\kms{} (Fig.~\ref{fitex}a, Table~\ref{elids}). 
The $\sim$900\,\kms{} blueshift of \nv{}
is similar to that of the narrow \civ\ component, the broad \aliii\ component, 
and one of the narrow components of \lya\ and \siivoiv, respectively.
Most of our line shifts are consistent with those of L97 to within the errors 
and the spectral resolution (Leitherer 1995; Tables~\ref{elids} and 
\ref{fefits}). 
We suspect the velocity shift differences $\lsim$ 1300\,\kms{} with L97 for 
\siii \,\lam \,1263, \siii +\oi \,\lam \,1306, \siivoiv \,\lam \,1400, and 
\niii \,\lam \,1750 can be explained by a combination of uncertainties in 
the line positions of $\lsim$ 0.5$-$1\AA{} (\ie 250$-$500\,\kms; conservative 
error), resulting from the heavy line blending present in both studies, and
the slightly degraded resolution in our G130H+G190H specta (\S~\ref{dataproc}).

L97 suggested an outflowing component is responsible both for the 
blueshifted line peaks (of both low and high ionization), the blueshifted 
($\sim$2000\,\kms ) emission (blue wing asymmetry) and the associated weak UV 
absorption in \lya , \nv , and \civ\ which they detect.

\subsubsection{Line Strengths \label{linestrengths}}

L97 use the doublet ratios of \mgii\ and \aliii\ to deduce the location of 
their emitting regions relative to the `outer BLR boundary'. 
As discrepancies are seen between our line measurements and those of L97, it 
is of interest to briefly discuss the reasons and the implications for their 
BLR size results.
At \lam\ $\lsim$ 1500\AA\ the main differences are due to the fact that L97 does
not scale the G130H spectrum to match the level of the G190H and G270H spectra,
as done here.

The line measurements at \lam\ $\geq$ 1500\AA\ are different in part due to
different Gaussian fitting techniques (cf. Laor \et 1994), but mostly due to
L97's use of a local continuum (A. Laor, 1997, private communication) as opposed 
to a global one, as done here except at the \feii\ bump, \lam 2300 --
3090\,\AA. The EW measurements in regions of isolated 
and unblended emission lines (where the continuum is well determined) agree well
(\cii \,\lam \,1335, \lam \,1345-feature, and \siii \,\lam \,1814). However, 
for the remaining lines (see Table\,1 in L97 and Table~\ref{elids}, this work), 
the EW measurements differ by $-$75\% to 180\% and the line fluxes by $-$70\% 
to 215\%, where the faint or blended lines deviate the most 
($| \Delta {\rm EW} |\,>$\,30\% and $| \Delta ({\rm line~flux}) |\,>$\,30\% in 
lines such as \siii \,\lam \,1260 [blended], \oiiisf \,\lam \,1664, \alii \,\lam 
\,1670, \feii\ \,\lam 1786, \aliii \,\lam \lam \,1854,\,1863, \nii \,\lam 
\,2141, \cii ]\,\lam \,2326).  
These discrepancies measured from the same data illustrate the significant 
uncertainties associated with line and continuum fitting in \qso\ spectra.

We can not easily confirm the suggestion by L97 that \aliii \,\lam \lam 
\,1854,\,1863 can probe the BLR size, as it appeared only marginally thermalized
(ratio = 1.25:1). Our measurements  of the narrow \aliii\ doublet\footnote{The
\aliii\ doublet components have equal widths to within the spectral resolution.}
(ratio = 0.9:1) suggests it is thermalized, possibly being emitted somewhat
closer to the continuum source, \ie inside the BLR outer radius.
We also find \mgii{} to be (entirely) thermalized (doublet ratio = 1:1); L97
measure a ratio = 1.2:1. Contrary to L97 we find a much better overall fit
to the narrow \mgii\ emission using more than two components; four components
are required to ensure an equal line width in each doublet (Table~\ref{elids}).
Thus it appears that neither \mgii{} nor \aliii\ is a suitable probe of the
outer boundary of the BLR (see discussion by L97).

\subsection{The \ciii\ Complex \label{C3_complex}}

The density-sensitive lines, \ciii \,\lam \,1909 and \siiii ]\,\lam \,1892 are 
important BLR density diagnostics. At first sight their relative strengths in
\izw\ are highly unusual: \ciii\ is $\sim$10 times weaker and \siiii ] 
significantly stronger\footnote{The critical density of \siiii ] is 
1.1$\times$10$^{11}$ \cmcub\ (Baldwin \et 1996; L97).} than usual leading L97 to 
argue for unusual high densities ($\sim$10$^{11}$ \cmcub).
However, in reality this comparison is difficult to make. Published measurements
of \ciii\ line strengths (\eg Wilkes 1986, Francis \et 1991, Baldwin \et 1995) 
are generally measurements of the entire \ciii\ complex including \siiii ], 
\aliii, and \feiii{} emission, which are all strong in \izw .  In typical \qsos\ 
with much larger intrinsic line widths these lines are so heavily blended that 
detailed deblending techniques often cannot uniquely determine the relative
contributions from individual transitions. Note, if \siiii] and \feiii{} are 
both strong at high densities the complex would not necessarily appear asymmetric 
at FWHM $\gsim$ 3500\,\kms. By artificially broadening the \izw\ spectrum to 
FWHM of 3000\,\kms{} and 5000\,\kms, a typical range for \qsos, we estimate the 
\ciii\ line strength measurements (line flux and EW) in a similar, broad-lined 
source would be overestimated\footnote{The \aliii\ line is excluded in the 
comparison.} by a factor of $\sim$2.

Detailed modeling of the full UV spectra of other NLS1s also concludes that these 
sources have high density emitting regions ($\sim10^{11} - 10^{12}$ \cmcub,
Kuraszkiewicz \et 2000), lending support to L97's conclusions.  Given that the 
strong \feiii\ UV34 multiplet is severely blended with the \ciii\ and \siiii] 
emission lines, we note that it is likely that both L97 and Kuraszkiewicz \et 
overestimate the strength of the \siiii] line as they underestimate the strength 
of the \feiii\ UV34 1895\AA{} transition.  The combination of a resulting smaller 
\siiii]/\ciii\ ratio and strong \feiii\ and \aliii\ contributions will in the 
framework of photoionization models (\eg Fig.~3e, Korista \et 1997) then further 
strengthen their conclusions that the BLR densities, at least in NLS1s, are very high.

A theoretical estimate of the relative triplet strength for the \feiii\ UV34 
\lam \lam 1895\,1914,\,1926 transitions for AGN physical conditions is not 
readily available.  The relative significances of the possible excitation 
mechanisms are not 
fully established in part due to the complexity of the \feii\ and \feiii\ 
emission. The \feiii\ UV34 excitation mechanism is probably not due to 
electron impacts owing to the high energy levels of the multiplet 
transitions (several eV) compared to the (typical) plasma temperature of 
$\sim$10$^4$\,K ($\sim$ 1\,eV).  The multiplet could be due to fluorescence 
and/or photoionization/recombination, as is likely to be the case for the UV 
\feiii\ emission in general (A.\ Pradhan 2000, private communication). 
In the optically thick regime all the transitions should have equal strengths 
due to thermalization (\eg L97). Hartig \& Baldwin (1986; hereafter HB86) 
estimate the relative \feiii\ UV 34 multiplet strengths by fitting the \lam 
\lam 1895, 1914, 1926 transitions in the spectrum of H0335$-$336 where this 
multiplet dominates the \siiii] and \ciii\ emission.  Assuming H0335$-$336 
has no \siiii] and \ciii\ emission at all, they find a multiplet ratio of 
0.9:1:0.3, while including \siiii] and \ciii\ in their fits yields a relative 
flux ratio of 0.9:1:0.7.  
The \izw{} spectrum is consistent with either of these ratios, as we discuss next. 

In order to estimate the likely contribution to the \ciii\ complex 
from the \feiii\ UV34 triplet, we modeled this multiplet with special 
emphasis on the relative transition strength, on varying its contribution 
and noting the effect on the \siiii] and \ciii\ model fits (see 
Table~\ref{modelfits}).  We adopted the approach by 
HB86, who use the emission profile of \feii\ UV191 to model the \feiii\ 
triplet (J. A. Baldwin 2000, private communication). We note that due to the 
strong blending in the \ciii\ line complex and the non-orthogonality of 
Gaussian functions, the model fits are not unique.  This is also clear 
from the fact that a range of reasonable model fits to the \feiii\ UV34 
emission can be made.  Assuming there is no 
\siiii] and \ciii\ emission at all (or broad emission thereof at least) 
a triplet ratio of 0.91:1:0.46 can be fit (not shown). However, the 
significant residuals at \lam \,1890 and \lam \,1907 strongly argue that 
\siiii] and \ciii\ emission is present.
Several fits were made both where the triplet ratio was fixed at 1:1:1 and
where it was allowed to vary freely. Sample fits are shown in Figure~\ref{fitex}e
and Table~\ref{modelfits} (Note that the triplet is located at 
\lam \lam 1893\,1912,\,1924).
The best fits were found for triplet models where the \lam\,1914\,\AA\ 
transition is the strongest. 
We cannot distinguish between the various fits nor between the optically 
thick and thin cases. More advanced modeling is required. 
A reasonable approach in fitting the \feiii\ in this region until more is 
known about the UV34 triplet may be as follows. If \aliii\ is strong, \feiii\ 
(including UV34) is highly likely to be present and similarly strong (HB86). 
If so, two fits can be made: one ignoring the presence of \feiii\ UV34, and 
one in which its contribution is maximized. This allows an upper limit 
to the fitting uncertainty involved to be estimated.

In Figs.~\ref{template}, ~\ref{feiiimodel}, and~\ref{fitex}d the \feiii\
template is shown with the UV34 triplet ratio 0.375:1:0.425 (model B)
for illustrative purposes only.
We note that other strong \feiii\ transitions may be present in the \ciii\
complex in addition to \feiii\ UV34, as indicated in Figure~\ref{feids}.

\subsection{The \siivoiv \,\lam \,1400 and \civ\ Features \label{L1400strength}}

The \siivoiv \,\lam \,1400 feature is strong relative to \civ\ 
(\lam 1400/\civ{} $\sim$1.3; Table~\ref{elids}), and the \civ /\lya{} 
ratio (0.11) is also low compared to the typical values for \qsos:  
0.1\,--\,0.6 and 0.2\,--\,0.6, respectively 
(Wilkes 1986; Francis \et 1991; Baldwin \et 1995). We note that \lam 1400/\civ\ 
measured from the non-scaled G130H spectrum (\S~\ref{scaling}) is also unusually 
strong ($\sim$1).
The line lists by Nahar (1995) and Nahar \& Pradhan (1996) indicate a 
(conservative) upper limit to the iron emission contribution to the \lam 1400
feature of $\sim$20\%. 
Thus, the \lam 1400/\civ{} line ratio indicates a density 
$n_e \gsim {\rm 10}^{\rm 11} {\rm cm}^{-3}$ according to the models 
by Rees, Netzer \& Ferland (1989), consistent with the estimate by
L97 of $n_e \,\sim {\rm 10}^{11}$\cmcub\ and typical NLS1 
densities (Kuraszkiewicz \et 2000).  
At high densities a simultaneous strengthening of the Silicon lines and 
weakening of the Carbon lines occurs (\eg Rees \et 1989). 

Another contributor to the large  \lam 1400/\civ\ intensity ratio may 
be an ionization effect due to the unusually red spectral energy distribution
of \izw\ [\eg similar to the explanation by Zheng \& Malkan (1993) of the 
Baldwin effect (Baldwin 1977)].
The luminosity brightening relative to the {\it IUE} data measurements 
(\S~\ref{dataproc})
appears to have increased these line ratios in agreement with Zheng \& Malkan 
(1993), with the \civ\ complex flux increasing the least ($\sim$75\,\%; \lya\ 
increased by $\sim$110\,\%; Wu \et 1983; Wang \et 1996; Rodr\'{\i}guez-Pascual \et 
1997).

The broad component of \civ\ is relatively weak compared to the other UV 
lines (Table~\ref{elids}; Fig.~\ref{fitex}c). 
Marziani \et (1996) define a strong broad \civ\ component in the {\it HST} 
spectropolarimetry data\footnote{Note, their flux level is different from our
spectrum and that of L97.} of \izw , but they fitted the component to emission 
which we identify as \siii , possibly \feii\ and blueshifted \civ. Given the
lack of deblending or detailed line identification their broad component fit has
an EW\,=\,21\,\AA{}, compared to ours of $\sim$9\,\AA{}.

\subsection{Is Iron Emission Associated with Outflows? \label{BaldwinABC}}

Baldwin \et (1996) study the nature of the kinematic components in the emission
line spectra of seven \qsos\ and, based on one of them, Q0207$-$398, propose 
that three different components are present in the line profile. They argue that
Q0207$-$398 is a ``misaligned'' broad absorption line (BAL) \qso\ as the 
blueshifted line emission in Q0207$-$398 is typically absorbed in BAL \qso\ 
spectra.  Baldwin \et also connect this blueshifted emission and absorption to 
the expanded photospheres of (bloated) stars close to the central source and 
argue that \aliii \,\lam \,1857 emission is an indicator of the presence of 
these stars.

The UV spectrum of Q0207$-$398 bears a remarkable resemblance to that of \izw\ 
(narrow line cores, strong \aliii , \feii , and \feiii\ emission, 
high density emitting regions, and blueshifted emission in the high ionization 
lines), and the different kinematical components in Q0207$-$398 are similar to
the emitting regions at different redshifts in \izw\ (Phillips 1976; van 
Groningen 1993), providing further support for the argument that \izw\ is (also)
a ``misaligned'' BAL \qso\ (L97). 
Baldwin \et (1996) connect the \aliii\ emission to outflows (and stars).  So as 
the \aliii\ and \feiii\ emission strengths appear to be connected (\eg HB86) we 
speculate that the iron emission itself is somehow related 
to the presence of the outflows and/or to the stars. The fact that the iron 
line profiles do not appear asymmetric or are significantly blueshifted 
indicates that this emission does not originate in the outflowing gas itself.  
In any case, strong iron (\feii{} and \feiii) emission may be connected with 
high densities (HB86; Joly 1991; Baldwin \et 1996; Lawrence \et 1997;
Kuraszkiewicz \et 2000).

\subsection{Is the Iron Spectrum of \izw\ Typical? \label{FespecTyp}}

An underlying assumption for using the \izw\ spectrum as an iron template is 
that the iron emitting mechanism in this target is similar to that of typical 
\qsos\ and that the spectrum provides a good representation of the iron 
transitions and iron line ratios observed in AGN and \qsos .  We therefore 
address the question of how typical the emission lines, including iron,
in \izw\ may be.

\izw\ has a number of unusual properties in addition to strong \feii\ emission.
It has strong \caii\ emission at \,\lam \lam \,8498,\,8542,\,8662 (the {\it 
infrared triplet}, permitted lines) and 
\lam \lam \,7291,\,7312 (forbidden lines) (Phillips 1976, van Groningen 1993, 
Persson \& McGregor 1985), strong \cstar \,\lam \,1175 emission (Laor 
\et 1997a; L97), strong \siiii ]\,\lam 1892, weak \ciii \,\lam \,1909, strong 
IR emission (\eg Rieke \& Low 1972; Rieke 1978) plus strong \feiii\ emission 
transitions in its spectrum 
(Figs.~\ref{emissmodel} $-$~\ref{feiiimodel}).  

Strong \caii\ emission is observed in $\sim$30\% of all \qsos\ and AGN 
(Netzer 1990) and is generally thought to be emitted in the deepest interior of 
BLR clouds (Persson \& McGregor 1985; van Groningen 1993) with very high
column densitites (N$_H > {\rm 10}^{\rm 24.5} {\rm cm}^{-2}$). 
Strong \caii\ emitters, however, do not otherwise appear different from those of 
the average AGN population (Netzer 1990).

\cstar \lam 1175 is not a commonly detected metastable transition of 
C\,{\sc iii} in AGN and has previously most often been detected in absorption 
(\eg Bromage \et 1985, Kriss \et 1992), perhaps due to the difficulty in
identifying very broad, but weak features (\S~\ref{FeCorrection}). See \eg 
Laor \et (1995) and Hamann \et (1997) for weak and marginal detections in 
emission.  The line may result from the enhanced density (see discussion by 
L97) of the emitting medium already deduced for \izw. Several studies show a 
trend toward strong \feii, \feiii, and/or \aliii{} at higher BLR densities 
(\S~\ref{BaldwinABC}).  If high densities are a common property
of strong iron emitters, the iron spectrum is not expected to be unusual.

Strong \feiii\ features are more prevalent in \izw{} (Table~\ref{feids}) than
in previous AGN studies, although \feiii\ features at 2070\,\AA{} and 
$\sim$2420\,\AA{} have been reported in the past (HB86; L97).
Our study suggests, however, this difference is not real but due to
misidentifications in the past.
Francis \et (1991) note unidentified features in the \lam \lam 2000$-$2200\AA\ 
and \lam \lam 2900$-$3200\AA\ regions, consistent with \feiii\ lines according 
to our identifications (see references to Table~\ref{feids}). 
The presence of both ions is potentially important for deriving the physical 
conditions in the iron-emitting regions, because the details of the iron 
emission spectrum are highly sensitive to the physical conditions (\eg Netzer 
1980; D.\ Verner, 1997, private communication).

Lipari \et (1993) and Lipari (1994) link {\em extremely} 
strong optical \feii\ emission to starburst activity. Though \izw\ is currently
undergoing vigorous star formation (based on Barvainis \et 1989; Eckart \et 
1994; Sanders \& Mirabel 1996), the iron emission observed in AGN is not 
characteristic of emission from star forming regions.
Pure starburst galaxies (\eg NGC\,7714) do not emit permitted \feii\ 
emission, only IR forbidden lines (\eg [\feii ]\,1.6\,$\mu$m; L.\ Ho, 1998,
private communication) mainly because the stellar ionizing continuum is too 
soft. 
To obtain the permitted lines the hard, non-thermal continuum from a central AGN 
source is needed to penetrate to the high-density interior of the BLR clouds 
(\eg Netzer 1990).

Based on the above discussion, we find no strong evidence that the iron emission
in \izw\ is unusual compared with that of the general \qso\ population.
This is confirmed by our successful fitting of the iron emission in several \qso{} 
spectra (\S~\ref{sample_Fecleaning}). 

\section{Summary and Conclusions \label{summary}}

We have presented a UV iron template based on {\it HST} archival data of \izw , 
and described the method with which the template was generated. 
Compared to previous empirical templates (Corbin \& Boroson 1996) this covers
a large range (\lam \lam 1250 $-$ 3090\AA), was generated by careful fitting
of the non-iron emission and absorption lines, and allowed the generation of
separate \feii\ and \feiii\ templates.
We have demonstrated its application to fit and remove the \feii\ (and \feiii)
emission in spectra of several \qsos, including 3C273, allowing subsequent 
studies of weak and heavily blended emission lines (\eg \ciii ) free from the 
large uncertainties otherwise associated.
This shows that the iron emission in \izw\ is sufficiently similar to that 
in other broad-lined \qsos\ to be useful.  Although it has limitations, 
the iron template fitting process is an important tool to eliminate and
study the iron emission in active galaxies, at least until accurate
theoretical models are developed.

We confirm previous results reporting the presence of blueshifted emission (\ie 
blue profile asymmetry) in the spectrum of \izw , especially in the 
high-ionization lines (L97), and of several emission regions of 
different (absolute) redshift, line width and physical conditions (Phillips
1976; van Groningen 1993).
The presence of these regions along with the UV spectral properties are 
consistent with the picture suggested by Baldwin \et (1996) in which the 
blueshifted emission originates in the same region giving rise to the BAL 
troughs in BAL \qsos . This may connect the NLS1s with BALs, perhaps through 
source orientation.  We also argue that (strong) iron emission may be connected 
with high densities and to outflowing material.

We discuss the fact that the \siiv +\oiv \,\lam \,1400 feature is very strong  
relative to \civ. This can be explained by a simultaneous weakening of the 
Carbon lines and strengthening of the Silicon lines, common to Narrow Line 
Seyfert 1s (Kuraszkiewicz \et 2000), probably due to a combination of high 
densities and a low ionization parameter.

Based on earlier {\it IUE} measurements we find \izw\ to have brightened a 
factor $\sim$2 in its continuum emission since 1978\,--\,1982.

\acknowledgments

We are grateful to 
Dr.\ Adam Dobrzycki for help and guidance on the \izw{} {\it HST} data processing, 
Drs.\ Luis Ho and Ari Laor for comments on early versions of the paper, and
Drs.\ Kirk Korista, Ari Laor, Anil Pradhan, Dima Verner, and Beverly Wills
for useful discussions. We also owe thanks to Bev Wills for kindly providing 
digital data of the 3C273 UV-optical spectrum and the associated \feii\ models 
presented by Wills \et (1985).  We are grateful to an anonymous referee 
for very helpful comments, leading to significant improvements of this paper.
Many thanks are also due to the IRAF help desk, in particular Frank Valdes and 
Mike Fitzpatrick, for invaluable help with and guidance through IRAF problems.
MV is very pleased to thank the Smithsonian Astrophysical Observatory for their 
hospitality and gratefully acknowledges financial support from the Danish 
Natural Sciences Research Council (SNF-9300575), the Danish Research Academy
(DFA-S930201), a Research Assistantship at Smithsonian Astrophysical 
Observatory (NAGW-4266, NAGW-3134, NAG5-4089; P.I.: Belinda Wilkes), and  
the Columbus Fellowship at The Ohio State University.  BJW 
gratefully acknowledges financial support from NASA contract NAS 8-39073 
(Chandra X-ray Center).

\appendix
\section{Comments on Individual Spectral Features in \izw{} 
\label{linecomments}}

Unless otherwise noted, the identifications of \feii\ and \feiii\ emission
features are made using the iron line lists available on the web and from recent
publications (Fuhr \et 1988; Penston \et 1983 ; Ekberg 1993; Giridhar \& 
Arellano Ferro 
1995 ; Nahar 1995; Nahar \& Pradhan 1996; Quinet 1996; Quinet \et 1996; Kurucz 
\& Bell 1995).
\\ \noindent
When (iron) residuals are said to be removed or
excluded (\ie subtracted) from the template, it is done in order to prevent an 
overestimation of the iron emission strength and subsequent overcorrection for 
the iron emission when the template is applied to \qso\ spectra.
In addition, some line fits may require a local continuum level (different
than the global continuum setting). However, the EW measurements are always
with respect to the global continuum.
\vskip 0.2cm \noindent
{\bf \lam \lam 1120$-$1135 emission}: \feiii\ UV1 emission is clearly detected, 
but is heavily affected by absorption (Figs.~\ref{emissmodel} 
and~\ref{template}). Due to the uncertainty in correcting for
this absorption in the noisier end of the spectrum, we chose not to include
the region \lam 1075$-$1135\AA\ in the template. 
\vskip 0.01cm \noindent
{\bf \cstar \lam \,1176 emission}: This emission line 
(Figs.~\ref{emissmodel},~\ref{template}, and~\ref{fitex}a) is an excited 
meta-stable level of C\,{\sc iii} and rarely detected in AGN spectra [but see
Laor \et (1995); Hamann \et (1997), and see for absorption detections Kriss 
\et (1992); Bromage \et (1985)].  
Its implications are discussed in \S~\ref{FespecTyp}, and Laor \et (1997a, 
1997b) 
also discuss possible excitation mechanisms of this transition.
\vskip 0.01cm \noindent
{\bf \lam \lam 1150-1245 range:} Faint emission of \feii\ is expected in this 
\wav\ range (Figs.~\ref{emissmodel} and~\ref{template}). The leftover residuals 
from subtracting the fits to the very strong, non-iron lines (especially \lya ) 
are rather noisy and do not obviously match expected \feii\ transitions. We 
therefore choose to remove these residuals from the template 
(Fig.~\ref{feiiimodel}). 
\vskip 0.01cm \noindent
{\bf \lya \lam 1216 emission}: Two narrow components are necessary to reproduce
the shape of the narrow line core (Fig.~\ref{fitex}a). The weaker of the two 
components is 
blueshifted ($\sim$900\,\kms ) relative to the peak position of the stronger 
component. The high S/N data constrain the fit well. Although the \lya\ profile 
does not closely resemble that of the Balmer lines (cf.\ \ha\ and \hb\ 
modeled by BG92, and L97), the profile shapes of all three
are well reproduced by one broad and two narrow Gaussian components, one of 
which is blueshifted resulting in an asymmetric profile. 
\vskip 0.05cm \noindent
{\bf \ov \lam 1218 emission?}: 
The feature clearly appears once the fit to \lya\ is subtracted and the line 
position is well defined at 1218\AA\ (Fig.~\ref{fitex}a).
The identification (Penston \et 1983) is uncertain as one would expect a
high blueshift of the line similar to the other high-ionization lines.
\vskip 0.05cm \noindent
{\bf \nv \,\lam \lam 1238,\,1243 emission}: When the absorption is corrected, a 
blue asymmetric profile (Fig.~\ref{fitex}a) is left, similar in 
appearance to the non-deblended profiles of \siivoiv\ and \civ . The doublet 
profile is fitted well 
with three Gaussian components, one of which is considerably blueshifted 
($\sim$2500\,\kms )
relative to the \lam 1243 line. The remaining doublet feature is 
blueshifted $\sim$900\,\kms\ similar to the blueshifted \lya\ emission, relative
to the rest frame.
\vskip 0.05cm \noindent
{\bf \lam \lam 1230,\,1234 absorption}: L97
identify this absorption (Fig.~\ref{fitex}a) as due to ($\sim$2000\,\kms ) 
blueshifted \nv\ absorption intrinsic to \izw, 
supported by their detection of weak associated absorption in \lya\ and \civ\ 
also at this blueshift. 
\\
The absorption lines cannot be due to \oi \,\lam 1306 and \siii \,\lam 1309,
although the \wavs\ coincide very well with their laboratory \wavs, as these
transitions originate in slightly excited states which 
are not common in the ISM (Savage \et 1993). 
\vskip 0.05cm \noindent
{\bf \lam 1257 -- 1268 emission}: This feature is identified as \siii \lam 1263
(Fig.~\ref{fitex}a).
Due to the detection of \feii\ UV9 emission at \lam \lam 1270$-$1280, \feii\ 
(UV9 \lam \lam 1250$-$1280\AA ; Fig.~\ref{feids}) is also expected to be 
blended with 
this \siii\ multiplet.  The \wavs\ of the Gaussian components (Table~\ref{elids})
fitted to the \siii\ line blend are uncertain, due 
in part to the short \wav\ emission being blended with
the \lya\ fit (and hence partly subtracted) and in part to blending with \feii .
\vskip 0.01cm \noindent
{\bf \siiii $^{\star}$ \lam 1297 emission?}: The identification of this feature 
is not confirmed (\eg Fig.~\ref{template}). It is not clear whether the emission
is due to \siiii $^{\star}$ emission or to a blueshifted component of \oi 
\,\lam1306\,+\,\siii \,\lam 1309, similar to that found in the 
high-ionization lines. In the latter case the blueshift is $\sim$2000\,\kms , 
in agreement with the blueshift of the high-ionization lines. 
A third possibility is faint \feiii \,\lam 1298.6 and \feii \,\lam \lam 
1297,\,1299 emission which is expected based on the iron line lists, 
although the observed emission feature appears too strong.
\vskip 0.01cm \noindent
{\bf \oi \,\lam 1304 emission}: Faint \feii\ appears, especially in the blue 
wing, but the iron emission could not be deblended (Fig.~\ref{template}).
\vskip 0.01cm \noindent
{\bf \lam 1343 feature}: The identification is uncertain (Figs.~\ref{emissmodel}
and~\ref{template}). No obvious \feii, 
\feiii\ or other commonly observed \qso\ broad emission line  matches the \wav\ 
position. Candidate identifications are O\,{\sc iv}, \caii, \feiii\ and \fev. A 
coincidence occurs with some \feiii\ transitions, as noted in Table~\ref{elids},
but it is not clear whether an isolated feature, as observed, is expected.
\caii\ is also observed in the IR (\S~\ref{FespecTyp}), but this \lam 1343
feature is redshifted with respect to the expected \caii\ position 
(Table~\ref{elids}).
\vskip 0.01cm \noindent
{\bf \siivoiv \,\lam 1400 emission}: Faint \feii\ emission is expected 
at \lam 1393 and \lam 1408. A faint \feiii\ feature is expected at \lam 1395,
which is identified in the residuals after subtracting the \siivoiv\ fits. The 
strength of \feiii{} \lam 1395 is likely to be underestimated: the feature is
heavily blended with the \siivoiv\ emission and the data do not permit 
sufficient constraints to be placed on each contributing emission component.
\\
Blueshifted emission is clearly present in this line complex 
(Fig.~\ref{fitex}b). A deblending of 
each of the components is not straight forward due to the severe blending of
the components emitted in the restframe with the blueshifted line emission.
Hence, each individual component has a larger uncertainty than the complex fit
as a whole. We interpret the complex as follows: the ``restframe'' emission has
a general blueshift of $\sim$900\,\kms\ relative to the \mgii\ line peak
(consistent with \lya , \nv , \civ , and \aliii ) and the excess blueshifted 
emission has a projected velocity of 
$\sim$\,--1100\,\kms\ relative to the former (``restframe'') emission and a 
total
blueshifted velocity of $\sim$2000\,\kms\ relative to the true restframe. The
complex clearly has a broad component, which displays the strongest blueshift of
2000\,\kms\ (Table~\ref{elids}).
\\ 
The line complex as a whole is strong relative to \civ , most likely because 
\civ\ is weak and/or the density is high and the ionization parameter is low 
(see discussion in 
\S~\ref{L1400strength}). Alternatively, the iron emission may be stronger than 
we currently suspect, making the \siivoiv\ emission appear stronger and 
contributing to the high \lam 1400/\civ\ line ratio.
\vskip 0.01cm \noindent
{\bf N\,{\sc iv}] \lam 1486 emission}: This emission line is not detected 
(Figs.~\ref{template} and~\ref{fitex}c). 
It may, however, be weak and heavily blended with iron emission.
\vskip 0.01cm \noindent
{\bf \lam \lam 1515,\,1517 dips}: These features (Fig.~\ref{fitex}c) appear very
similar to some of the 
Galactic absorption doublets, but comparison with the low dispersion {\it IUE} 
spectrum (Lanzetta \et 1993) clearly show that this dip is a lack of emission 
surrounded by line emission, which we identify as \feiii\ (Table~\ref{feids}).
\vskip 0.01cm \noindent
{\bf \siii \,\lam \lam 1527,\,1533 emission}: This doublet transition is almost 
certainly present (Fig.~\ref{fitex}c).  The lines are, however, strongly blended
with \civ\ and possibly iron emission, 
and the individual line profiles and their strengths cannot be determined with
confidence. Due to the apparent weakness of the broad \civ\ component we suspect
the contribution from the \siii\ doublet to be \mbox{significant.} An 
alternative 
interpretation is that \civ\ has a strong and rather blueshifted component of 
$\sim$6000\,\kms\ relative to the restframe. In that case, there must be several
blueshifted emitting regions as the observed emission does not appear to reflect
the doublet flux ratio which is estimated for the narrow \civ\ components 
(Table~\ref{elids} and Fig.~\ref{fitex}c).
\feiii\ UV84 and weaker \feiii\ features are also expected to be present (see
below).
\vskip 0.01cm \noindent
{\bf \civ \,\lam \lam 1548,\,1550 emission}: Blueshifted emission appears 
present (Fig.~\ref{fitex}c) but, due to the heavy blending, a well constrained 
Gaussian component fit (deblending)
to determine the blueshift and flux is not possible. 
The data only support a weak broad component
(Fig.~\ref{fitex}c; Table~\ref{elids}).
Narrow profiles were fitted to the peaks of the \siii \,\lam \lam 1527,\,1533 
and \civ \,\lam \lam 1548,\,1550 doublet lines which are clearly resolved. 
The best fit to the \civ\ narrow emission has slightly different widths of
the transitions (Table~\ref{elids}). If the line width is fixed and a flux
ratio of 2:1 is approximated a Gaussian residual appear 
(see the alternative fit to the narrow \civ\ emission in Table~\ref{elids}).
In addition, a two-component feature at \lam 1536 and \lam 1539, coinciding with 
\feiii\ UV84 emission (see below and Fig.~\ref{feids}), is required to fit the 
complex well. 
An alternative fit has two narrow components with a flux-ratio of approximately 
2:1 (Table~\ref{fefits}) suggesting an additional blueshifted 
($\sim$2300\,\kms ) component of \civ\ emission.  The likelihood that this 
emission is \sii\ is probably rather low as expected strong transitions at 
$\sim$\lam 1542, \lam 1547, \lam 1553, and \lam \lam 1550\,--\,1600\,\AA\ 
(Morton 1991) are not seen. 
\vskip 0.01cm \noindent
{\bf \feiii{} \lam 1520 $-$ 1540 emission}: The UV84 multiplet of \feiii\ among 
other \feiii\ lines of detectable strengths (\lam \lam \,1515.5, 1516.2, 1518.8, 
1524.5, 1525.0, 1526.0, and 1527.0; see Table~\ref{fefits} and Fig.~\ref{feids}).
are expected close to the positions of \siii\ and \civ; the 
most probable lines are expected at \lam \lam 1526, 1527, 1531--1532, 1538--1539, 
and 1550--1551 (Nahar \& Pradhan 1996). If
these transitions are present \siii\ and \civ\ may be even
weaker relative to the other resonance UV lines.
The {\it IUE} spectrum presented by Lanzetta \et (1993) does not show any signs 
of strong \feiii\ emission in this region. But since \izw\ has brightened by a 
factor of $\sim$2 (\S~\ref{dataproc}) in continuum flux and both the line
strengths and the continuum slope have changed since the {\it IUE} data were 
taken, this is not a strong argument against the presence of \feiii\ in the 
{\it HST} spectrum. 
Due to the uncertainty in the identification and strength of this possible
iron emission feature, it is not included in the \feiii\ and \feii\ templates.
\vskip 0.01cm \noindent
{\bf \lam 1807-1875 emission}: \feiii\ (UV97,117 at \lam \lam 1830$-$1855\AA\ 
and UV53,63 at \lam \lam 1850$-$1871\AA ; see Fig.~\ref{feids}) is the most 
likely identification, but fainter \feii\ (UV65,66) emission is also consistent 
with some of the residuals in the range \lam \lam \,1807\,--\,1875\AA\ 
(Fig.~\ref{fitex}d). \feii\ is, however,  expected to dominate in the range \lam 
\lam \,1807\,--\,1836\AA\ (most of the weaker, blended iron multiplets are not 
marked in Fig.~\ref{feids}).
Due to the heavy blending, no attempt were made to separate the individual 
\feii\ and \feiii\ contributions. A compromise was 
made: the emission at \lam \lam \,1837\,--\,1872\AA\ is treated entirely as 
\feiii\ emission (and included in the \feiii\ emission model and \feiii\ 
template) while the \lam \lam \,1802\,--\,1836\AA\ \wav\ range is adopted as 
pure \feii . This is clearly an approximation.
\vskip 0.01cm \noindent
{\bf \aliii \,\lam \lam 1854,\,1863 emission}: The doublet is relatively strong,
and well resolved in this spectrum (Fig.~\ref{fitex}d). A broad base component 
is clearly present along with a number of iron emission transitions. 
We detect no obvious excess blueshifted emission and find that the most likely 
identification of the non-\aliii\ emission 
in the \lam 1850$-$1875\AA\ region is that of iron (Figs.~\ref{feids} 
and~\ref{fitex}d).
\vskip 0.01cm \noindent
{\bf \siiii ] \,\lam 1883 emission:} This was identified by L97 as possible 
\siiii ] emission (Fig.~\ref{fitex}d). The feature is observed at \lam 1880.3\AA 
. 
The \feii\ UV126 multiplet is also expected (and consistent with the emission)
at \lam \lam 1864.6,1864.7,1880.97, providing an alternative identification.
The 1864\AA\ lines are blended severely with the \aliii\ emission and the 
identification cannot be confirmed via line strength arguments. When in doubt we
prefer to underestimate the iron emission (see the introduction to this section 
and \S~\ref{fitmethod}), so we do not include this feature in the template. 
\vskip 0.01cm \noindent
{\bf \feiii \,\lam 1892-1906 emission:} 
The residuals between \lam 1898 and 1907 were not fitted with Gaussian 
components (Fig.~\ref{fitex}d). However, they were isolated by subtraction of 
fitted non-iron lines and then included in the \feiii\ template 
(Fig.~\ref{feiiimodel}).  With the \feiii\ UV34 \lam \,1914 emission feature  
observed at $\sim$1912\AA{}, the 1895\AA\ transition is expected at 1893\AA, the 
position of a weak feature (Fig.~\ref{feids}). Given the apparent weakness of
the feature at \lam \,1893\AA\ the \feiii\ UV34 multiplet is not likely to 
dominate the strength of \siiii] \,\lam1892, detected at $\sim$1890\,\AA\
(Fig.~\ref{fitex}e).
The UV34 triplet (1895\AA, 1914\AA, 1926\AA) may have optically thin line 
ratios 0.9:1.0:0.3 (\S~\ref{C3_complex}) while all transitions have 
equal strengths in the optically thick limit. As discussed in 
section~\ref{C3_complex}, our modeling of the \feiii\
UV34 triplet show that no tight constraints can be placed on the relative
line ratios. However, this modeling does indicate an optically thin ratio
(cf.\ models B and C in Fig.~\ref{fitex}e).
The spectrum indicates that this triplet is emitted in the optically thin
region given the apparent relative weakness of the 1895\AA{} and 1926\AA{} 
features. The relatively strong \siiii]/\ciii\ line ratio combined with the
relatively strong \feiii\ emission indicates rather high BLR densities.
\vskip 0.01cm \noindent
{\bf \ciii\ \lam 1909 emission complex}: The emission feature 
is a complex blend of emission lines, but the spectral resolution and the narrow widths 
of the line cores permit a separation of the narrow-line cores of the \siiii ] 
\,\lam 1892, \ciii \,\lam 1909 and \feiii \,\lam \lam 1914,\,1926 lines 
(Fig.~\ref{fitex}d). Two broad components of \siiii ] and \ciii\ could be 
fitted (Table~\ref{elids}), although their individual fits, including their
strengths, are not well constrained. A number of other \feiii\ transitions are 
detected in and around this line complex (Fig.~\ref{feids} and 
Table~\ref{fefits}).
\vskip 0.01cm \noindent
{\bf \feiii \,\lam 1914 emission}: The emission at \lam \lam 1907$-$1918
is consistent with contributions from a number of intermediate to strong \feiii\
multiplets; among them, the strongest are UV34, 101, and 83 and the weaker ones 
are UV57, 135, and 108. This heavy blending is probably responsible for the 
residual UV34 (\lam 1914) feature appearing stronger than, \eg UV68 \lam 1952, 
and UV50 \lam \lam 1987 -- 1996 (Fig.~\ref{feids}).  If UV51 (\lam \lam 
1915.1,1922.8,1930.4,1937.3,1943.4) is present, it must be faint. 
See section~\ref{C3_complex} for discussion of the relative \feiii\ UV34
contribution.

\vskip 0.01cm \noindent
{\bf \nii \,\lam 2141 emission}: The pseudo-continuum was used as the local 
continuum level (Fig.~\ref{template}).
\vskip 0.01cm \noindent
{\bf \lam 2160-2190 region:} 
The absolute strengths of the emission and absorption lines are not 
straightforward to determine in the $\sim$2100$-$2250\AA\ region due to the 
presence of dust absorption features (\S~\ref{dataproc}) and the uncertainty in 
the placement of the absolute continuum level (Fig.~\ref{template}).
According to the line lists of Nahar (1995), the expected 
\feii\ transitions are faint between 2100 and 2200\AA . 
\vskip 0.01cm \noindent
{\bf \lam \lam 2250-2290 region}: The possible \feiii\ UV73,153 (\lam \lam 2258.1, 
2274.7, 2277.6, 2285.7) emission is not fitted and not included in the \feiii\ 
emission model as the identification is uncertain, due to poor \wav\ 
coincidence. 
\vskip 0.01cm \noindent
{\bf \feiii\ UV47 \lam \lam\ 2418,2438 emission:}  The profile was fitted with 
three Gaussian components, one of which accounts for a broad component and the 
weaker of the two narrow components is blueshifted $\sim$500\,\kms\ with respect
to the stronger one (Fig.~\ref{fitex}f). The choice of including a broad 
component was based on the appearance of the \ha\ profile (Phillips 1977; L97).
\\
Unfortunately, the spectrum does not offer good constaints on the absolute
strengths of the broad \feiii\ UV47 component fit. A good fit requires that the 
actual continuum level for the fitting be placed 
$\sim$6\% below the observed flux level at the apparent line base.
\\ 
The profile is also consistent with a fit of two 
narrow Gaussian components (and no broad component) and the actual continuum
level (for the fitting) is placed at the observed flux level, but with the known
similarity between the (Balmer) hydrogen lines and the \feii\ lines, 
the presence of a broad \feiii\ component is much more likely (see also L97).
\vskip 0.01cm \noindent
{\bf \lam 2435 emission feature}: The identification is uncertain. Possibilities
include the (slightly blueshifted) O\,{\sc ii} (UV18) \lam \lam 2433.6, 2444.4, 
2445.6. Its blueshift ($\sim$775\,\kms ) is consistent with that of \oii \lam
2470. Some Si\,{\sc i} transitions are also expected nearby (Si\,{\sc i} UV45 
and UV2; Table~\ref{elids}). The feature was fitted with the pseudo-continuum 
as the local continuum. The best fit was obtained with two Gaussians 
(Fig.~\ref{fitex}f); it is not clear how the individual Gaussian parameters 
(width and position) relate to those of the multiplet components.
\vskip 0.01cm \noindent
{\bf \oii \,\lam 2470 emission}: None of the lists of \feii\ and \feiii\ 
transitions (including that by Giridhar \& Arellano Ferro 1995) predict strong 
transitions at this \wav . L97 identify the feature as \oii\ 
(Figs.~\ref{template} and~\ref{fitex}f). 
The pseudo-continuum acts as the local continuum level in the component fitting.
\vskip 0.01cm \noindent
{\bf \lam 2481 emission feature}: The identification is uncertain. No obvious
strong iron emission is expected at this position. \ci \,\lam \,2478 is a 
possible
identification. The fit is based on the pseudo-continuum (Fig.~\ref{fitex}f).
\vskip 0.01cm \noindent
{\bf \mgii \,\lam \lam 2795,\,2803 emission}: 
The systemic redshift was measured with a two-Gaussian component fit (one
doublet; \S~\ref{fitmethod}) to the narrow line core early in the course of
this work. A more detailed and improved line fit reveals a second doublet
with slightly larger line widths blueshifted 400\,\kms{} relative to the
stronger doublet at rest in the quasar frame (Fig.~\ref{fitex}g and Table~\ref{elids}).  
Each doublet has a thermalized line ratio (=\,1:1).

The fit displays evidence for excess blueshifted emission similar to that in
\lya , \nv , \heii\ and possibly \siivoiv , and \civ .
L97 fitted the entire doublet (broad and narrow components) using
\ha\ as a template profile. Two such profiles with relative strengths 1.2:1
fitted the line well, and they argued for higher electron densities in the 
\mgii\ line gas based on the thermalized doublet ratio ($\sim$1:1). 

It is clear from Figure~\ref{fitex}g that no iron emission is left in the 
\lam 2770$-$2820\AA\ range after subtraction of the \mgii\ fit.  \feii\ 
emission is not expected to be strong at the position of \mgii\ (D.\ Verner, 
1997, private communication), but is not entirely 
absent either (see Fig.~\ref{feids} and Verner \et 1999). 
As the data cannot constrain the strength of the broad component well, we 
prefer to underestimate the \feii\ strength to prevent an overcorrection of the 
iron emission when the iron template is applied (see \S~\ref{fitmethod}).
This choice of the local underlying continuum is consistent with that found
by L97.

\section{The 3C273 {\it HST} Data \label{3c273data}}

In \S~\ref{sample_Fecleaning} \HST\ archival data, covering the full UV
range from \lya\ to beyond \mgii, of the nearby radio-loud quasar, 3C273, 
were used to demonstrate the applicability of the \izw\ based \feii\ and 
\feiii\ templates to the iron emission in other AGNs. The 3C273 UV spectrum 
was generated by combining data from the FOS and the Space Telescope Imaging 
Spectrograph (STIS). We used the 1\arcsec\ aperture G130H, G190H, and G270H 
FOS grating spectra from 1991 January 16, 14, and 15, respectively, which 
only partially cover the \feii\ bump around \mgii. 
The spectral range was extended longward of \mgii\ by combining these spectra 
with a 52$\times$2 aperture G430L STIS grating spectrum observed on 1999 January 
31.  The latter spectrum was multiplied by a factor 1.142 to coincide with the
average flux level of the FOS spectra. The G270H spectrum was truncated at
3200\AA\ (observed) before being combined with the G430L spectrum, which then
provided the \mgii\ line profile. There is a $\sim$50\AA{} gap between the G130H 
and G190H spectra. An interpolation was performed using the average continuum
flux levels in the $\sim$40\AA\ ends of each spectrum, avoiding absorption lines,
to facilitate the fitting of the iron emission.
The spectra were resampled to a common dispersion of 2.37\AA/pix in the rest frame
based on the G430L spectrum. The lower resolution of the G430L spectrum did not 
affect the iron fitting in any way. This was confirmed by a simultaneous fitting 
to the higher resolution spectrum made from the FOS spectra only (0.44\AA/pix, 
rest frame; no fitting to the full `small iron bump' was possible, however) 
which showed no significant differences in the fitted models. This is most likely 
due to the relative broadness ($\sim$4000\kms) of the lines in the spectrum 
(\S~\ref{sample_Fecleaning}).

\newpage
%

\clearpage

\figcaption[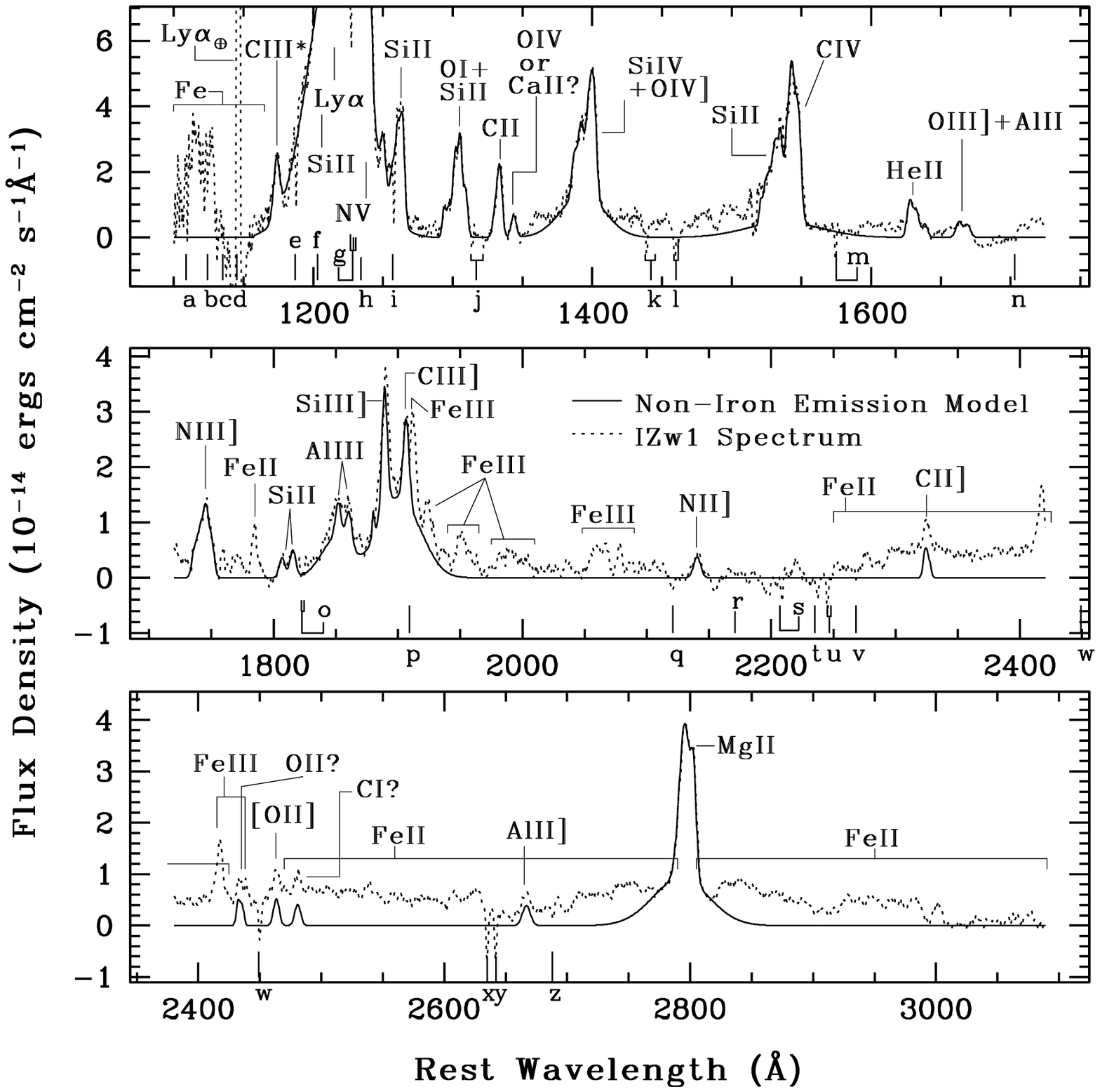]{The non-iron emission model (solid line) 
overplotted on the continuum-subtracted \izw\ spectrum (dotted line).  
The assumed continuum of the {\it HST} spectrum is a broken power-law, 
F$_{\nu}\,\sim\,\rm \nu^{-\alpha}$ with a break at \lam 1716\AA{} (see text
for details). 
The absorption lines are marked using letters a--z below the spectrum and listed
in Table~\ref{alids}.  The detected emission lines are labeled.
Geocoronal \lya\ is seen at \lam 1145\AA{} (restframe). 
\label{emissmodel}}

\figcaption[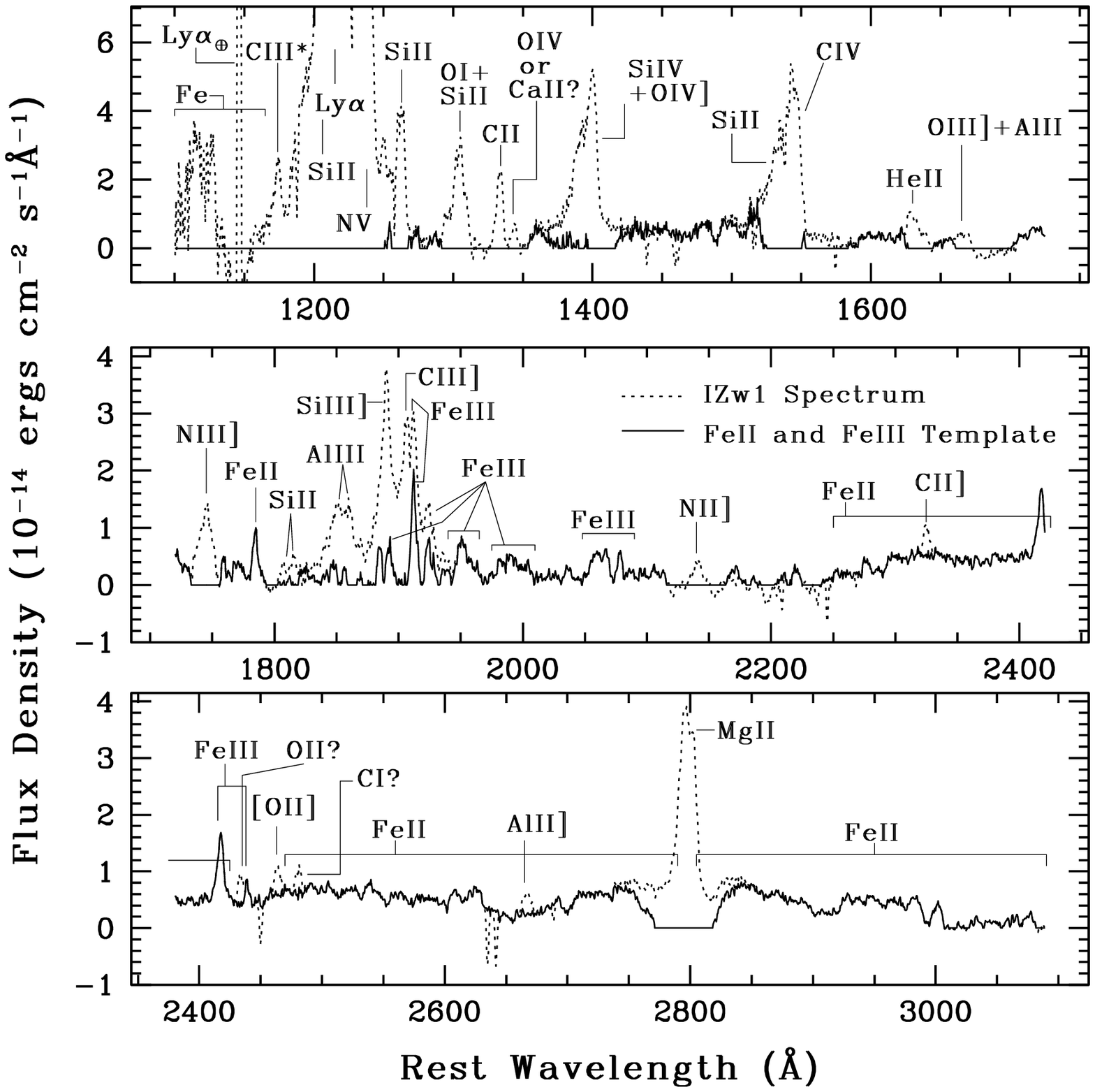]{The iron template (solid line; sum of \feii\ and 
\feiii ) overplotted on the \izw\ spectrum (dotted line). 
The detected emission lines are labeled.
\label{template}}

\figcaption[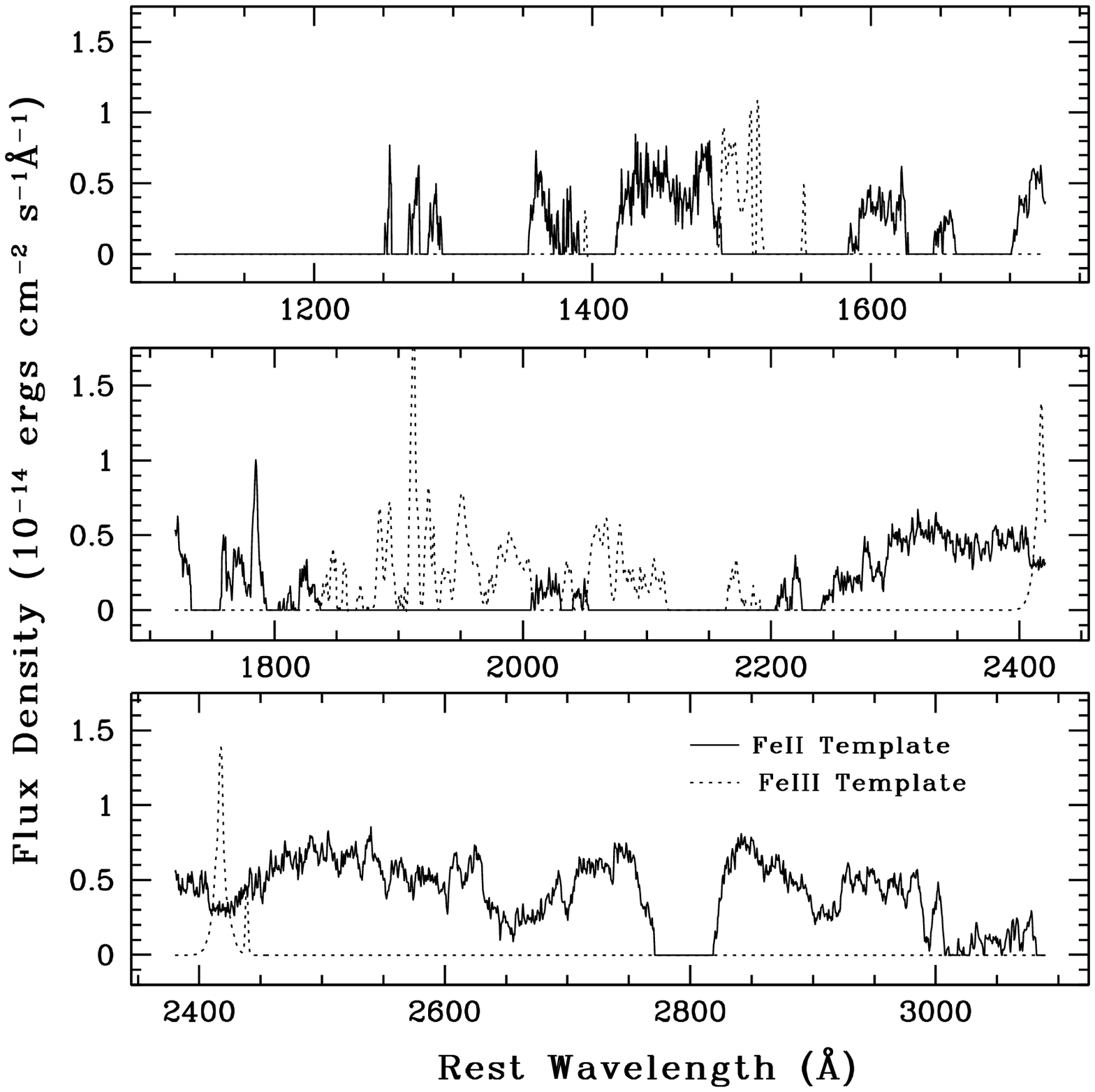]{The \feiii\ template spectrum (dotted line) 
overplotted on the \feii\ template spectrum (solid line).  Most of the
\feiii\ features are fitted with Gaussian profiles. \label{feiiimodel}}

\figcaption[ftab4.eps]{
Suggested identifications of individual, mostly unblended UV \feii\ and 
\feiii\ emission multiplets based on the positions and (rough) strengths
listed by Moore (1950). Each label consists of the ionizatation level
and the multiplet number separated by a hyphen. Multiplets with labels in
parentheses are slightly blended. Square brackets denote multiplets which
are suggested present by the fitting process (see Appendix~A; `2-10' is
expected only).  Heavily blended multiplets are not marked. 
Multiplets with labels ending with an `m' have relative transition strengths 
inconsistent with the optically thin laboratory strengths. 
Note that the positions in the spectrum of the multiplet transitions are
sometimes blueshifted by 1-2\,\AA{} relative to the laboratory wavelengths
marked. This is particularily evident for \feiii \,UV34 at $\sim$1912\,\AA{}
(labeled `3-34m'). See \S~\ref{C3_complex} and Appendix~A regarding the 
relative strengths in this multiplet.
\label{feids}}

\figcaption[f5.eps]{Examples of line fitting. Each individual component 
is shown as well as their sum (dot -- dashed line) superposed on the data (solid line). 
The non-iron emission is dotted while the fitted iron emission is dashed. 
Note, the fit is sometimes so good it entirely coincides with the spectrum. 
{\bf a)} \lya\ complex; note the strongly blue-asymmetric \nv\ profile. 
This is also seen in the non-deblended \civ\ and \mgii\ profiles 
(App.~\ref{linecomments}), {\bf b)} \siiv +\oiv \,\lam \,1400 blend, {\bf c)} 
\civ\ complex; `?': the identification of this feature is uncertain; see text, 
{\bf d)} \feii\ UV191, and \ciii\ complex [Note, the \znii\ ISM absorption at 
1910\AA. Both the sum of the non-iron emission line fits and the sum of 
non-iron and iron fits are shown in dot-dashed curves. The latter (full) sum 
coincides so well with the data in many places that it is not easily seen, \eg 
around \siiii ], \ciii, and \feiii\ redward of \ciii .
The fit to \feii\ UV 191 also coincides well with the data].
The fit to \feiii\ UV34 is model B.
{\bf e)} Other sample model fits to \feiii\ UV34 as discussed in text. The
individual fits to \siiii] and \ciii\ are not shown, but the residuals are. 
{\bf f)} \feiii\ UV47 \lam \,2418,\,2438, and the non-iron features nearby, and 
{\bf g)} \mgii . 
\label{fitex}}

\figcaption[f7.eps]{The {\it HST} spectrum of 3C273 (Q1226$+$023) and the best
fitted iron `models'. The upper, large panel in (a) shows the original spectrum
(shifted for clarity; see \S~\ref{sample_Fecleaning}) with the continuum
overplotted (top) compared to the residuals (middle) after subtracting the
best fitted iron `model' (bottom). The latter is displayed separately below
the large panel for clarity. Panel (b) shows the original spectrum, two iron `model'
fits, and the corresponding spectrum residuals. Neither of the iron `models' 
fits the $\sim$2500\AA\ multiplets very well, but they fit reasonably well to 
the blue and red part of the ``multiplet'', respectively. Panel~(c) shows 
these iron `models' more clearly. See \S~\ref{sample_Fecleaning} for more details. 
The flux density is in units of 10$^{-13}$ \ergsA.
\label{FeCleanfig3C273}}

\figcaption[f6ab.eps]{
Sample fitting of the iron emission in typical high-redshift quasar spectra. 
In the upper, large panels the original spectrum (top) with the power-law 
continuum fit, is compared to the residual spectrum (continuum superposed; 
middle) after subtracting the best fit iron emission model (bottom). 
In the lower, smaller panels the iron models are plotted separately 
for visibility.  Q0020$+$022 and Q0252$+$016 display pronounced 
`pseudo-continuum' in the \lam \lam 1500 $-$ 1900 \AA\ range.  The data are 
from M. Vestergaard \et (2001, in preparation) with flux density in units 
of 10$^{-16}$ \ergsA.
See \S~\ref{sample_Fecleaning} for continuum parameters and details on the
relative spectrum shifts valid for each \qso. 
\label{FeCleanfigs}}

\setcounter{figure}{0}
\newpage
\epsscale{1.10}
\plotone{f1.eps}    
\figcaption[f1.eps]{The non-iron emission model (solid line)
overplotted on the continuum-subtracted \izw\ spectrum (dotted line).
The assumed continuum of the {\it HST} spectrum is a broken power-law,
F$_{\nu}\,\sim\,\rm \nu^{-\alpha}$ with a break at \lam 1716\AA{} (see text
for details).
The absorption lines are marked using letters a--z below the spectrum and listed
in Table~\ref{alids}.  The detected emission lines are labeled.
Geocoronal \lya\ is seen at \lam 1145\AA{} (restframe).
}

\clearpage
\epsscale{1.10}
\plotone{f2.eps}    
\figcaption[f2.eps]{The iron template (solid line; sum of \feii\ and
\feiii ) overplotted on the \izw\ spectrum (dotted line).
The detected emission lines are labeled.
}

\clearpage
\epsscale{1.10}
\plotone{f3.eps}    

\figcaption[f3.eps]{The \feiii\ template spectrum (dotted line)
overplotted on the \feii\ template spectrum (solid line).  Most of the
\feiii\ features are fitted with Gaussian profiles. 
}
\newpage
\hfill
\figcaption[ftab4.eps]{
Suggested identifications of individual, mostly unblended UV \feii\ and
\feiii\ emission multiplets based on the positions and (rough) strengths
listed by Moore (1950). Each label consists of the ionizatation level
and the multiplet number separated by a hyphen. Multiplets with labels in
parentheses are slightly blended. Square brackets denote multiplets which
are suggested present by the fitting process (see Appendix~A; `2-10' is
expected only).  Heavily blended multiplets are not marked.
Multiplets with labels ending with an `m' have relative transition strengths
inconsistent with the optically thin laboratory strengths.
Note that the positions in the spectrum of the multiplet transitions are
sometimes blueshifted by 1-2\,\AA{} relative to the laboratory wavelengths
marked. This is particularily evident for \feiii \,UV34 at $\sim$1912\,\AA{}
(labeled `3-34m'). See \S~\ref{C3_complex} and Appendix~A regarding the
relative strengths in this multiplet.
}

\setcounter{figure}{3}
\newpage
\epsscale{0.92}
\plotone{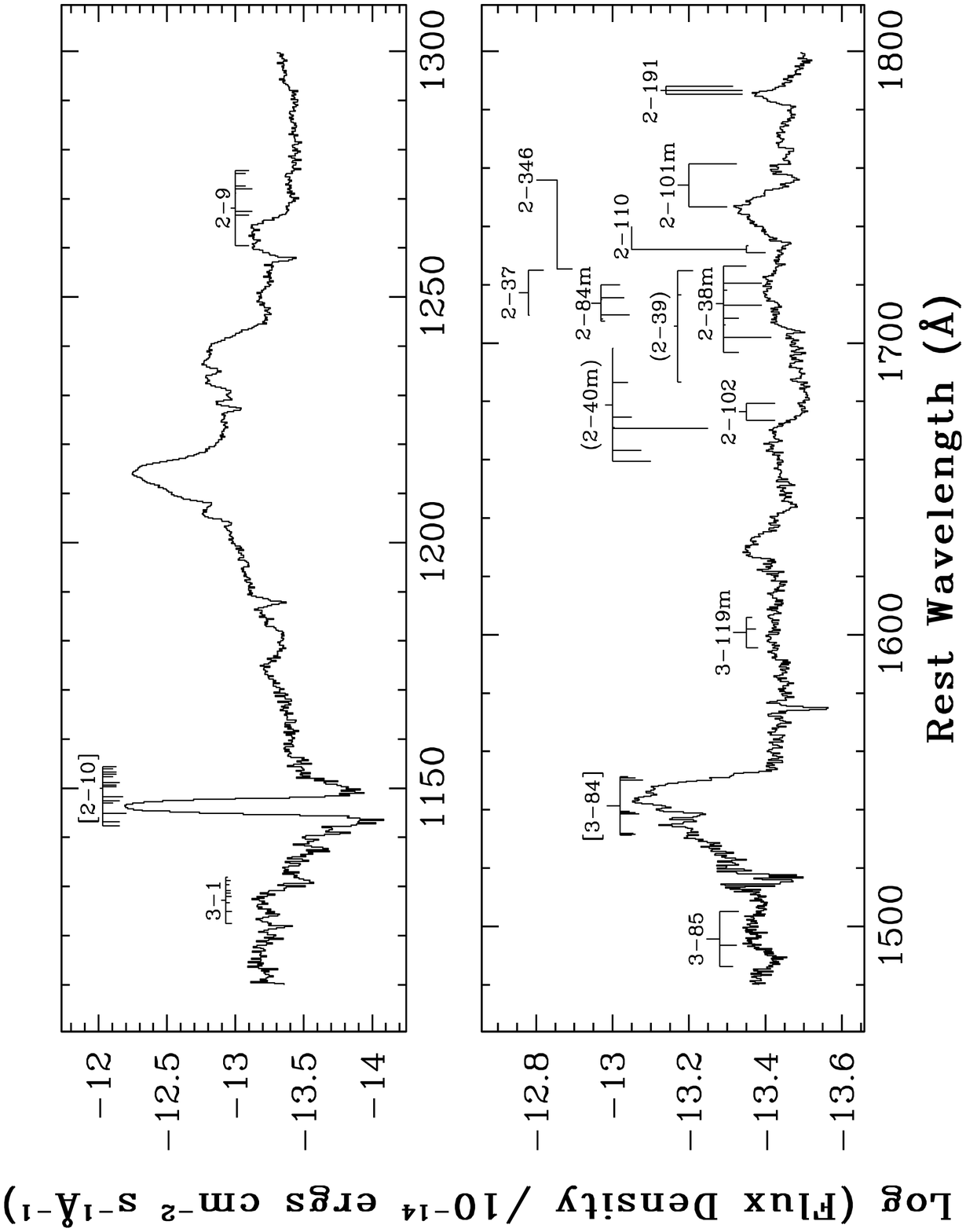}    
\figcaption[f4ab.eps]{(See caption above)}

\newpage
\setcounter{figure}{3}
\plotone{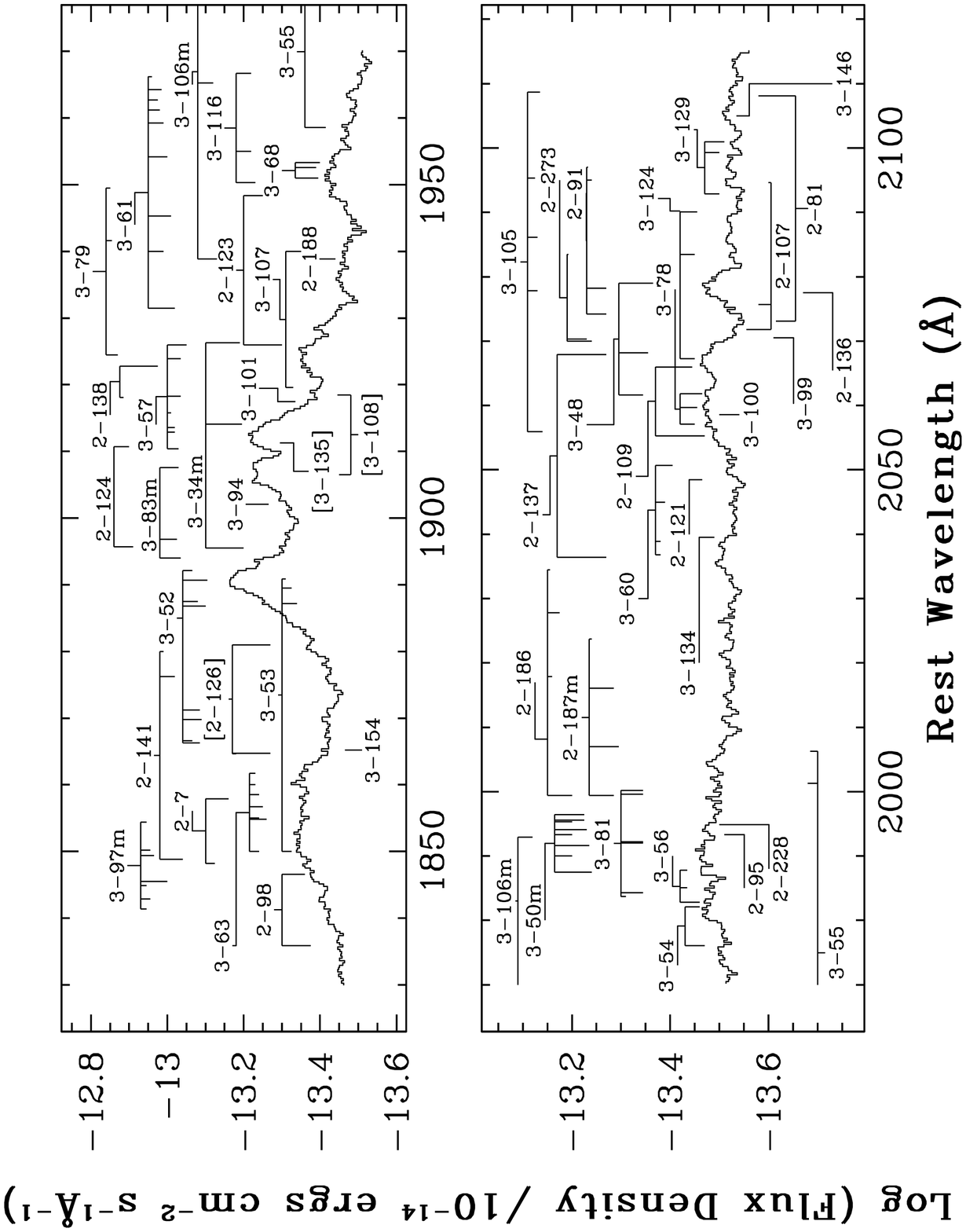}    
\figcaption[f4cd.eps]{continued}
\newpage
\setcounter{figure}{3}
\plotone{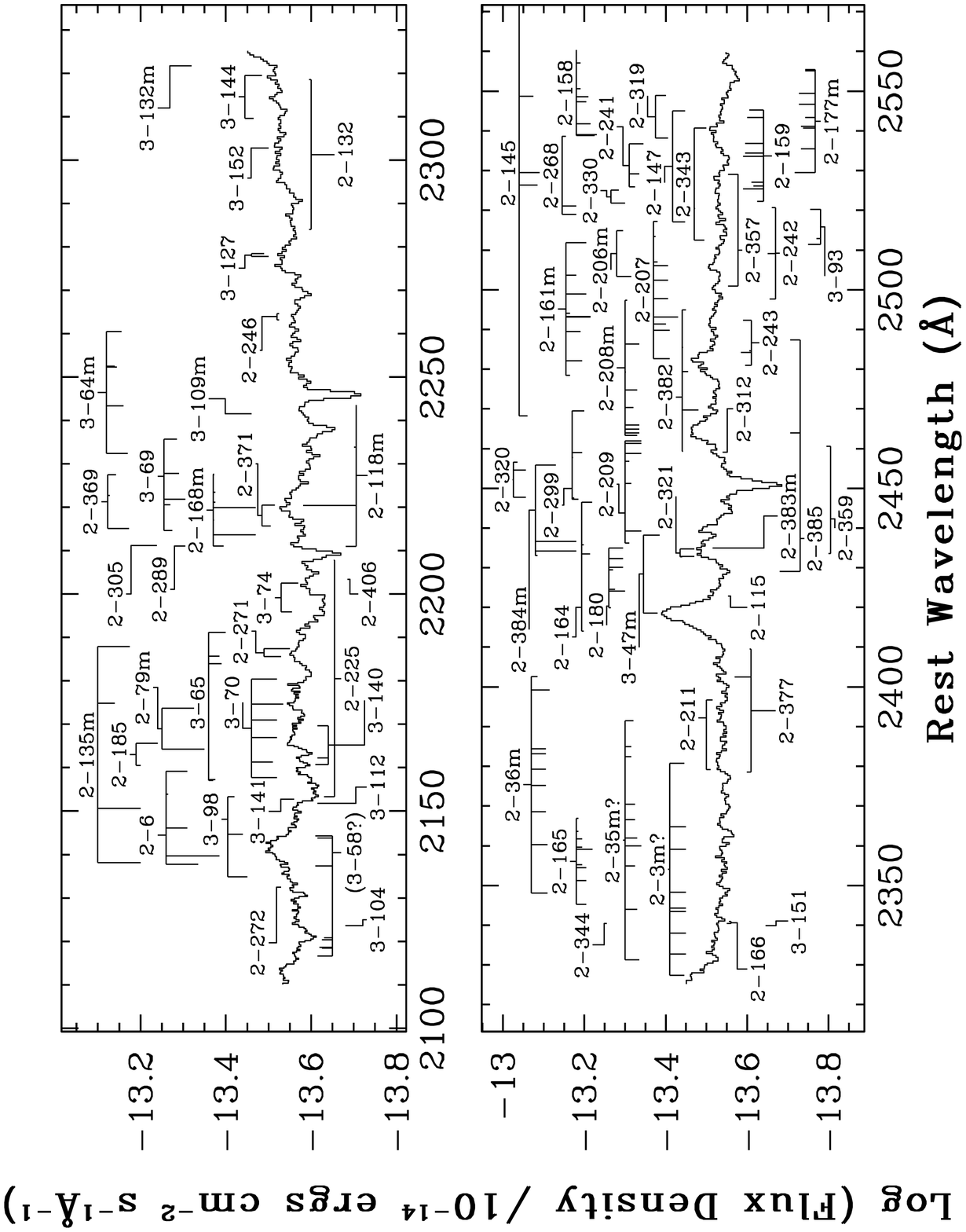}    
\figcaption[f4ef.eps]{continued}
\newpage
\setcounter{figure}{3}
\plotone{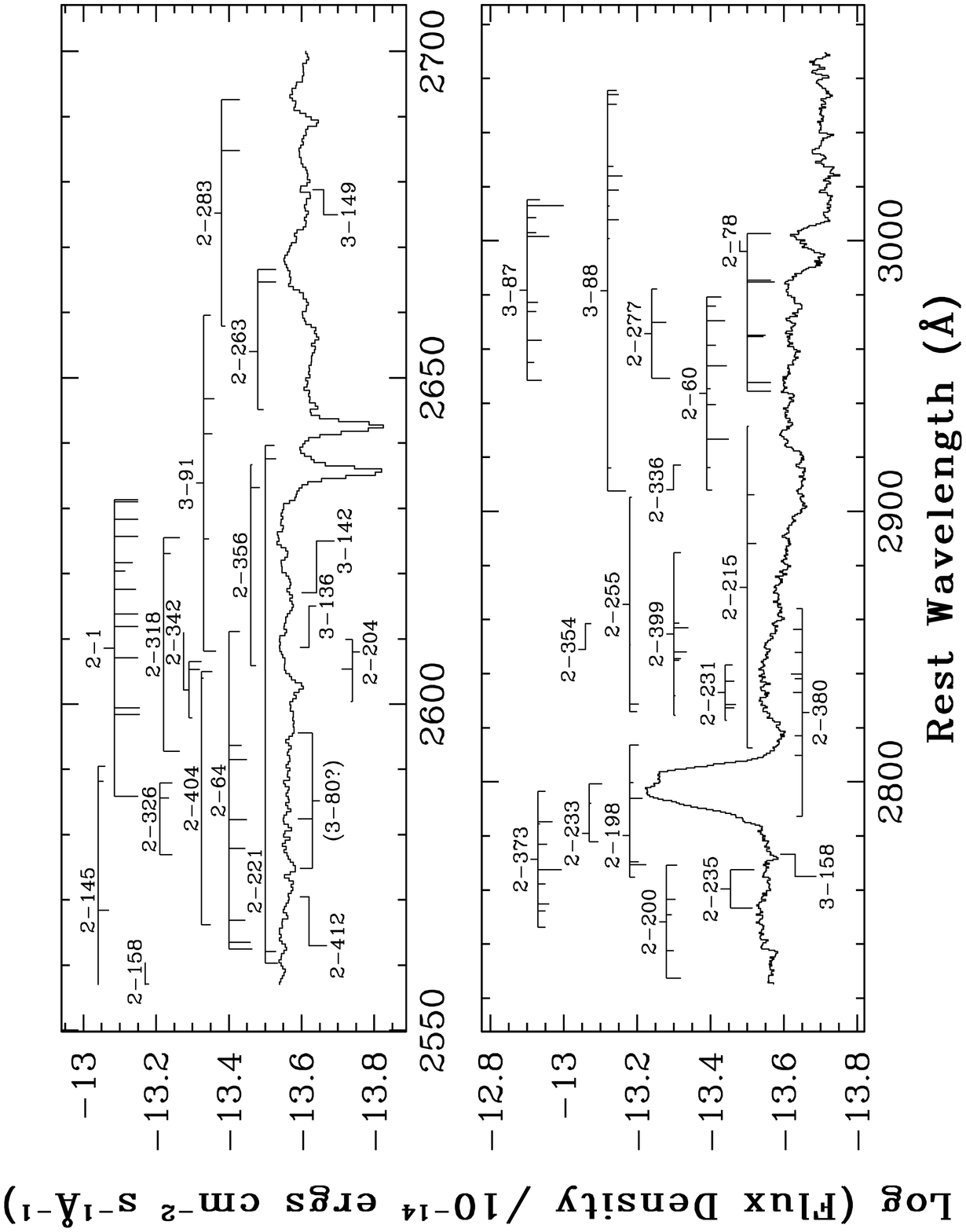}    
\figcaption[f4gh.eps]{continued}

\newpage   
\setcounter{figure}{4}
\hfill
\figcaption[f5.eps]{Examples of line fitting. Each individual component
is shown as well as their sum (dot -- dashed line) superposed on the data (solid line).
The non-iron emission is dotted while the fitted iron emission is dashed.
Note, the fit is sometimes so good it entirely coincides with the spectrum.
{\bf a)} \lya\ complex; note the strongly blue-asymmetric \nv\ profile.
This is also seen in the non-deblended \civ\ and \mgii\ profiles
(App.~\ref{linecomments}), {\bf b)} \siiv +\oiv \,\lam \,1400 blend, {\bf c)}
\civ\ complex; `?': the identification of this feature is uncertain; see text,
{\bf d)} \feii\ UV191, and \ciii\ complex [Note, the \znii\ ISM absorption at
1910\AA. Both the sum of the non-iron emission line fits and the sum of
non-iron and iron fits are shown in dot-dashed curves. The latter (full) sum
coincides so well with the data in many places that it is not easily seen, \eg
around \siiii ], \ciii, and \feiii\ redward of \ciii .
The fit to \feii\ UV 191 also coincides well with the data].
The fit to \feiii\ UV34 is model B.
{\bf e)} Other sample model fits to \feiii\ UV34 as discussed in text. The
individual fits to \siiii] and \ciii\ are not shown, but the residuals are.
{\bf f)} \feiii\ UV47 \lam \,2418,\,2438, and the non-iron features nearby, and
{\bf g)} \mgii .
}
\newpage
\setcounter{figure}{4}
\epsscale{0.92}
\plotone{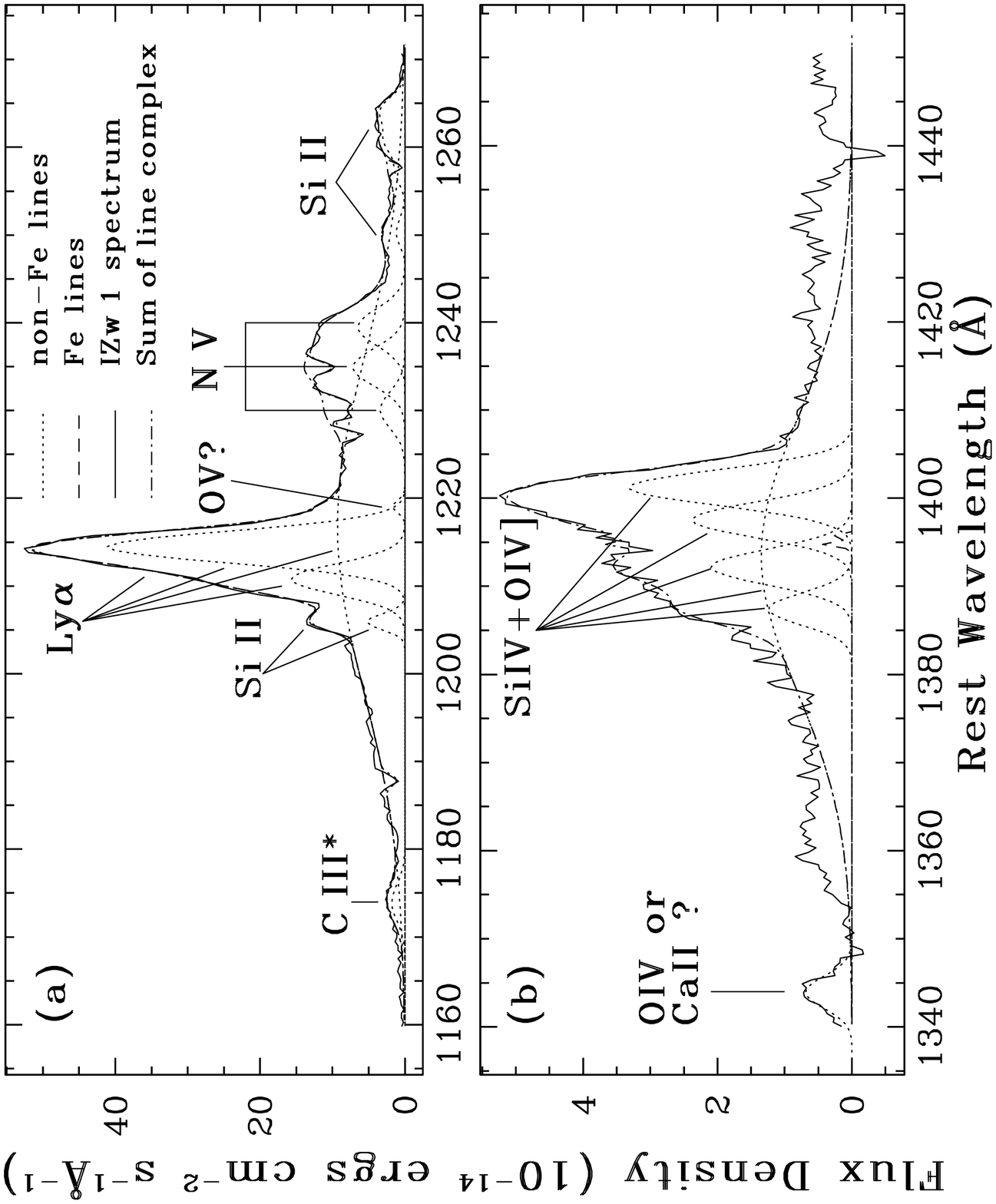}    
\figcaption[f5.eps]{(See caption above)}

\newpage
\setcounter{figure}{4}
\plotone{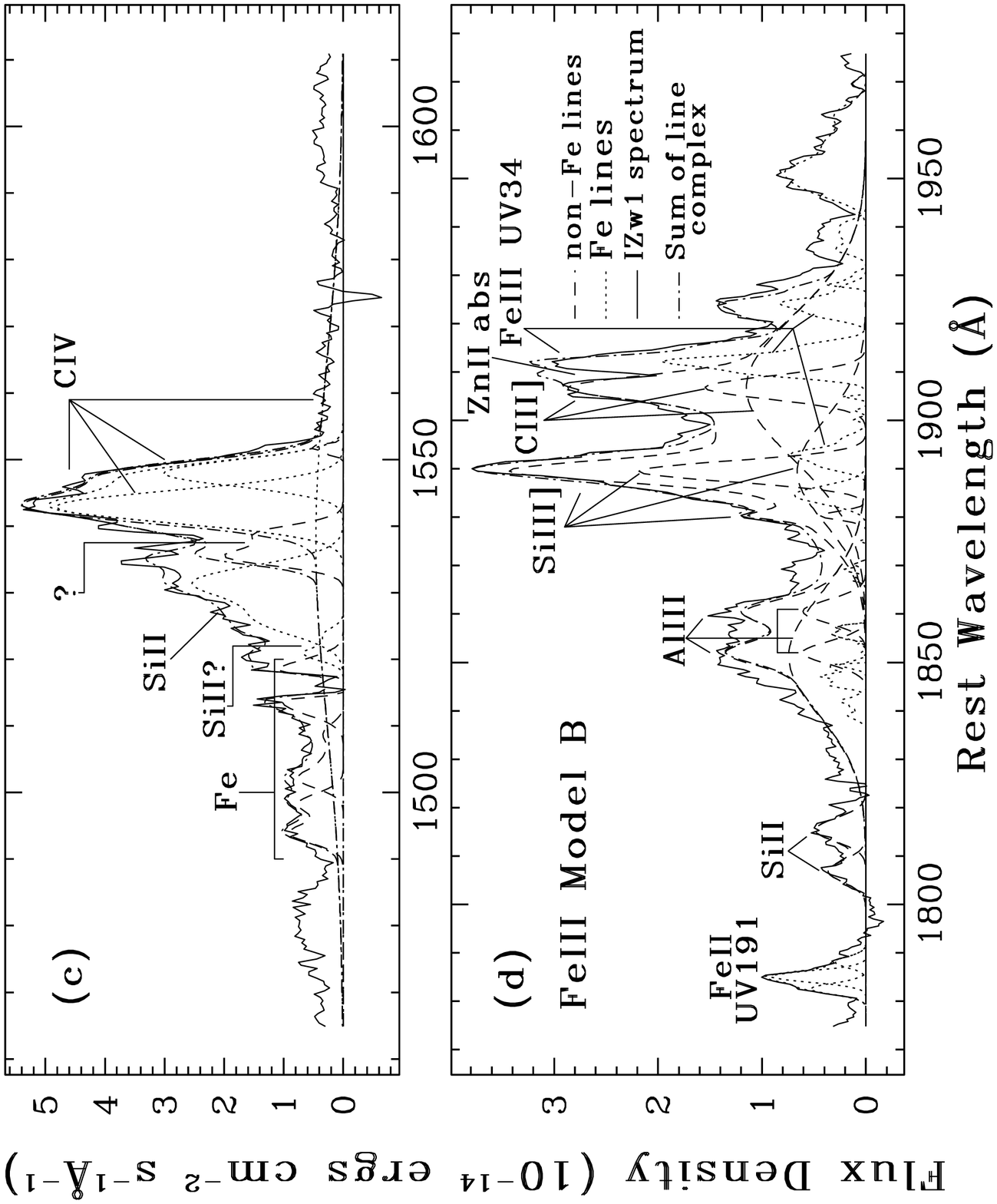}    
\figcaption[f5cd.eps]{continued}
\newpage
\epsscale{1.10}
\setcounter{figure}{4}
\plotone{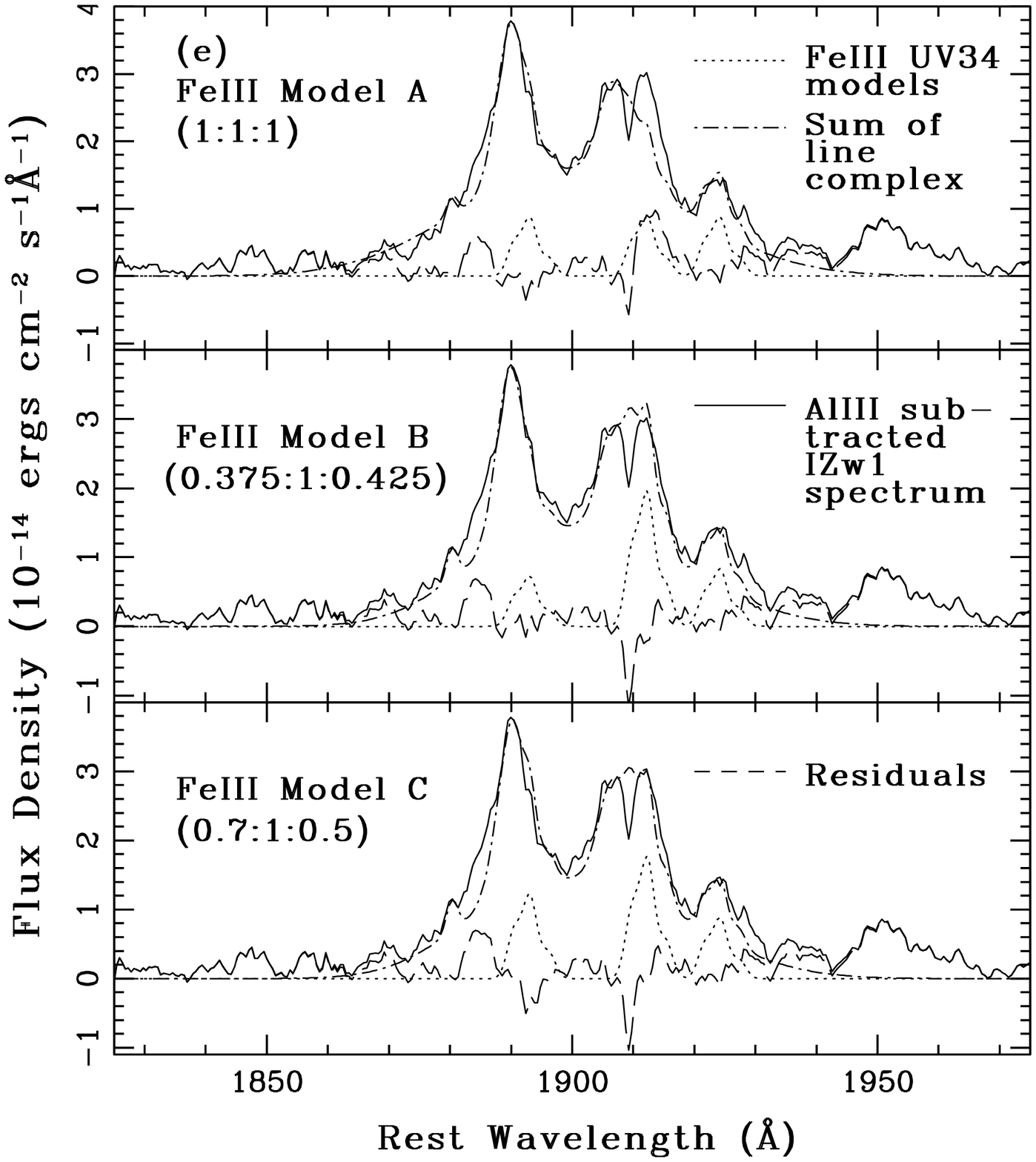}    
\figcaption[f5e.eps]{continued}
\newpage
\epsscale{0.92}
\setcounter{figure}{4}
\plotone{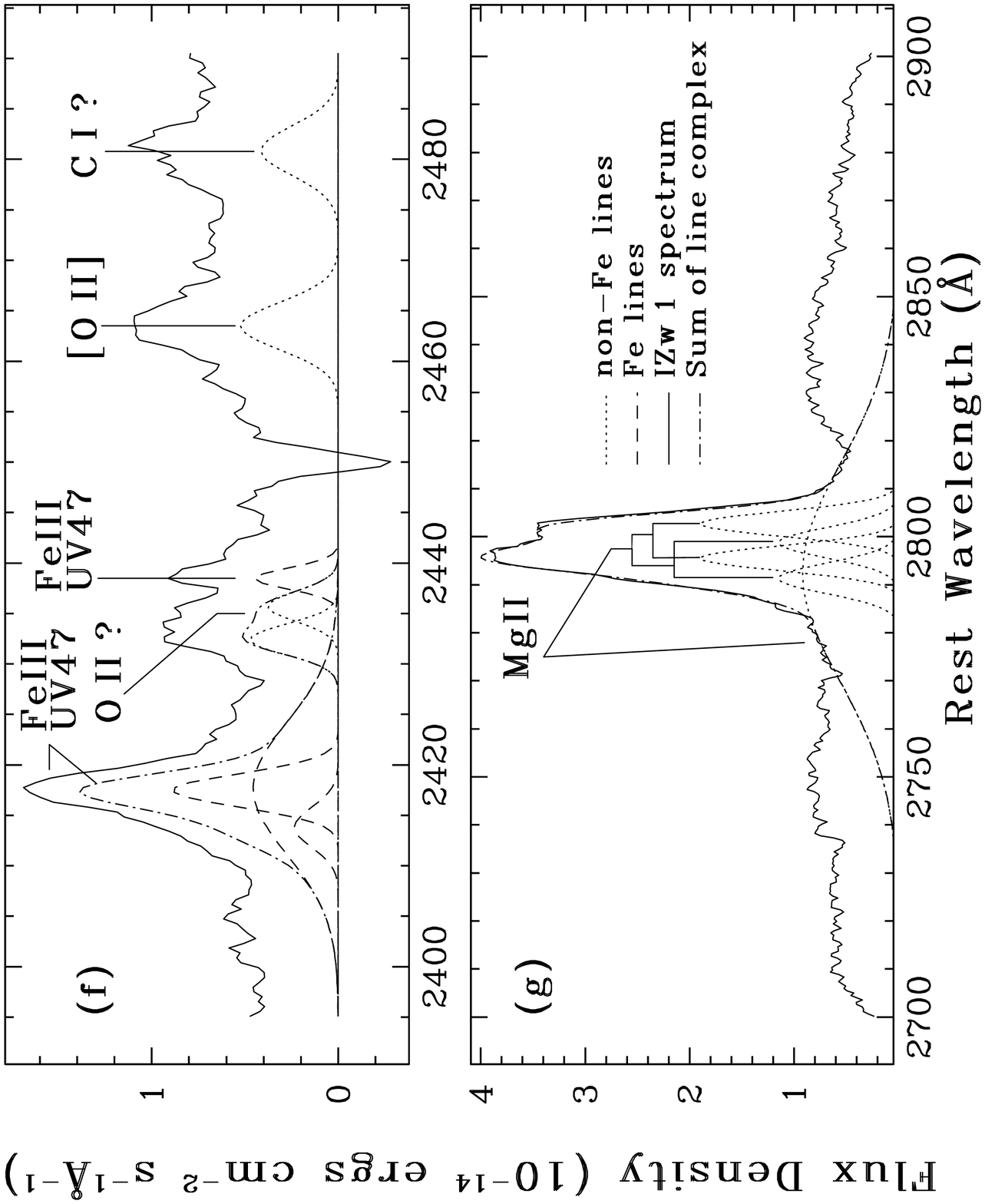}    
\figcaption[f5fg.eps]{continued}
\newpage
\epsscale{1.00}
\plotone{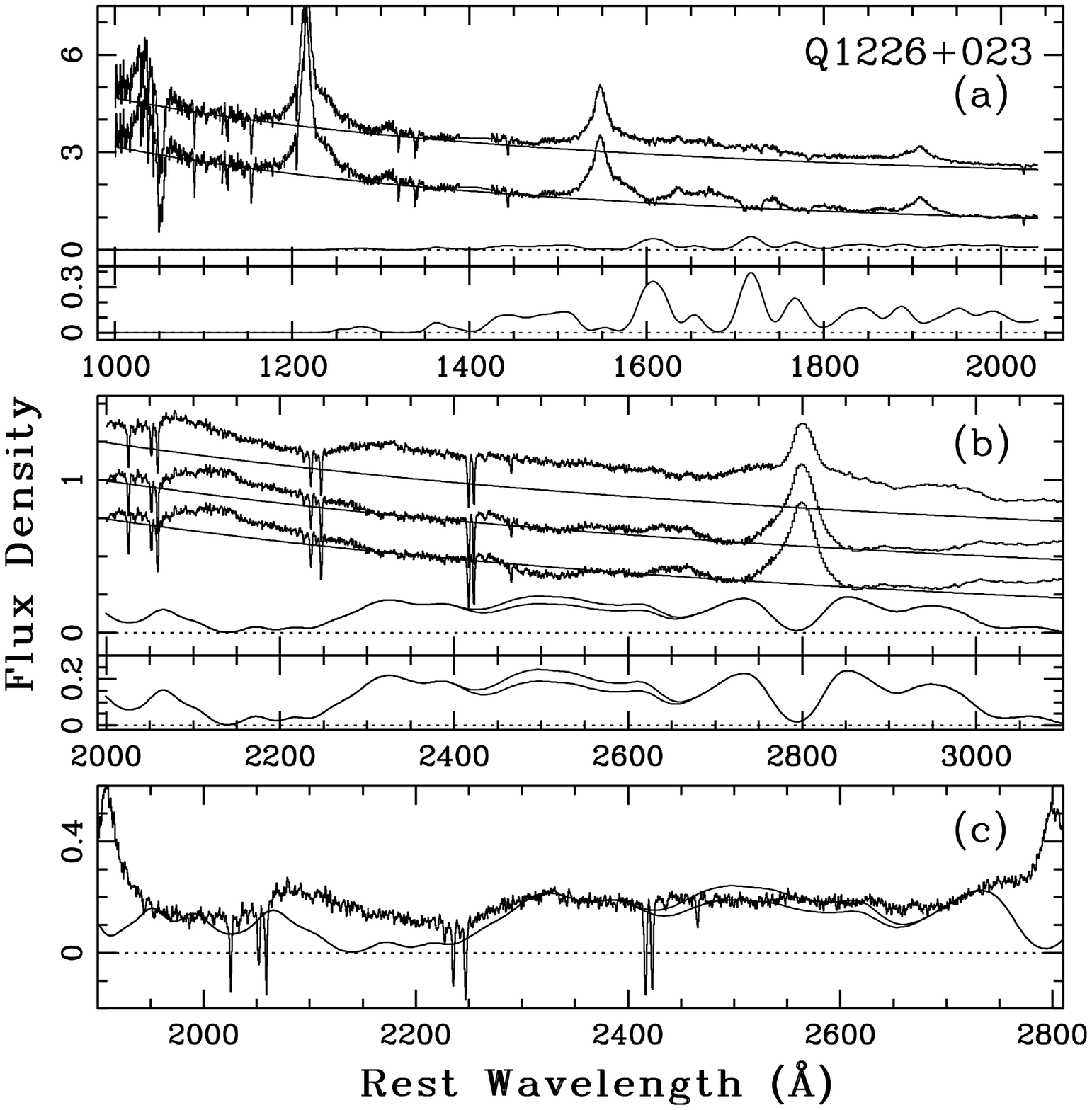}
\figcaption[f6.eps]{The {\it HST} spectrum of 3C273 (Q1226$+$023) and the best
fitted iron `models'. The upper, large panel in (a) shows the original spectrum
(shifted for clarity; see \S~\ref{sample_Fecleaning}) with the continuum
overplotted (top) compared to the residuals (middle) after subtracting the
best fitted iron `model' (bottom). The latter is displayed separately below
the large panel for clarity. Panel (b) shows the original spectrum, two iron `model'
fits, and the corresponding spectrum residuals. Neither of the iron `models'
fits the $\sim$2500\AA\ multiplets very well, but they fit reasonably well to
the blue and red part of the ``multiplet'', respectively. Panel~(c) shows
these iron `models' more clearly. See \S~\ref{sample_Fecleaning} for more details.
The flux density is in units of 10$^{-13}$ \ergsA.
}

\newpage
\setcounter{figure}{6}
\figcaption[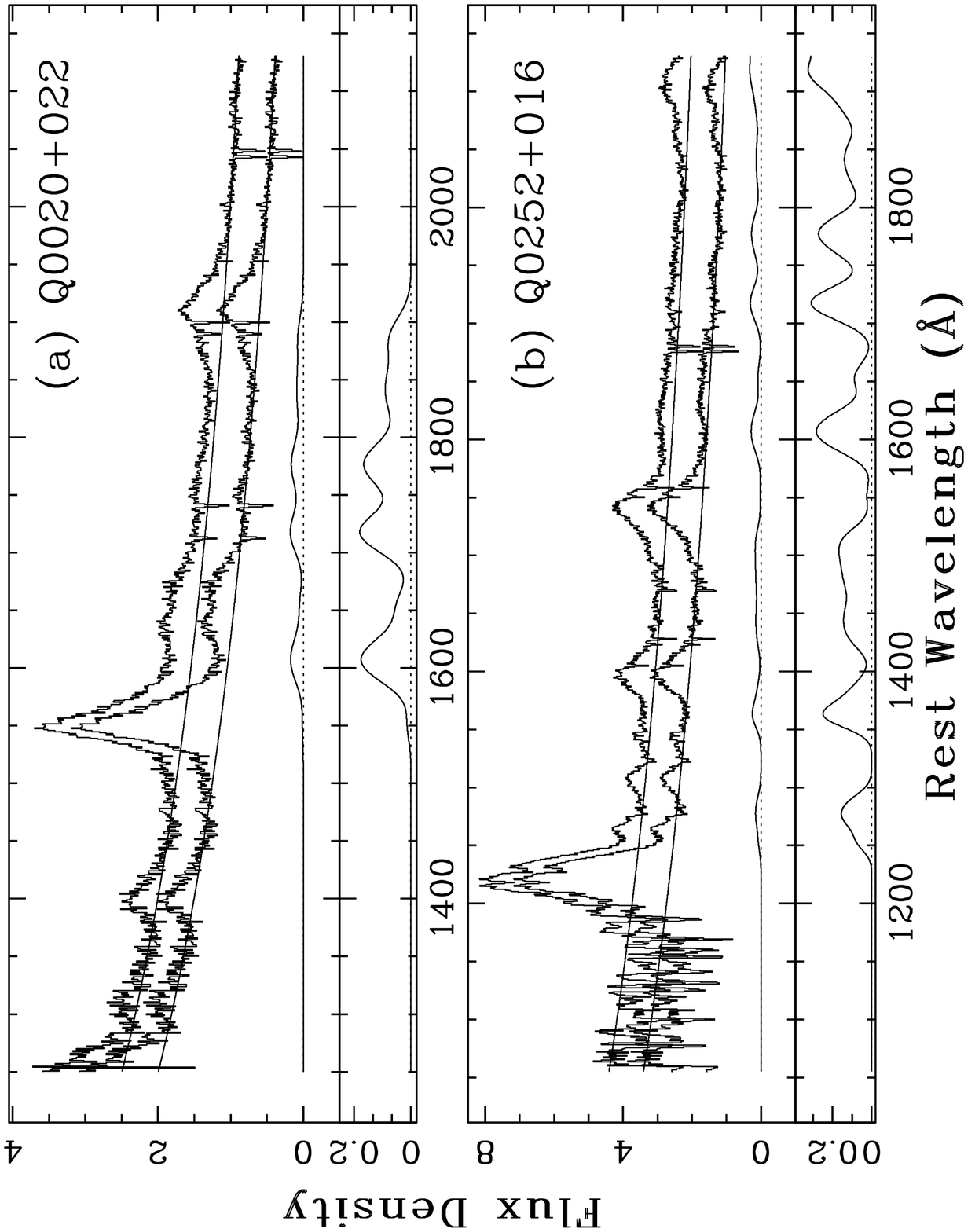]{
Sample fitting of the iron emission in typical high-redshift quasar spectra.
In the upper, large panels the original spectrum (top) with the power-law
continuum fit, is compared to the residual spectrum (continuum superposed;
middle) after subtracting the best fit iron emission model (bottom).
In the lower, smaller panels the iron models are plotted separately
for visibility.  Q0020$+$022 and Q0252$+$016 display pronounced
`pseudo-continuum' in the \lam \lam 1500 $-$ 1900 \AA\ range.  The data are
from M. Vestergaard \et (2001, in preparation) with flux density in units
of 10$^{-16}$ \ergsA.
See \S~\ref{sample_Fecleaning} for continuum parameters and details on the
relative spectrum shifts valid for each \qso.
}

\newpage
\setcounter{figure}{6}
\epsscale{0.92}
\plotone{f7ab.eps}
\figcaption[f7ab.eps]{(See caption above)}
\newpage
\setcounter{figure}{6}
\epsscale{0.92}
\plotone{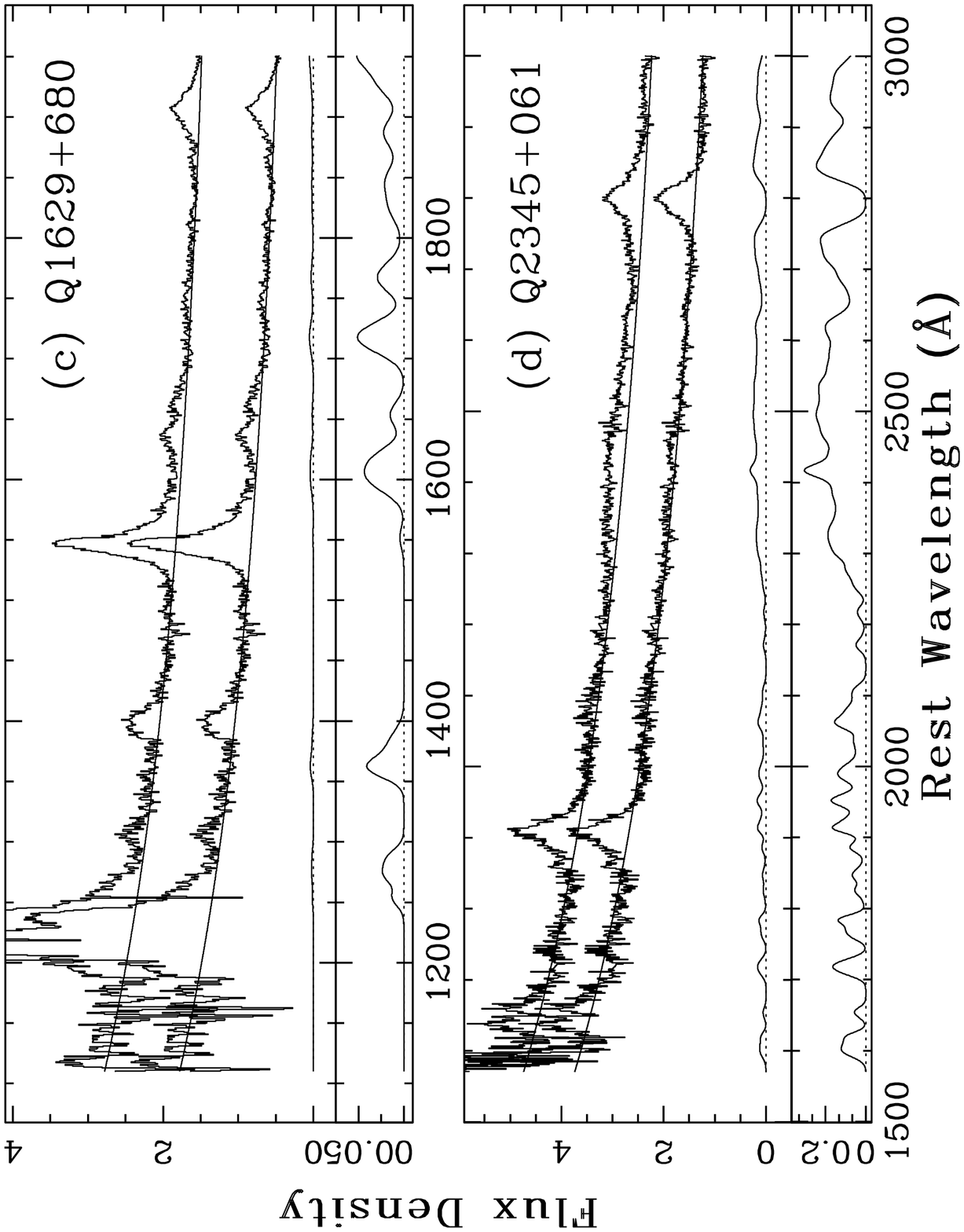}
\figcaption[f7cd.eps]{continued}
\clearpage

\begin{deluxetable}{ccccccccc}
\tablewidth{495pt}
\footnotesize
\tablecaption{Log of Observations of \izw\ (PG\,0050$+$124) \label{hstobs}}
\tablehead{
\colhead{Grating} & 
\colhead{Date} & 
\colhead{$\lambda$-range} &
\colhead{$\Delta \lambda $} & 
\colhead{Res\tablenotemark{a}} & 
\colhead{Detector} &
\colhead{Aperture} &
\colhead{Exp.\ Time} &
\colhead{$\lambda$ Offsets\tablenotemark{b} } \\
\colhead{\nodata} &
\colhead{(UT)} &
\colhead{(\AA )} &
\colhead{(\AA /pix)} &
\colhead{(\AA )} &
\colhead{\nodata} &
\colhead{(arcsec)} &
\colhead{(sec)} &
\colhead{(\AA )} 
}
\tablecolumns{9}
\startdata
G130H & 02/13/94 & 1087$-$1606 & 0.251 & 0.96 & blue  &0.86&29\,700&0.36230 \\
G190H & 09/14/94 & 1572$-$2312 & 0.359 & 1.39 & amber &0.86&6\,030&1.10670 \\
G270H & 09/14/94 & 2222$-$3277 & 0.511 & 1.97 & amber &0.86&2\,100&1.03755 \\
\enddata
\tablenotetext{a}{Spectral Resolution}
\tablenotetext{b}{Wavelength offset determined from ISM absorption lines; 
as described in \S\,\ref{dataproc}}
\end{deluxetable}

\clearpage

\begin{table}
\dummytable\label{alids}
\end{table}

\epsscale{1.00}
\plotone{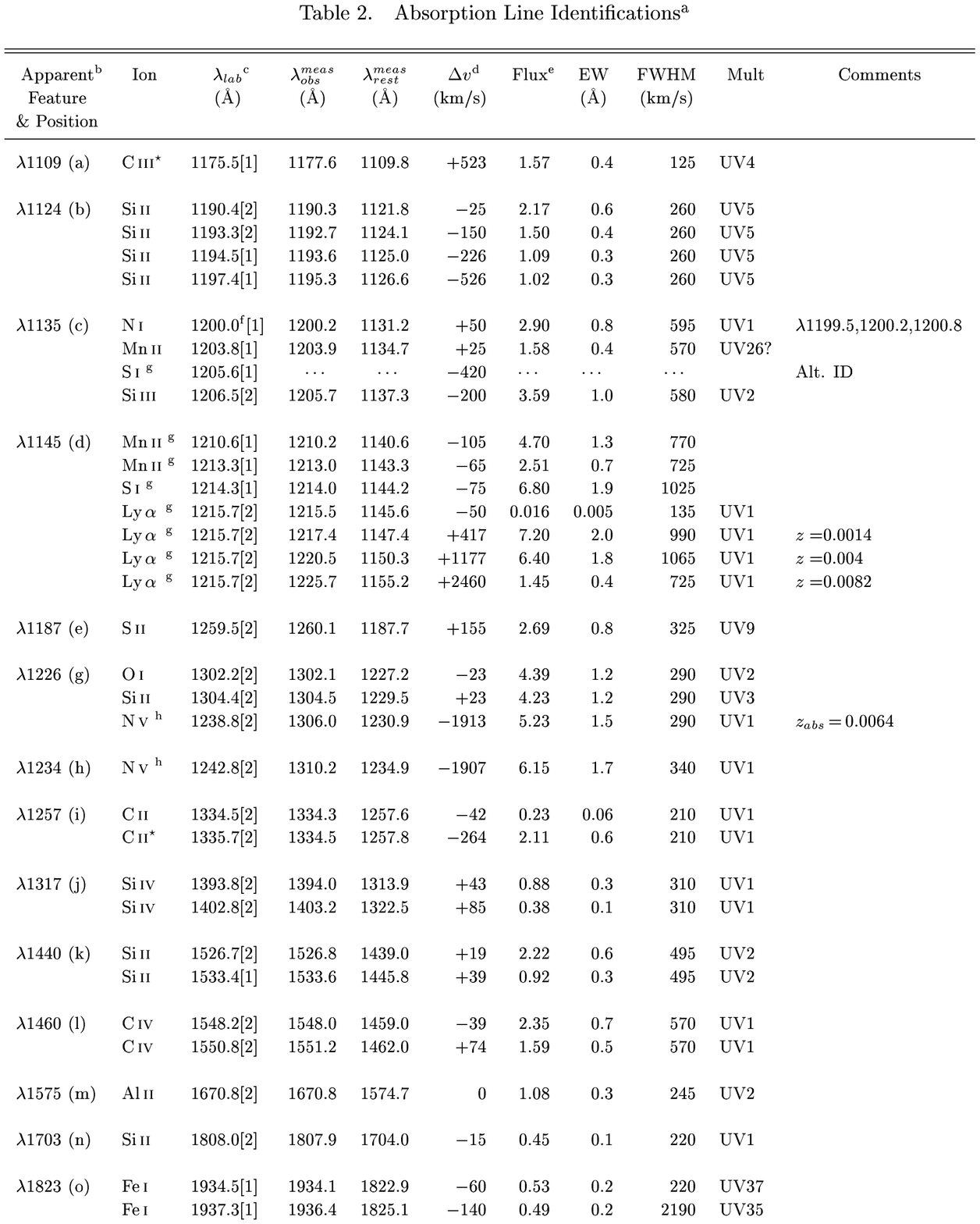}

\epsscale{1.00}
\plotone{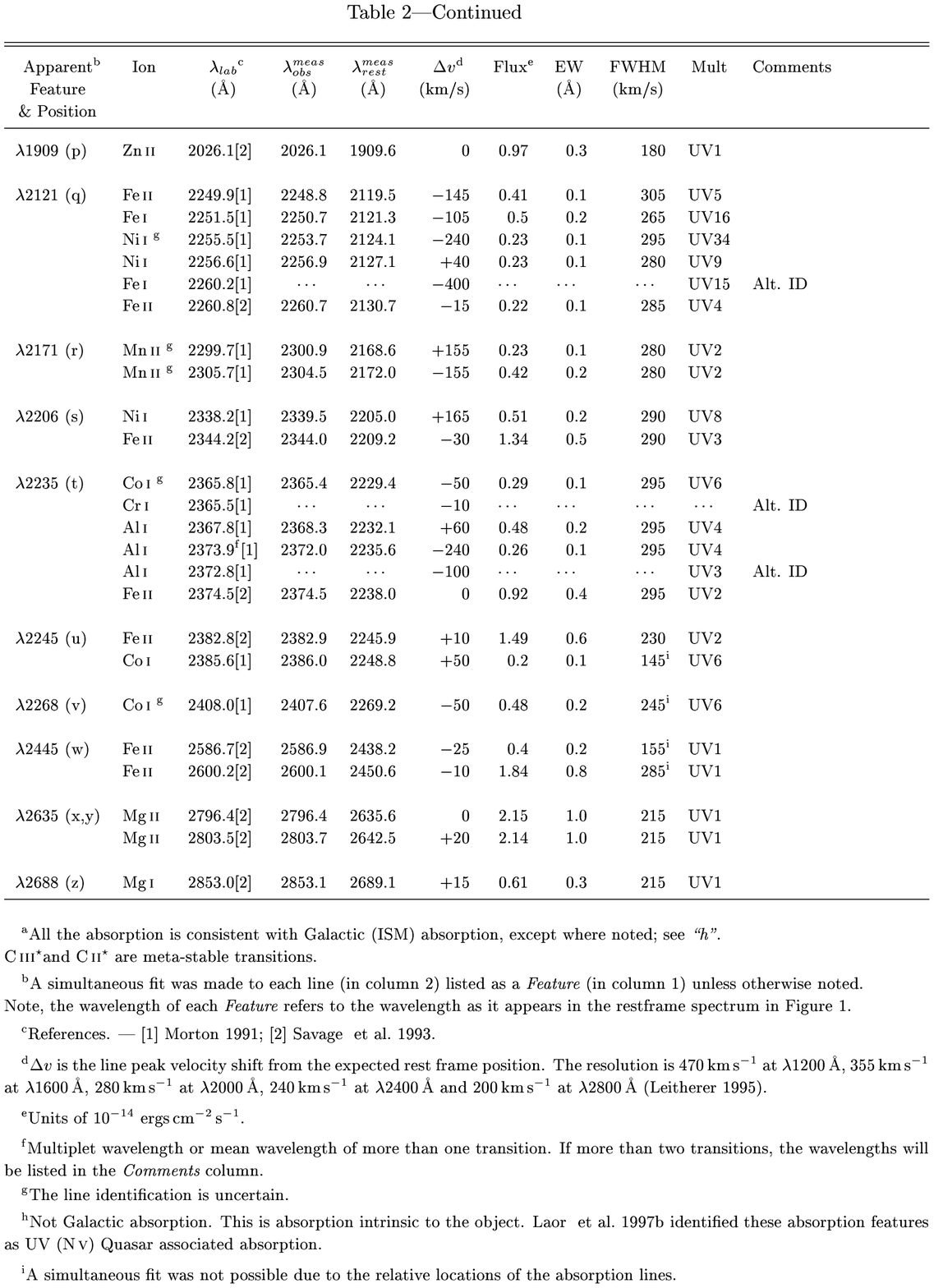}


\begin{table}
\dummytable\label{elids}
\end{table}

\epsscale{1.00}
\plotone{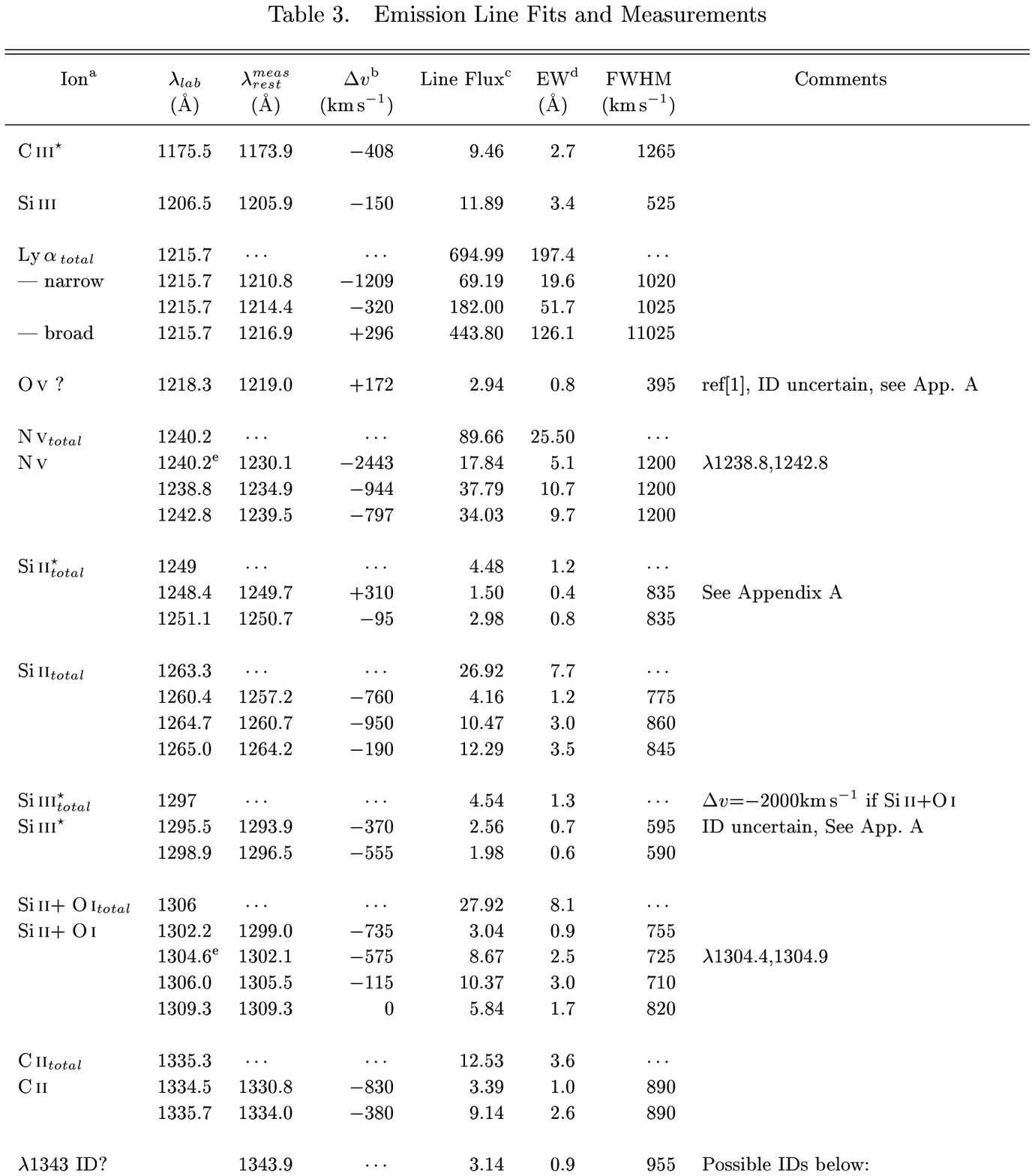}
\newpage
\epsscale{1.00}
\plotone{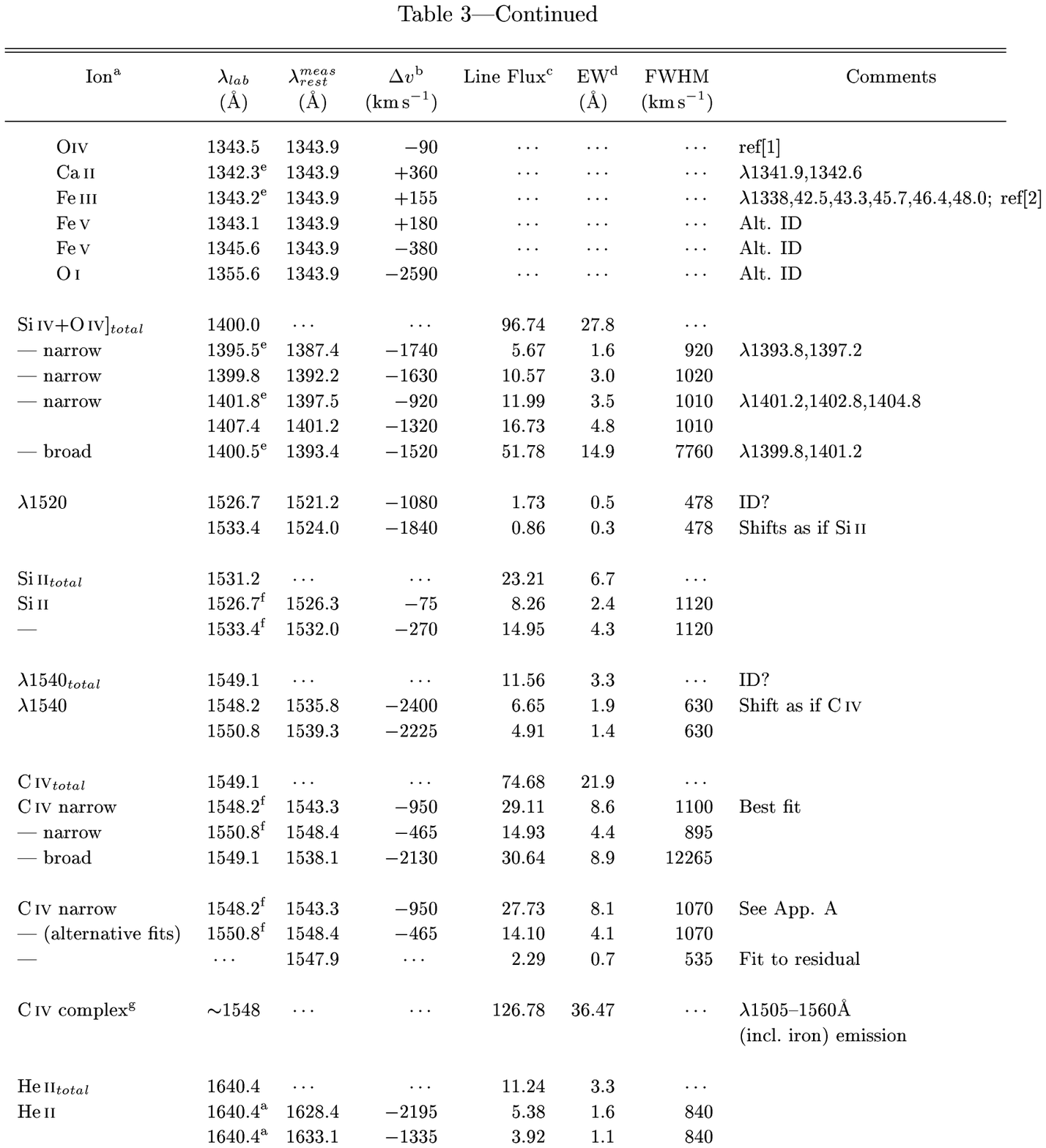}
\newpage
\epsscale{1.00}
\plotone{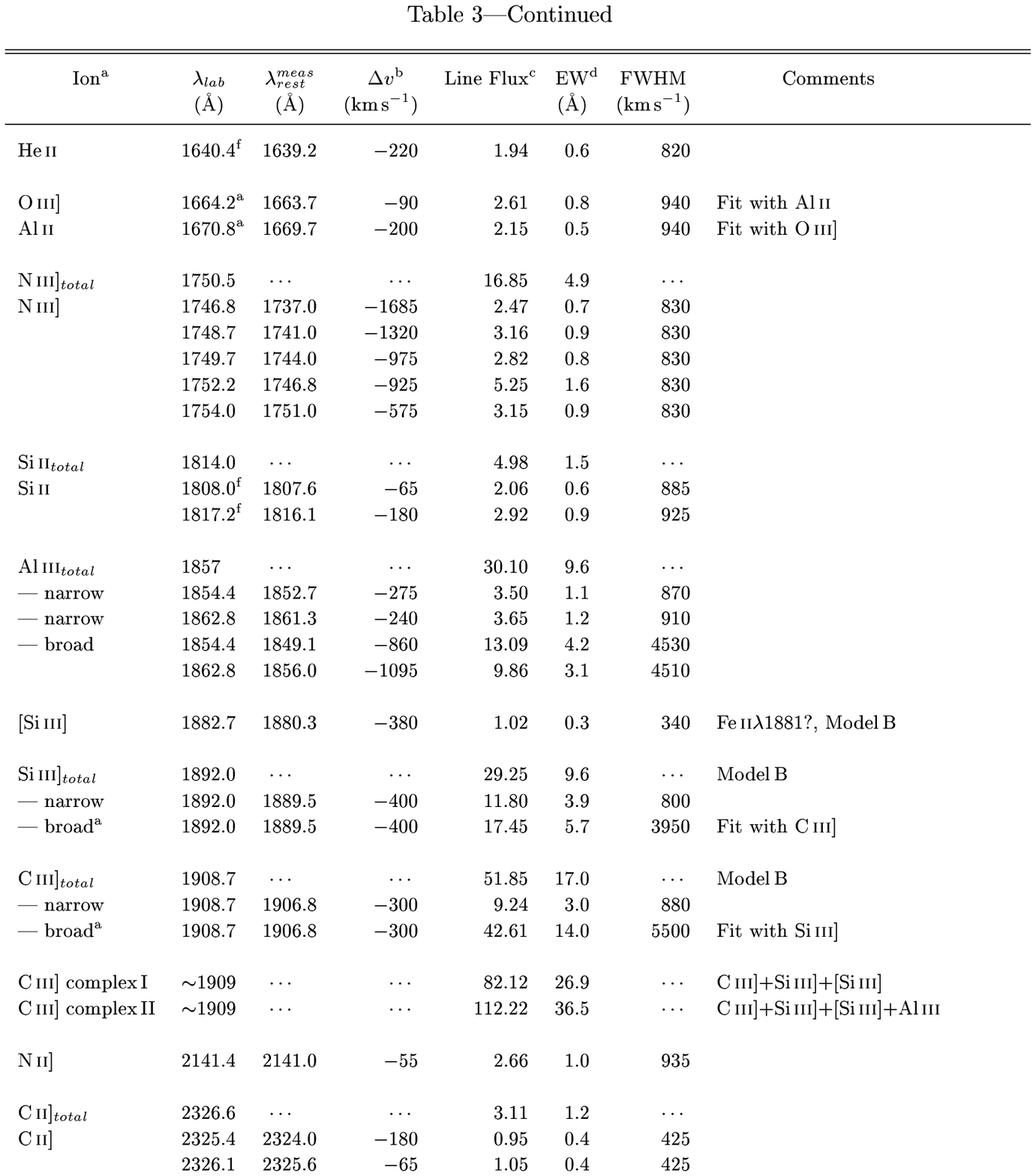}
\newpage
\clearpage
\epsscale{1.00}
\plotone{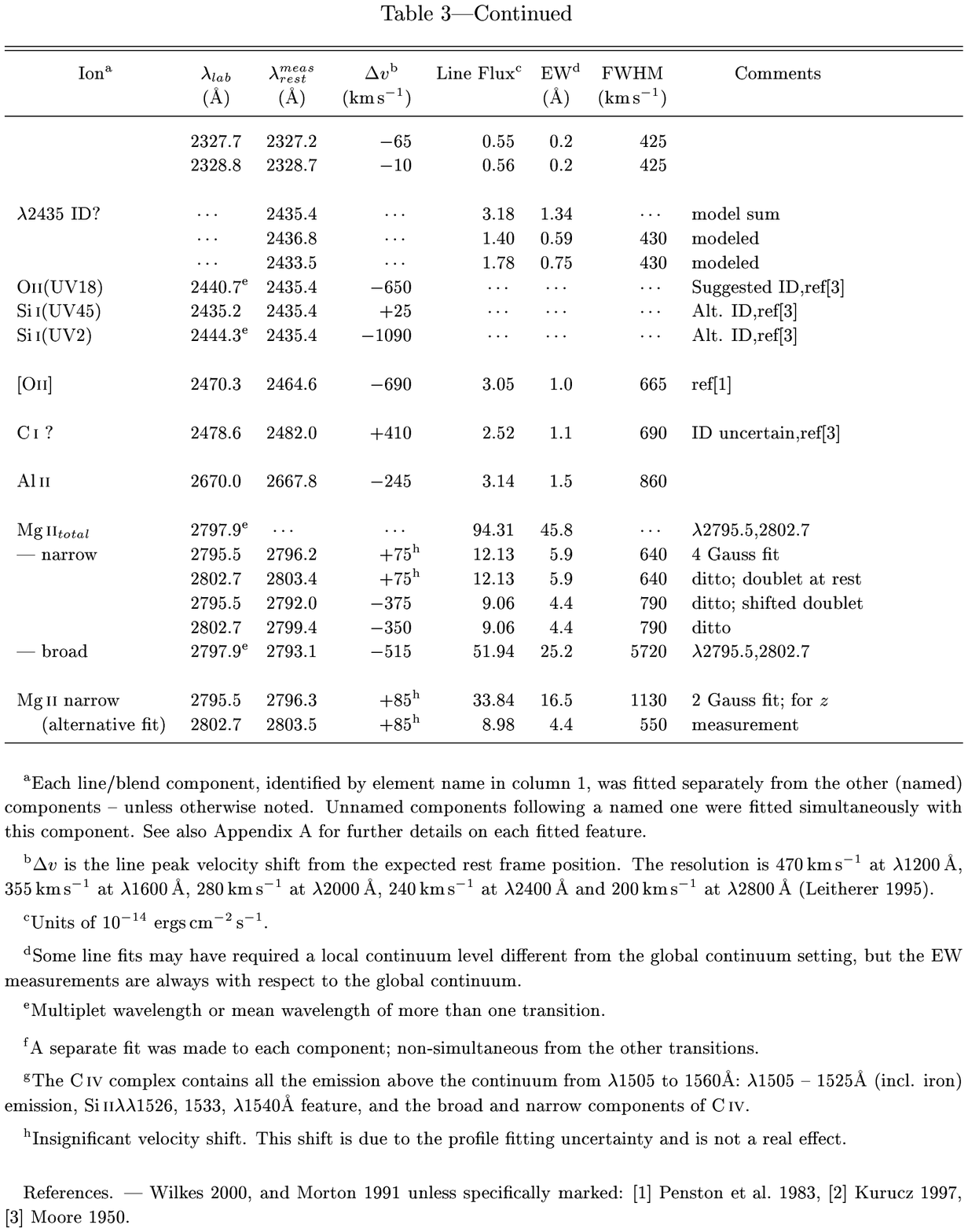}

\newpage

\begin{table}
\dummytable\label{fefits}
\end{table}

\epsscale{1.00}
\plotone{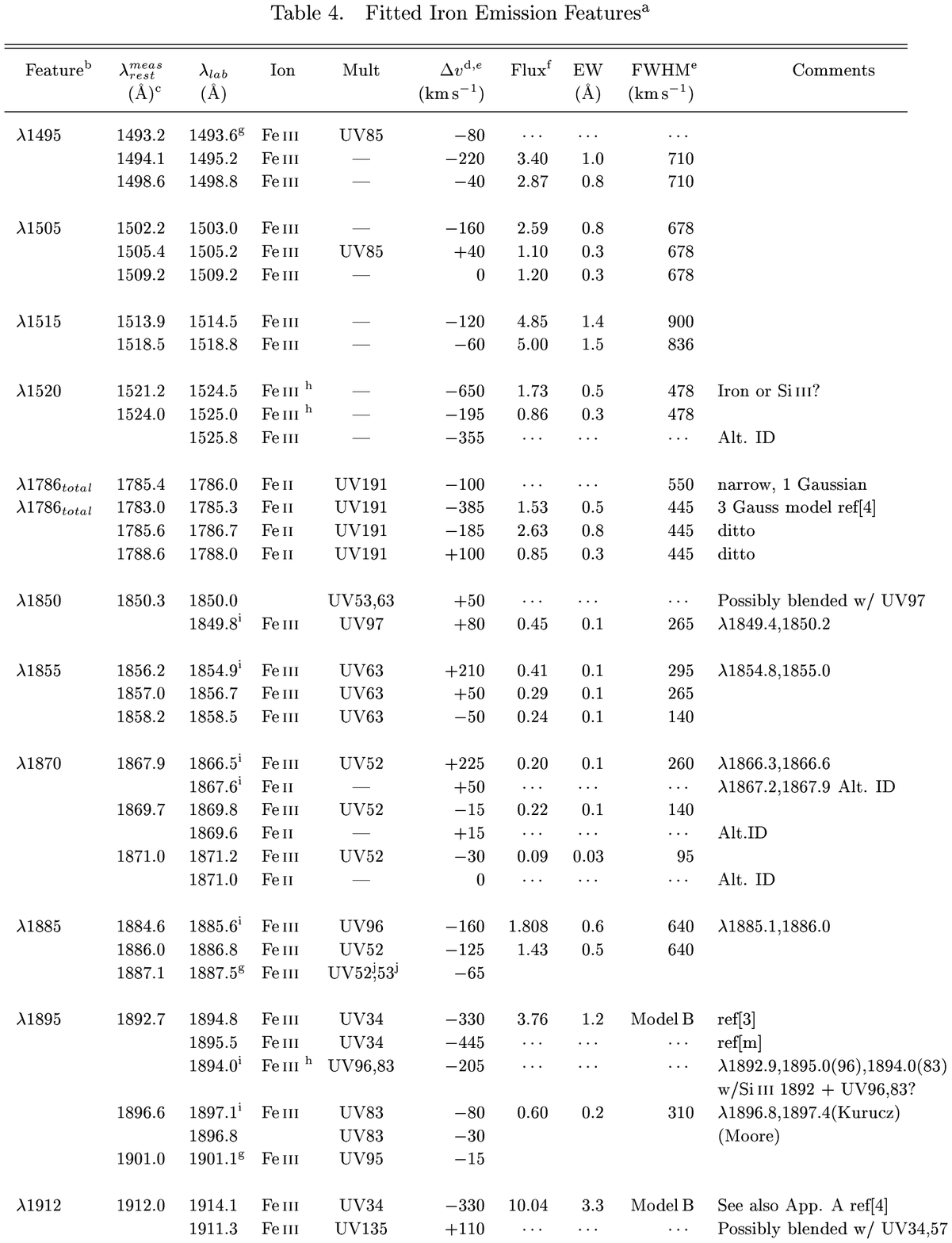}
\newpage
\epsscale{1.00}
\plotone{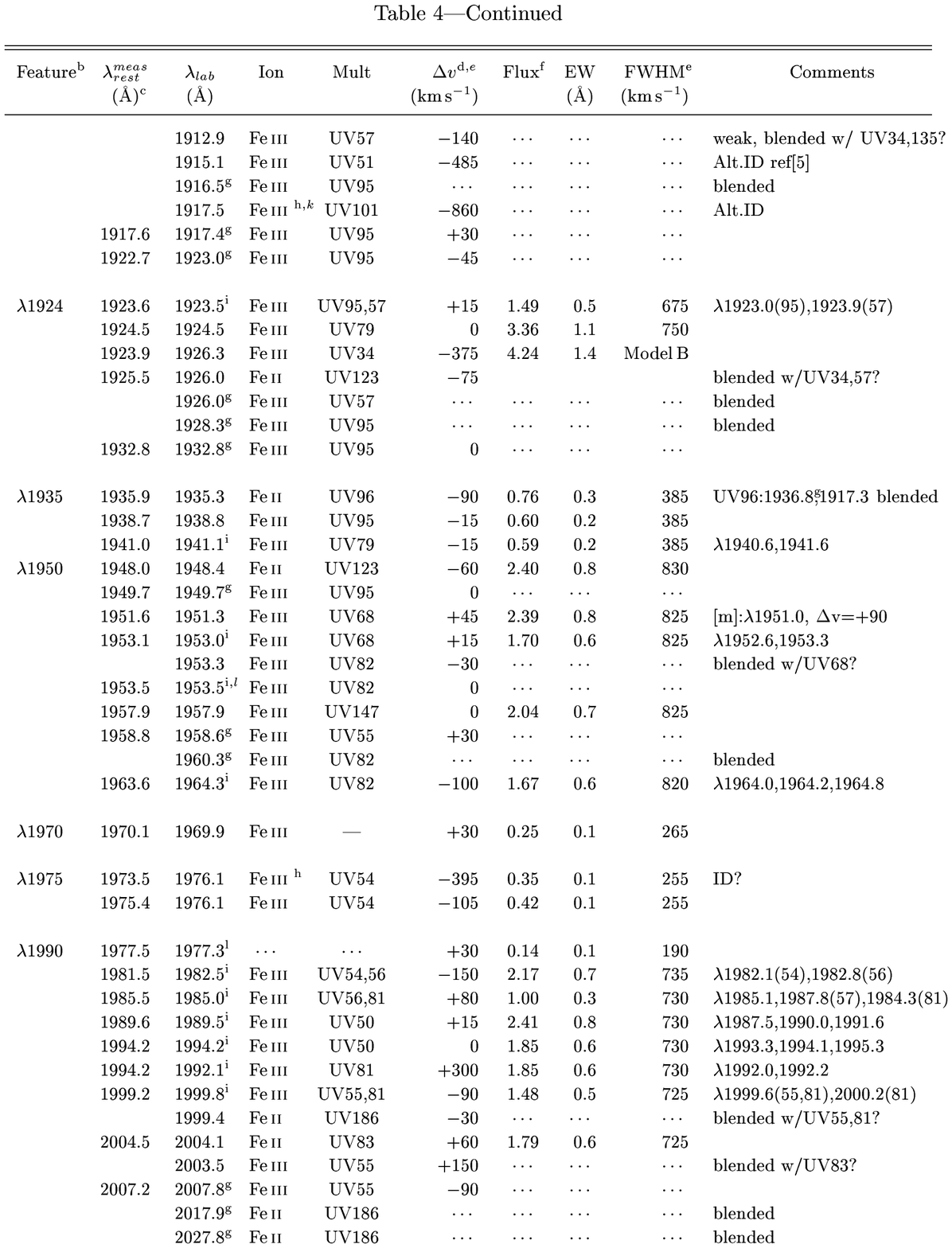}
\newpage
\epsscale{1.00}
\plotone{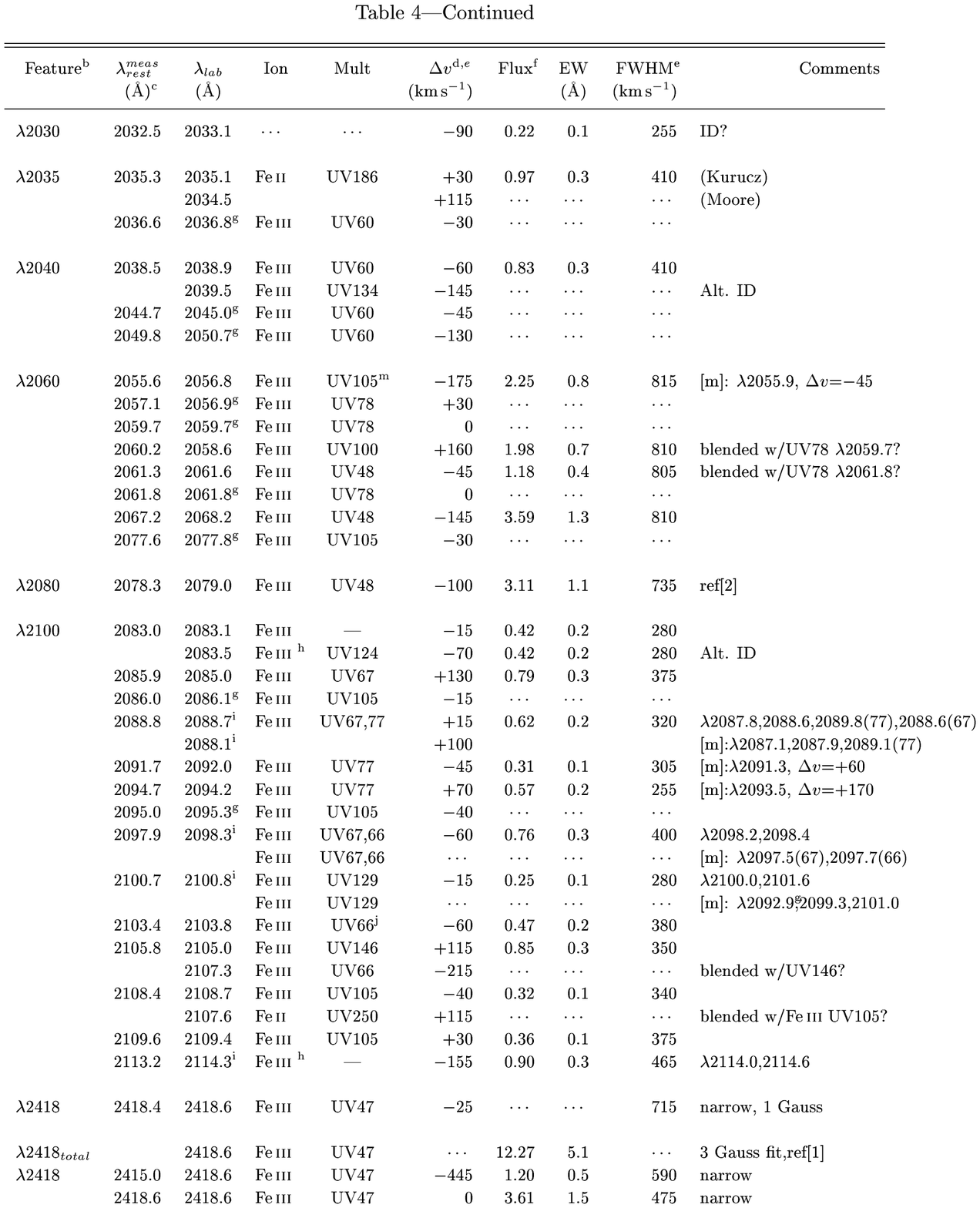}
\newpage
\epsscale{1.00}
\plotone{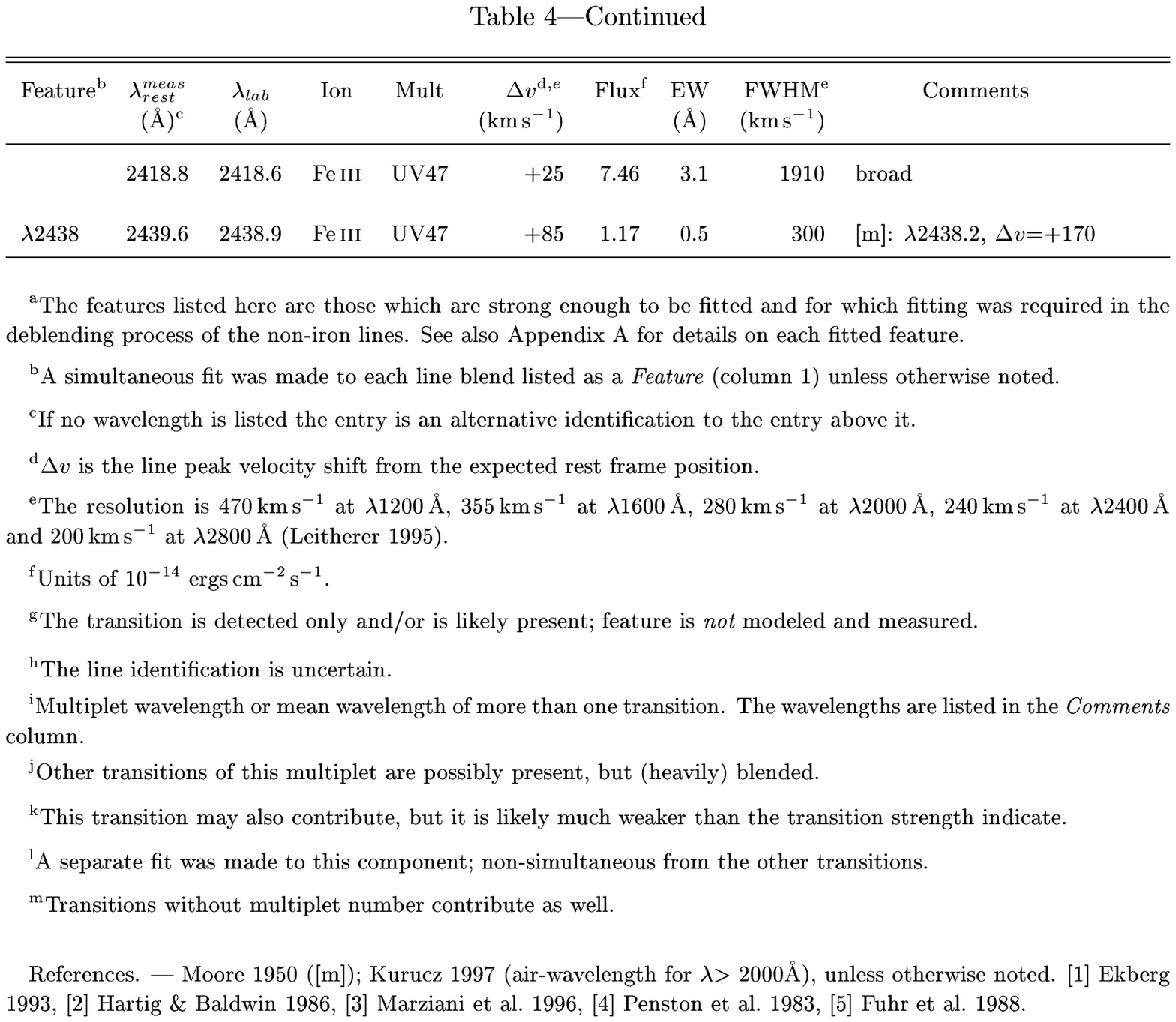}


\begin{deluxetable}{cccccccccccc}
\tablewidth{495pt}
\tablecaption{Pure Iron Emission Windows \label{fewindows}}
\footnotesize
\tablehead{
\colhead{ $\lambda$ limits\tablenotemark{a}} & 
\colhead{} &
\colhead{} &
\colhead{} &
\colhead{} &
\colhead{Range } &
\colhead{number} &
\colhead{} &
\colhead{} &
\colhead{} &
\colhead{} &
\colhead{} \\ 
\colhead{\nodata} &
\colhead{1} & 
\colhead{2} & 
\colhead{3} & 
\colhead{4} & 
\colhead{5} &
\colhead{6} &
\colhead{7} &
\colhead{8} &
\colhead{9} &
\colhead{10} &
\colhead{11} 
}
\tablecolumns{12}
\startdata
$\lambda_1$\,(\AA ) & 1350 & 1427 & 1490 & 1705 & 1760\tablenotemark{b}& 
1930\tablenotemark{c} ~or~ 1942 & 2250 & 2333\tablenotemark{d} & 2470 & 2675 & 
2855 \\
$\lambda_2$\,(\AA ) & 1365 & 1480 & 1505 & 1730 & 1800 & 2115& 
2320\tablenotemark{d}& 2445\tablenotemark{e} & 2625 & 2755 & 3010 \\
\enddata
\tablenotetext{a}{The precise limits vary from object to object}
\tablenotetext{b}{Avoid \niii \,\lam \,1750 line. The width and position may 
vary from object to object}
\tablenotetext{c}{Avoid \ciii \,\lam \,1909 broad base if present. If this base 
is strong, use the \lam\,1942\,$-$\,2115\,\AA\ range.} 
\tablenotetext{d}{Avoid \cii ] \,\lam 2326 } 
\tablenotetext{e}{Exclude [Ne\,{\sc iv}]\,\lam2419 if present}
\end{deluxetable}

\clearpage
\begin{table}
\dummytable\label{modelfits}
\end{table}
\vspace{25cm}
\epsscale{0.95}
\plotone{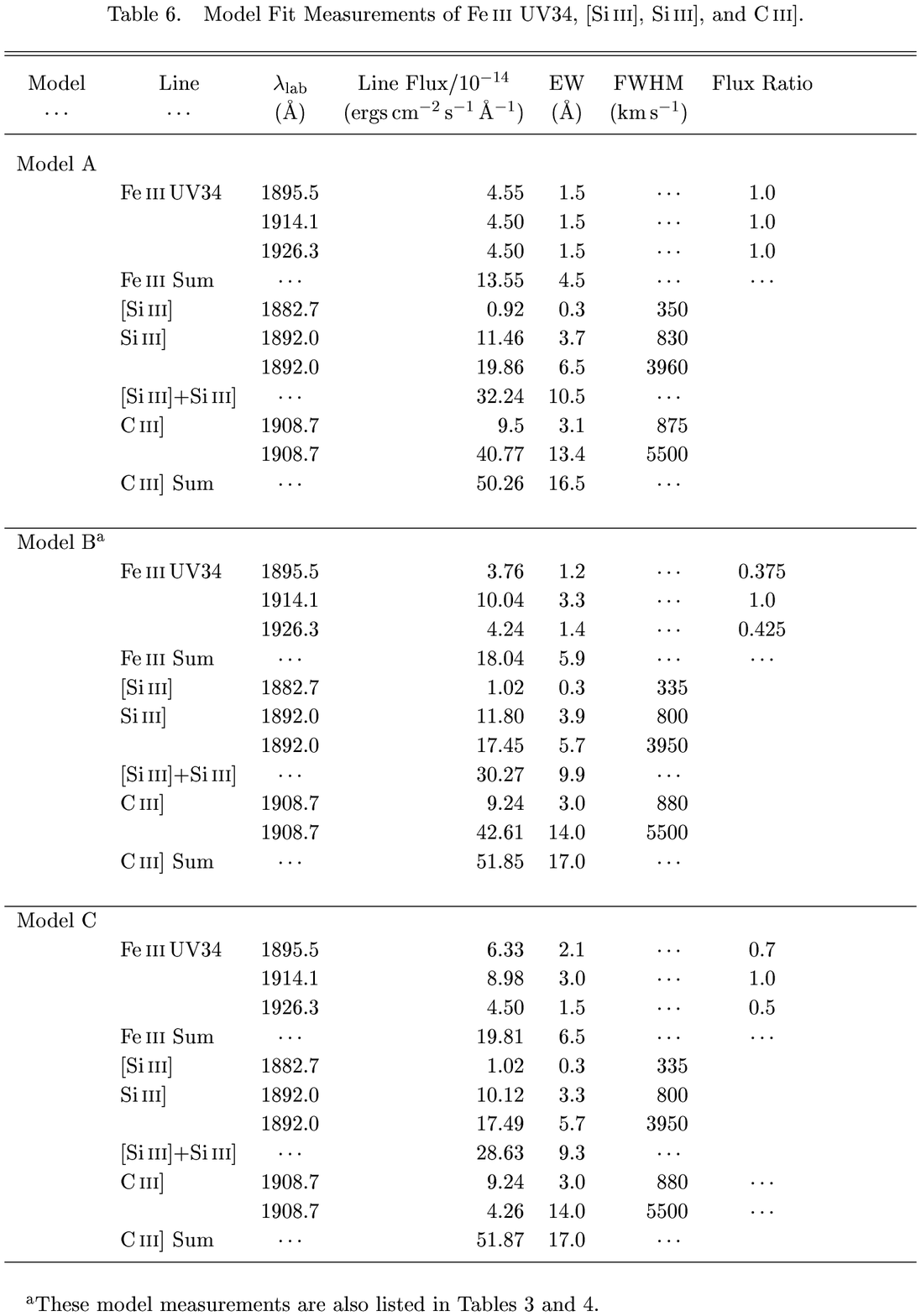}

\end{document}